\documentclass{article}
\usepackage{tabularx}
\usepackage{graphicx}
\usepackage{adjustbox}
\usepackage{verbatim}
\usepackage{booktabs} 
\usepackage[table]{xcolor} 
\usepackage[T1]{fontenc} 
\usepackage{cmap} 
\usepackage{rotating}
\usepackage{pdflscape}
\usepackage{pdflscape}
\usepackage{amsmath}
\usepackage{authblk}
\usepackage[mathlines,left]{lineno}
\usepackage[normalem]{ulem}
\usepackage{color}
\usepackage[utf8]{inputenc}
\usepackage{lmodern}
\usepackage{amsmath,amssymb}
\usepackage{physics}
\usepackage{siunitx}
\usepackage{subcaption}
\usepackage{float}
\usepackage{rotating}
\usepackage{refcount}
\usepackage{xcolor}
\usepackage{footmisc}
\usepackage[a4paper, top=2cm, bottom=2cm]{geometry}
\title{A Universal Telecom-Band Quantum Memory Enabled by Cross-Phase Modulation}

\author[1]{Domenico Compagnini\thanks{These authors equally contributed to this work}}
\author[2]{Noemi Tagliavacche $^*$\thanks{now at EGO Energy S.r.l, Via Felice Romani 9, Genova, Italy}}
\author[1]{Sara Congia}
\author[1]{Andrea Bernardi}
\author[2]{Marco Liscidini}
\author[2]{Matteo Galli\thanks{These authors jointly supervised this work}\thanks{matteo.galli@unipv.it}}
\author[1]{Daniele Bajoni$^\ddagger$\thanks{daniele.bajoni@unipv.it}}

\affil[1]{Dipartimento di Ingegneria Industriale e dell'Informazione, Università degli Studi di Pavia, via Ferrata 5, I-2700 Pavia, Italy}
\affil[2]{Dipartimento di Fisica "Alessandro Volta", Università degli Studi di Pavia, via Bassi 6, I-2700 Pavia, Italy}

\begin{document}

\maketitle

\begin{abstract}
Quantum memories capable of storing and retrieving photonic quantum states on demand are essential for quantum communication, entanglement distribution, and the synchronization of probabilistic quantum processes. All-optical loop-based memories offer broadband operation and compatibility with telecommunication infrastructure, but their performance is often limited by switching losses and the noise introduced by active control mechanisms. Here, we demonstrate a universal all-fiber quantum memory based on two Sagnac interferometers acting as dynamically controlled mirrors through cross-phase modulation. This all-optical switching mechanism enables an input/output transmission as high as 90\% (0.46 dB insertion loss), a round-trip transmission of 94.6\%, and a characteristic storage lifetime of \(18.54\pm0.21\) $\mu$s. The memory exhibits an operational spectral range exceeding 12.5 THz, while keeping noise to negligible levels across the whole telecom C-band. We experimentally demonstrate the preservation of time-bin-encoded heralded photons and time-bin entanglement, as well as compatibility with polarization- and frequency-bin-encoded optical states. Furthermore, we demonstrate the simultaneous storage of over 200 temporal modes and the selective retrieval of independently addressable modes. Our results establish low-noise switching as a promising approach to broadband quantum memories combining high transmission, encoding versatility, and programmable temporal-mode control.

\end{abstract}

\section{Introduction}
\label{Introduction}

Long-distance quantum communication relies on the faithful distribution and manipulation of non-classical states of light across optical fiber networks \cite{Kimble2008, Wehner2018}. In this context, quantum memories play a central role as they enable the synchronization of probabilistic photon sources, the temporal reordering of quantum information, the implementation of multiplexing strategies, and the realization of quantum repeater architectures \cite{Heshami2016}. The basic concept of quantum repeaters was introduced in Ref.~\cite{Briegel1998Repeater} and has since become a cornerstone of quantum network design. Quantum repeaters need quantum memories to synchronize probabilistic photon generation and entanglement swapping events between distant network nodes \cite{Sangouard2011}. To be viable for real-world deployment, such memories must satisfy a set of demanding requirements: they must operate at telecommunication wavelengths ($1550$~nm), exhibit high end-to-end efficiency, support broad bandwidths compatible with high-speed transmission, and operate with high fidelity at room temperature \cite{Ma2020}.

The performance requirements of a quantum memory depend strongly on its intended application. In quantum repeater architectures, storage lifetime is a particularly important parameter, as the memory must preserve quantum information while entanglement is established between distant nodes and successful events are communicated through classical channels. More generally, the ability to retain photonic quantum states for a controlled duration is essential for synchronizing probabilistic processes in quantum communication and distributed quantum information processing \cite{Heshami2016}. For local photonic applications, including the synchronization of heralded photon sources and the generation of multiphoton states, retrieval efficiency, operational bandwidth, and temporal multimode capacity become equally important. These requirements highlight the need to assess quantum memories through several complementary figures of merit, including storage lifetime, storage efficiency, state fidelity, and time-bandwidth product, rather than through a single architecture-specific parameter.

To date, photonic quantum memories can be broadly categorized into two classes: matter-based and all-optical \cite{Lvovsky2009,Heshami2016,Lei2023,Zhou2023,Ma2020}. Matter-based platforms, typically implemented in atomic vapors or rare-earth-ion-doped crystals, achieve extended storage lifetimes by mapping photonic states onto spin-wave excitations. Prominent protocols include electromagnetically induced transparency in atomic ensembles and the atomic frequency comb scheme in rare-earth-doped systems \cite{Bonarota2011,hsiao2018,parigi2015}. Within the telecommunications band, erbium-doped memories have successfully demonstrated the direct storage of entangled photons at 1550~nm \cite{Saglamyurek2015Erbium}. However, the practical deployment of these systems is currently hindered by narrow operational bandwidths, and the strict requirement for cryogenic temperatures \cite{SHINBROUGH2023297}. Current quantum storage technologies face inherent trade-offs between key performance metrics. While atomic vapor ensembles \cite{Afzelius2009,Heshami2016} and rare-earth-ion-doped crystals \cite{zhong2015} offer exceptional coherence times ranging from milliseconds to several hours, they are typically constrained by narrow operating bandwidths (in the MHz to GHz regime). Furthermore, because these platforms predominantly operate at visible or near-infrared wavelengths, interfacing them with standard fiber-optic networks often necessitates quantum frequency conversion, a process that introduces significant experimental complexity and noise \cite{Maring2017}.

In all-optical quantum memories, quantum information remains encoded in propagating optical fields rather than being transferred to stationary material excitations. This approach naturally supports broadband operation and avoids the spectral constraints associated with resonant light-matter interactions. Many practical implementations use optical switching elements capable of injecting and retrieving quantum states with high efficiency in free space optical cavities. Electro-optic switches and conventional loop-and-switch architectures have demonstrated short loop times capable of operating for single photon multiplexing experiments as well as longer loop time via the use of multiple reflections on optical mirrors  \cite{Kaneda2019,Fischer2026,Arnold2022Broadband}. However, increasing the cavity length in these systems to achieve microsecond delays within a reasonable physical footprint is challenging, as it requires either the insertion of optical fibers in the loop (with unavoidable additional losses) \cite{Lee2024FiberLoop,Cheng2025FiberCoupled}, or building highly complex multi-pass geometric folding (such as Herriott cells \cite{Arnold2022Broadband,Guo2026}) with advanced dielectric super-mirrors to mitigate diffraction and scattering. 

An alternative approach is to exploit the optical Kerr nonlinearity through cross-phase modulation, whereby a classical control field induces a phase shift on a quantum signal without requiring direct absorption or electro-optic modulation of the stored photons. When implemented in a fiber based optical cavity, this nonlinear phase shift can be converted into a controllable change in transmission and reflection, providing an all-optical switching mechanism \cite{Hall2011,Bustard2024,BonsmaFisher2024}. Nevertheless, its application to quantum memories requires careful consideration of the efficiency, spectral response, polarization dependence, and noise introduced by the nonlinear interaction, particularly when intense control pulses are employed at the single-photon level.

Beyond single-mode operation, modern high-rate photonic quantum networks increasingly leverage high-dimensional and multi-degree-of-freedom (DoF) encodings, such as time, frequency, and polarization bins \cite{Collins2007}. Effectively routing and synchronizing these states requires a storage medium that is universally compatible across these various spaces. However, a single quantum memory capable of simultaneously supporting time-bin, frequency-bin, and polarization encodings at telecommunications wavelengths has hitherto remained elusive.

Here, we report a fully fiber-integrated quantum memory operating within the telecommunications C-band that overcomes these limitations. The architecture utilizes a fiber cavity bounded by two Sagnac interferometers functioning as dynamically switchable mirrors. By exploiting the ultrafast Kerr nonlinearity in the optical fibers, the reflectivity of each Sagnac interferometer can be controlled on demand. This enables the deterministic, high-efficiency, and high-fidelity capture and release of single photons at discrete temporal intervals dictated by the cavity round-trip time. All operations are implemented entirely within a fiber-based platform, bypassing the need for free-space optics, cryogenic cooling, or narrowband atomic transitions.

We benchmark the performance of this quantum memory at the single-photon level, measuring a remarkably low input-output insertion loss of only 0.46~dB (90\% transmission). With the cavity round-trip time configured to 1.03~$\mu$s, we observe a characteristic storage lifetime of $\sim$18~$\mu$s, corresponding to a round-trip loss of just 0.24~dB. The memory faithfully preserves the spectral and temporal profiles of the stored photons, supporting the storage of qubits encoded across multiple degrees of freedom - time-bin, frequency-bin and polarization states - with quantum state fidelities exceeding 99\% for all configurations, thus proving its universality. 

The practical utility and high-capacity performance of this platform are fundamentally captured by its exceptionally large Time-Bandwidth Product (TBP). We demonstrate a TBP of $2.2 \times 10^8$, outperforming most state-of-the-art matter-based systems by several orders of magnitude. This remarkable figure of merit explicitly demonstrates that our all-optical approach circumvents the traditional trade-off between operational bandwidth and storage lifetime. Consequently, for broadband photonic networks, this enables high-capacity time-domain multiplexing with the simultaneous storage of more than 200 independent temporal modes under near-unity transparency. By pairing the inherent bandwidth and wavelength advantages of fiber optics with the high efficiency required for quantum protocols, these results establish a scalable pathway for high-speed quantum routing and synchronization, particularly for intranode multiplexing architectures.

Most significantly, we explicitly demonstrate the robust preservation of quantum entanglement using a memory-node configuration, where one photon of an entangled pair is dynamically stored within the memory while its partner freely propagates through a short fiber stretch. The retrieved photons maintain high-quality non-local correlations, yielding quantum interference visibilities of up to 98.8\%. By simultaneously achieving ultra-low loss, deterministic all-optical control, telecommunications compatibility, multi-degree-of-freedom versatility, and unequivocal entanglement preservation within a fully fiber-integrated architecture, this work establishes a truly universal quantum memory optimized for scalable quantum communication networks.

\begin{figure}
    \centering
    \includegraphics[width=1.0\linewidth]{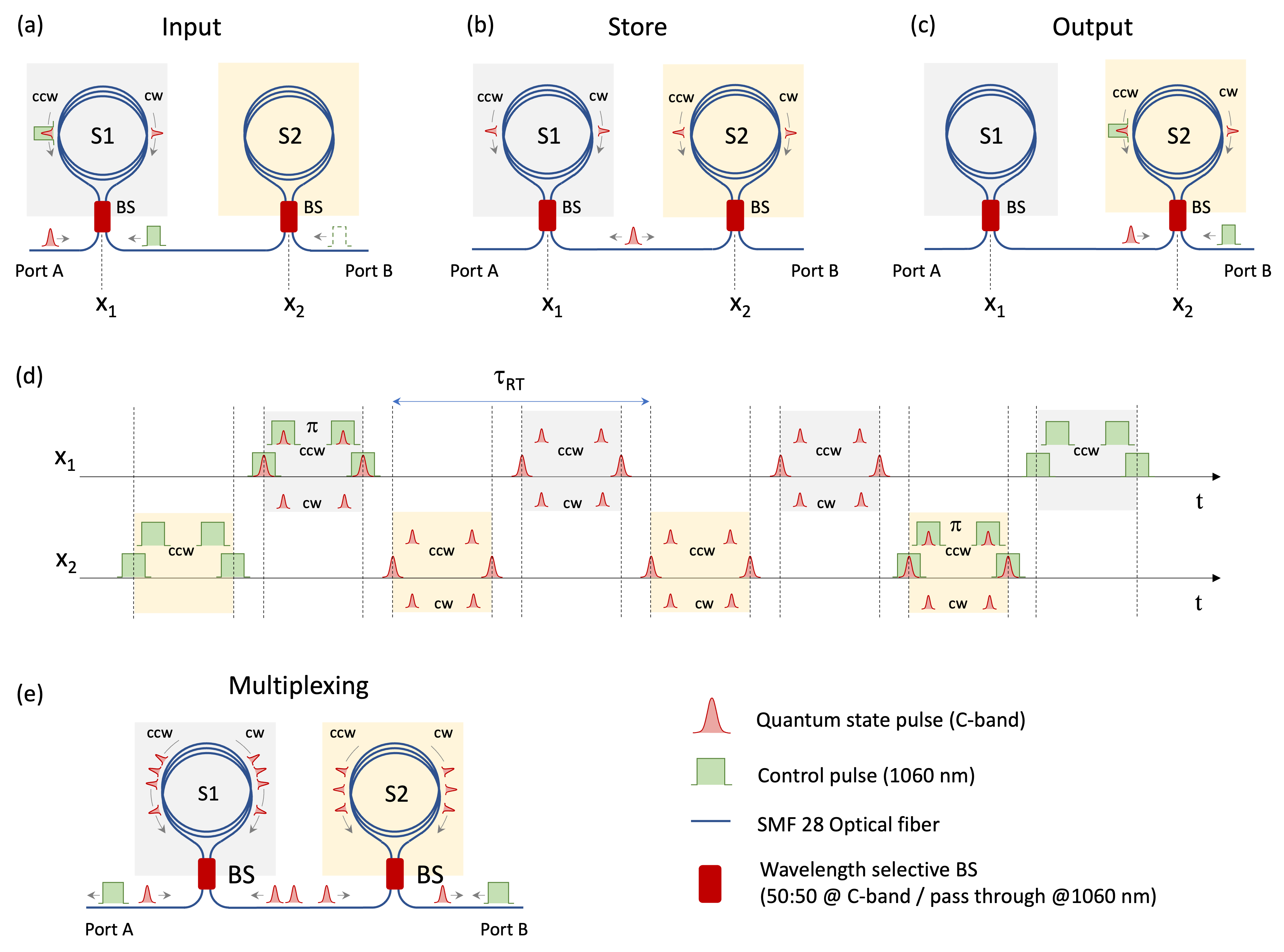}
    \caption{
    \textbf{Architecture and operational principle of the universal all-fiber quantum memory.} (a) Input operation: an incoming quantum signal pulse and a control laser pulse enter the memory constituted by Sagnac interferometers S1 and S2 through port A and port B, respectively. The beam-splitters (BS) at positions x$_1$ and x$_2$ divide the quantum pulse equally but do not split the control pulse which propagates only in the counterclockwise (CCW) direction. The pulses are precisely timed to overlap at the BS at x$_1$. The CCW component of the quantum pulse co-propagates with the control pulse in S1, acquiring a nonlinear phase shift of $\pi$ via cross-phase modulation. The subsequent constructive interference of the clockwise (CW) and CCW components renders S1 transmissive, dynamically injecting the quantum pulse into the cavity. (b) Storage: in the absence of control pulses, S1 and S2 act as highly reflective mirrors, trapping the quantum pulse within the low-loss fiber loop. (c) Output operation: a second control pulse is timed to overlap with the circulating quantum pulse at the BS at x$_2$. The identical physical mechanism used for injection renders S2 transmissive, deterministically extracting the quantum pulse through port B. (d) Timing diagram detailing the arrival of the signal and control pulses at positions x$_1$ and x$_2$ throughout a complete input-storage-output sequence. In this illustrative example, the quantum pulse is stored for three round-trips (loops). The gray and yellow shaded regions represent propagation within S1 and S2, respectively. (e) Schematic representation of multiplexing capability.}
    
    \label{fig:principle}
\end{figure}

\section{Working principle and linear characterization}
\label{sec:working_principle}

The working principle of the proposed quantum memory is illustrated in Fig.~\ref{fig:principle}. The core of the device consists of two Sagnac interferometers constructed from standard SMF-28 Ultra optical fiber each with a length of 100 m and connected as shown in Fig. \ref{fig:principle}. Both interferometers include an in-line polarization controller adjusted to ensure that they function as highly reflective mirrors for signal photons in the telecommunications C-band (around 1550~nm), with a measured reflectivity of 
(97.5 $\pm$ 0.2)\%. A crucial feature of this architecture is the wavelength-dependent behavior of the beamsplitters that close the Sagnac interferometers. Specifically, they operate as 50:50 splitters for telecom-band wavelengths - enabling the loop's mirror-like behavior - while acting as completely transparent pass-through elements for the control pulses at 1060~nm.

The operational protocol of the quantum memory proceeds as follows: an intense optical control pulse at 1060~nm enters the system via Port B (Fig.~\ref{fig:principle}(a)). Subsequently, a signal pulse encoding a quantum state enters the memory through Port A. The arrival times are synchronized such that the two pulses temporally overlap at the wavelength-dependent beamsplitter of the first Sagnac interferometer (Sagnac~1). This beamsplitter divides the quantum state pulse into counter-propagating (clockwise and counter-clockwise) components within the loop, whereas the control pulse is routed exclusively in the counter-clockwise direction (Fig.~\ref{fig:principle}(a)). As described in the Methods section, the optical and geometrical properties of our memory ensure that the CCW part of the quantum state pulse and the control pulse remain fully overlapped while propagating inside the Sagnac interferometer.

Exploiting the $\chi^{(3)}$ nonlinearity of the optical fiber, the control pulse imparts a phase shift exclusively to the co-propagating component of the quantum state pulse via cross-phase modulation (XPM)  \cite{Agrawal2023}. The XPM model used to estimate the required control-pulse power, including the polarization-insensitive configuration based on two orthogonally polarized control fields, is detailed in Supplementary Information, Sec.~S1.
By fine tuning the total peak power of the combined control field to 13.0~W,
the induced phase shift can be set to exactly $\pi$ (see Supplementary Information, Sec. S2). Under these conditions, Sagnac~1 becomes highly transmissive for the quantum signal, allowing it to enter the memory cavity, while the 1060~nm control pulse exits via Port A. The quantum state pulse is thus trapped and dynamically stored between the two highly reflective Sagnac interferometers (Fig. \ref{fig:principle} (b)). Retrieval of the stored state is achieved after an integer number of round-trips (loops) by injecting a second control pulse, which is timed to overlap with the circulating quantum state pulse in the second interferometer (Sagnac~2), as shown in  Fig.~\ref{fig:principle}(c). This switches the state of Sagnac~2, ejecting the quantum state pulse through Port B. Because this architecture relies on a ``loop-and-switch'' mechanism, the quantum state can only be retrieved at discrete temporal intervals dictated by the cavity round-trip time, hereafter referred to as loops. This operational timing can be engineered by adjusting the lengths of the Sagnac interferometers and/or the length of the connecting fiber stretch. In the present implementation, we use $\sim$ 100~m loops for the two Sagnac interferometers and a $\sim$ 5 m connecting fiber stretch, which sets the fundamental round-trip time to ($1030\pm0.05$)~ns.

The use of XPM for injection and retrieval operations, in combination with wavelength-dependent couplers, ensures a high overall efficiency for the quantum memory. Indeed, a direct measurement of the switching efficiency shows that the XPM process introduces negligible optical loss, yielding near-unity efficiency within experimental error, as reported in Section S2 of the Supplementary Information. Our architecture obviates the need for active electro-optic switching mechanisms \cite{Cardea2021} or intra-loop wavelength-division multiplexers \cite{maclean2018}, both of which would introduce significant insertion losses that degrade quantum memory performance.

The input-output efficiency of our device  $\eta_T$ is defined as the fraction of optical power successfully routed into and subsequently out of the cavity over the incident optical power, and approaches 90\% in our system (see eq. \ref{eq:efficiency} below). This value accounts for the losses of all the constituent optical components forming the two Sagnac interferometers, the connecting fiber stretch, as well as the external optical filters required to inject and extract the control pulses. The losses introduced by the optical filters used to spectrally clean the quantum signal at the output are also included in $\eta_T$. The value of 90\% therefore represents the net transmission of the quantum memory.
We benchmark this efficiency by referencing the memory's transmission to a reference bypass consisting of a 2~m segment of SMF-28 Ultra fiber. Because this bypass circumvents both Sagnac interferometers and all associated spectral filters, this normalization directly includes their cumulative losses into our system-level efficiency metrics.

These characterization experiments are performed using a broadband superluminescent diode (SLD) by injecting light into the memory and monitoring the output sequentially from the 1st to the 20th round-trip loop, following the procedure outlined in Fig. \ref{fig:principle}. A spectrometer coupled to a liquid-nitrogen-cooled CCD camera is employed for high-sensitivity detection. The resulting transmission spectra, depicted in Fig.~\ref{fig:linear}(a), can be used to simultaneously evaluate the memory's operational bandwidth and storage lifetime. The recorded spectra exhibit a broadband profile centered near 191~THz ($\sim$1570~nm) that decays exponentially with an increasing number of round-trips. 

At each frequency component, the transmission data are fitted using a zero-offset exponential decay function:
\begin{equation}
T(\#) = \eta_T \exp\left( -\frac{\#}{N_{C}} \right),
\label{eq:efficiency}
\end{equation}
where $\#$ denotes the extraction round-trip number, $\eta_T$ is the net input-output efficiency, and $N_{C}$ is the $1/e$ decay constant (the number of loops after which the transmission drops to $T(N_C)=\eta_T/e$). The corresponding storage lifetime is defined as $\tau = N_C \times 1030$~ns. The extracted parameters are plotted in Figs.~\ref{fig:linear}(b) and \ref{fig:linear}(c) for the efficiency and lifetime, respectively, where the shaded regions indicate the fitting uncertainties from the least-squares analysis.

The spectral range of this measurement is bounded between $\sim$200~THz ($\sim$1500~nm) by the edge-pass filters employed, and $\sim$187.5~THz ($\sim$1600~nm) by the combined response of the spectrograph and CCD camera. The spectral profile of the efficiency curve is primarily governed by the insertion loss of these optical filters and the alignment tolerances of the fiber U-bench used for their housing. Remarkably, the net efficiency $\eta_T$ peaks above 90\% at approximately 190.5~THz, and the 3~dB bandwidth exceeds the entire 12.5~THz investigated spectral range. The storage lifetime extracted from the fit of eq. \ref{eq:efficiency} to the experimental data, exceeds 18~$\mu$s throughout the entire telecommunications C-band. This lifetime is predominantly limited by intra-cavity propagation losses and to a round-trip loss of $\sim$0.24~dB per loop (a comprehensive breakdown of these losses is provided in the Supplementary section S5).

\subsection{Heralded Single-Photon Storage}
\label{Heralded Single-Photon Storage}

To validate the performance of our memory in the quantum regime, we conducted characterization experiments using a heralded single-photon source. Photon pairs were generated via spontaneous four-wave mixing (SFWM) in a silicon waveguide, as detailed in the Methods section. The signal and idler photons were subsequently separated using dense wavelength-division multiplexing (DWDM) filters. Detection of the idler photon by a Superconducting Nanowire Single-Photon Detector (SNSPD) provided a herald for a single-photon state in the signal channel, which was spectrally aligned with the International Telecommunication Union (ITU) Channel 21 ($\sim$1560.2~nm). 

The heralded single photons were routed into the memory and, upon on-demand retrieval, detected using a second SNSPD. This measurement protocol was repeated for successive round-trips, and the resulting coincidence counts were normalized to the transmission of the reference fiber bypass described previously. 

\begin{figure}
    \centering
    \includegraphics[width=1.0\textwidth]{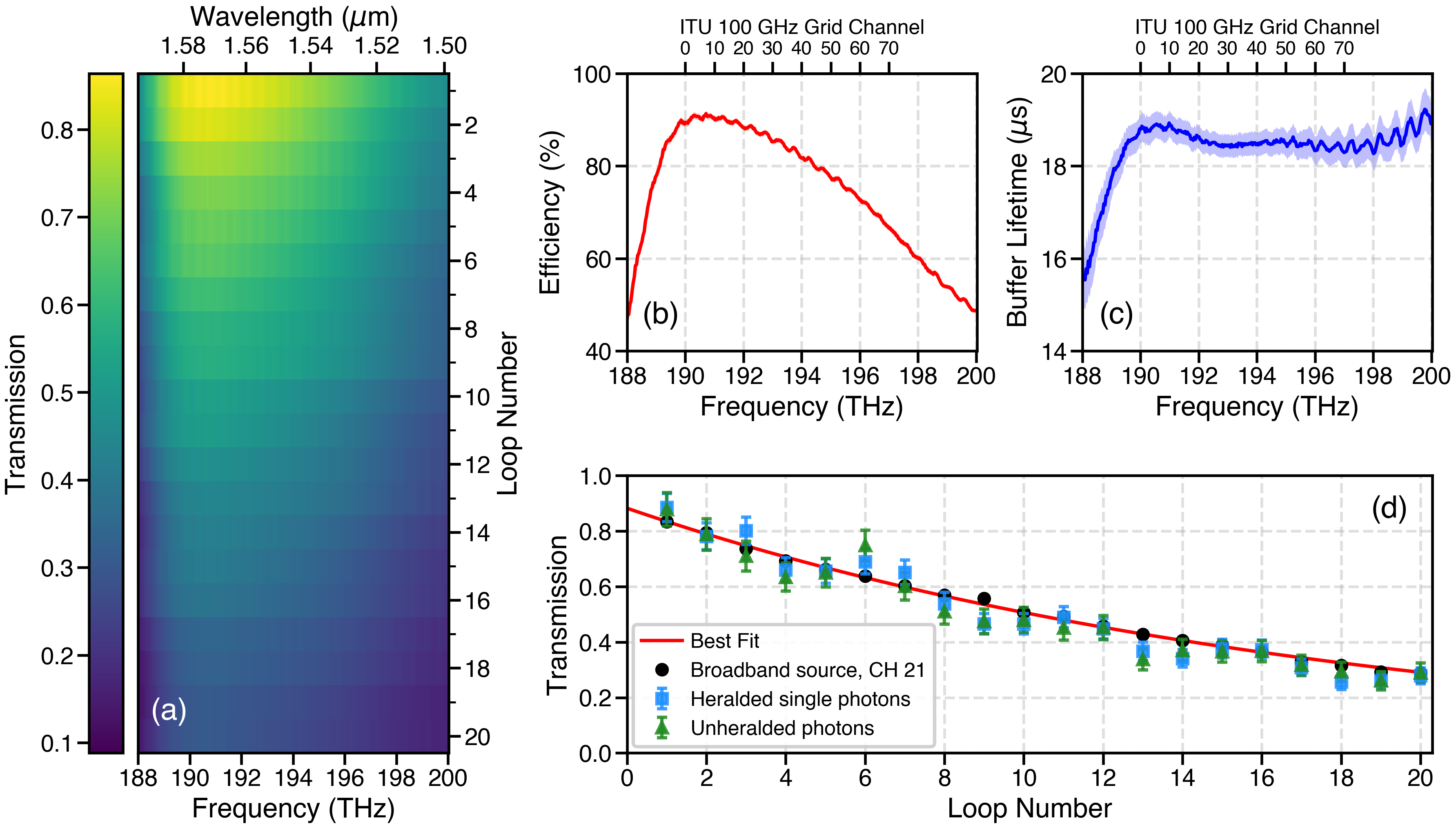}
    \caption{\textbf{Linear characterization of the quantum memory.} (a) Measured transmission (color scale) as a function of optical frequency and loop number during extraction, obtained using a broadband light source. (b) Extracted memory efficiency and (c) memory lifetime as a function of optical frequency, derived from the dataset in panel (a). (d) Measured transmission for ITU channel 21 as a function of the loop number, comparing data from the broadband source, heralded single photons, and unheralded single photons. The solid line represents an exponential fit to the data. }
    \label{fig:linear}
\end{figure}

The experimental results are presented in Fig.~\ref{fig:linear}(d), alongside the storage behavior of unheralded photons (where the source effectively serves as a weak photoluminescence background). The single-photon data exhibit excellent agreement with the classical characterization obtained using the broadband source. A least-squares exponential fit to the heralded data yields a $1/e$ decay constant of $18.0 \pm 0.2$ loops, corresponding to a characteristic storage lifetime of ($18.54 \pm 0.21)~\mu$s. Notably, owing to the high net efficiency of the architecture, approximately 50\% of the injected single photons are successfully retrieved after $10~\mu$s of storage, with 30\% remaining retrievable even after $20~\mu$s.

\begin{figure}
    \centering
    \includegraphics[width=\columnwidth]{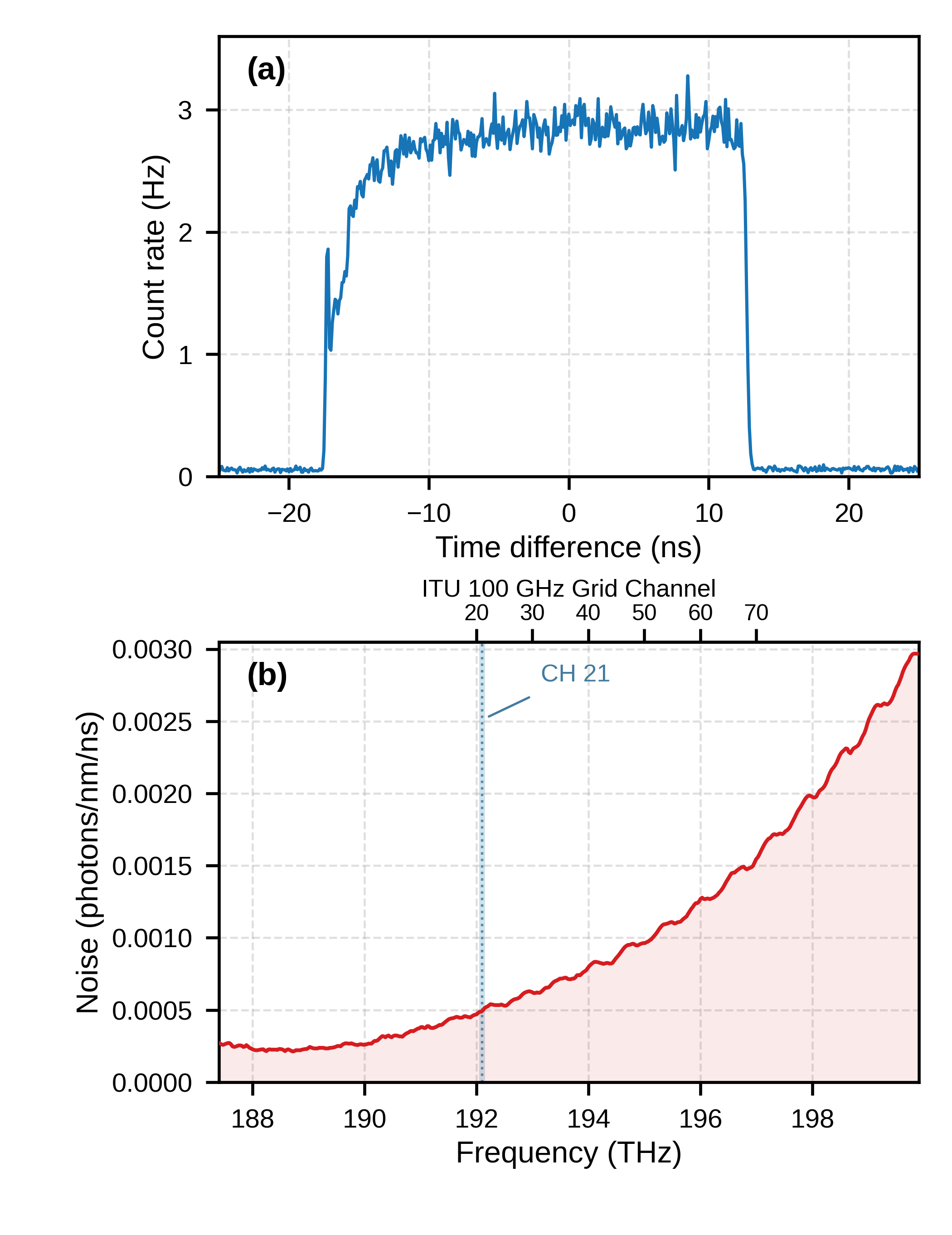}
    \caption{\textbf{Noise characterization.} (a) Noise induced by 30 ns long control pulses at 1060 nm within the 100 GHz bandwidth of ITU channel 21. The integration time was 300 s and the repetition rate 100 KHz. (b) Spectrum of the noise in the same spectral range as the data of Fig. \ref{fig:linear}. The dotted line represents ITU CH 21.}
    \label{fig:Ramannoise}
\end{figure}

A potential concern regarding the intense optical control pulses used to drive the cross-phase modulation switching mechanism is the generation of parasitic noise via spontaneous Raman scattering or secondary photoluminescence within the silica fiber. 
In order to characterize this noise contribution, we operated the quantum memory without injecting any quantum signal, performing a time-resolved characterization with an input/output rate of 100 kHz while filtering the output with a 100 GHz spectral filter centered at ITU CH21 before the detectors (with 88\% transmission). The results, acquired with an integration time of 300 s and a time resolution of 100 ps, are shown in Fig. \ref{fig:Ramannoise} (a). A weak optical signal was detectable in temporal correspondence with the arrival of the control pulses: the average counts in Fig. \ref{fig:Ramannoise} (a) on the plateau were $850.6 \pm 34.8$. After accounting for the filtering losses, detector efficiency (82\%), repetition rate and integration time, this noise level corresponds to $(4.9\pm0.2)\times 10^{-4}$ photons per nm of bandwidth per ns. We also studied the wavelength dependency of this small noise contribution, as reported in Fig. \ref{fig:Ramannoise} (b), finding that it remains below $\sim 1.5 \times 10^{-3}$~photons~nm$^{-1}$~ns$^{-1}$ across the entire telecommunications C-band. This exceptionally low noise floor is a direct consequence of our memory design: by selecting a control wavelength near 1060~nm, the energy separation from the telecom band ($\sim 3000~\text{cm}^{-1}$) places the quantum signals far beyond the dominant high-order Raman transitions of amorphous silica \cite{Mamun2018}. Consequently, as reported below, the memory preserves an excellent signal-to-noise ratio, single-photon purity and preserves interference in entanglement experiments without the need for aggressive, lossy spectral filtering or complex temporal gating.

The spectral dependence of the noise was measured by operating the memory without injecting any signal as described above but using the spectrometer and the CCD camera instead of the SNSPD. Results are shown in Fig. \ref{fig:Ramannoise} (b). The noise remained below $1.5\times 10^{-3}$ across the entire telecom C-band (ITU channels 16 to 60) and below $3\times 10^{-3}$ over the whole investigated range. Such low levels are compatible with quantum experiments over the entire operation bandwidth of the quantum memory. They were not expected to limit the single-photon fidelity, entanglement visibility, or multiplexing measurements reported in this work. Accordingly, no Raman-noise subtraction was required in the  quantum experiments data reported below.

\section{Single qubit fidelity and preservation of entanglement}
\label{Single qubit fidelity and preservation entanglement.}

\begin{figure}
    \centering
    \includegraphics[width=1\columnwidth]{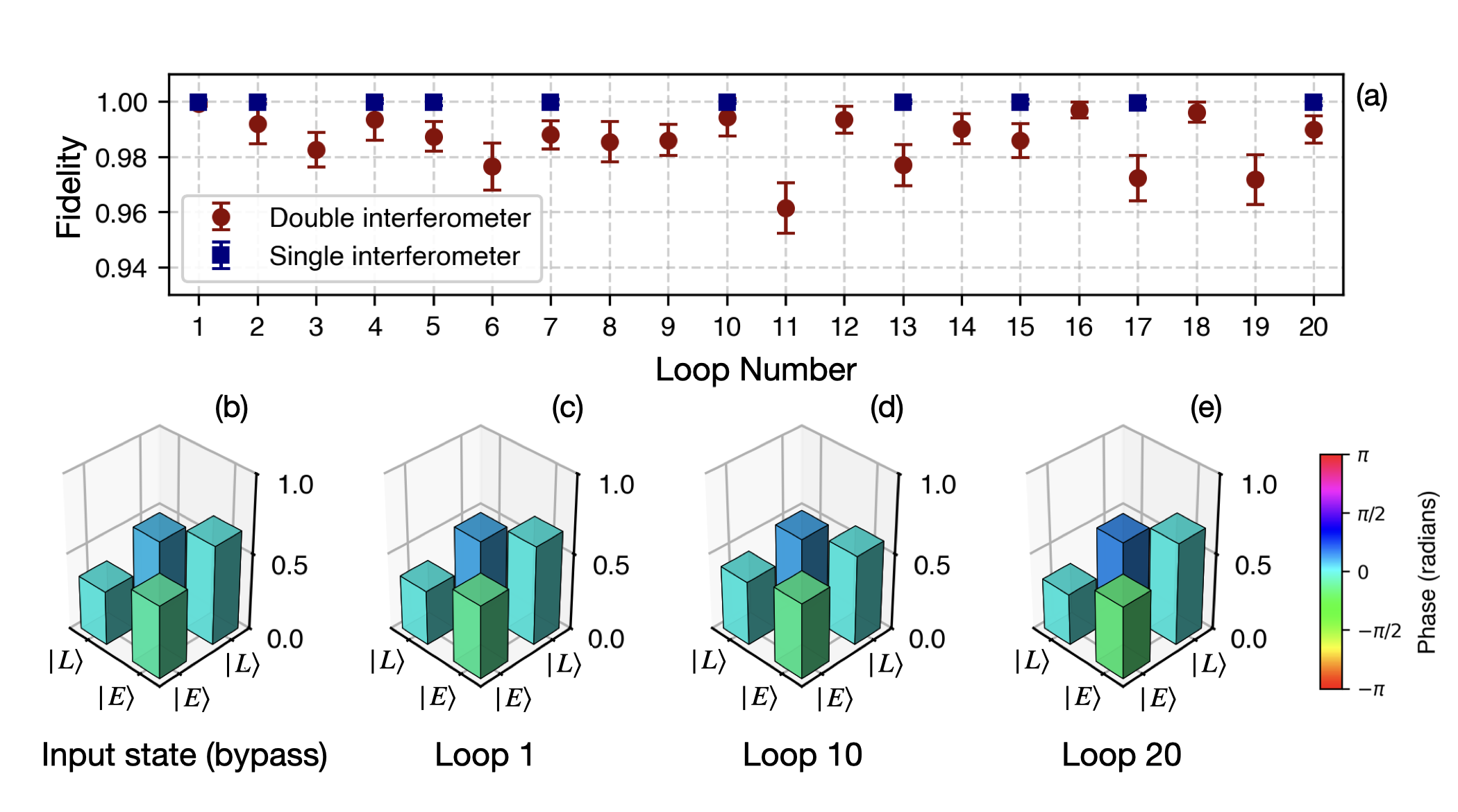}
    \caption{\textbf{Preservation of time-bin encoded photons.} (a) Measured fidelity to the initial state as a function of storage loop number (from 1 to 20) for an arbitrary time-bin input state (red dots) and a self-referenced $|+\rangle$ state (blue dots). (b)-(e) Reconstructed density matrices obtained via quantum state tomography for the initial input state (b), and the retrieved states from the memory output after 1 (c), 10 (d), and 20 (e) storage loops.}
    \label{fig:fidelity_time_bins}
\end{figure}

\subsection{Heralded Single-Qubit Storage}

The capacity of a quantum memory to faithfully preserve input quantum states is a fundamental requirement for quantum communication and networking protocols. To evaluate the state-preservation performance of our quantum memory, we utilized the same heralded single-photon source described in section \ref{Heralded Single-Photon Storage} paired with two unbalanced Mach-Zehnder interferometers (MZIs) to encode and analyze the quantum states \cite{Singh2025}. By controlling the relative phase and amplitude within the preparation MZI, we generated the time-bin qubit state $\cos(\vartheta)|E\rangle + e^{i\varphi}\sin(\vartheta)|L\rangle$, where $|E\rangle$ and $|L\rangle$ represent the early and late temporal basis states, respectively. For this characterization, the state parameters were arbitrarily configured to $\vartheta = 53^\circ$ and $\varphi = 15^\circ$. 

To establish the baseline fidelity of the input state, the quantum memory was initially bypassed using the 2~m reference fiber. The resulting state fidelities following storage are displayed as red circles in Fig.~\ref{fig:fidelity_time_bins}(a). The quantum state fidelities remain remarkably high, exceeding 95\% across all measured loops, with no observable evidence of state degradation as a function of storage lifetime. 

In this cross-interferometer configuration, the measured fidelity was ultimately bounded not by the performance of the quantum memory, but by residual phase instabilities between the physically distinct encoding and analyzing interferometers (despite active feedback stabilization applied to both systems). To circumvent this technical limitation and isolate the intrinsic contribution of the quantum memory, we transitioned to a self-referencing architecture that leverages a single physical MZI for both state preparation and analysis. This configuration implements the state $|+\rangle = \left(|E\rangle + |L\rangle\right)/\sqrt{2}$, where ambient thermal phase fluctuations are automatically common-mode rejected because the encoding and decoding paths are identical. 

The results from this self-referencing benchmark are plotted as blue circles in Fig.~\ref{fig:fidelity_time_bins}(a). Under these conditions, the state fidelities remain strictly greater than 99.9\% for all storage loops. This exceptional performance indicates that the memory introduces negligible decoherence or distortion to the time-bin encoded qubits. A comprehensive analysis and the complete dataset for all quantum state fidelity configurations are provided in the Supplementary Information Sec.~S6.1. The same experiments based on self referencing were also carried out using broadband (20 nm) photons and are also reported in the same section of the Supplementary Information. For these pulses the TBP was 4.63 $\times$ 10$^7$.

\subsection{Entangled photon pairs}

A fundamental benchmark for any quantum memory is its capability to preserve multi-photon quantum coherence, specifically, the entanglement between a stored photon and an external idler photon. We evaluate this performance by generating time-bin entangled photon pairs \cite{Marcikic2002} using the same pump laser and silicon waveguide configuration as in the preceding measurements. The experimental setup utilizes the same pair of Mach-Zehnder interferometers; however, in this configuration, the preparation interferometer is positioned to encode the time-bin states directly into the pump beam. 

\begin{figure}
    \centering
    \includegraphics[width=1\textwidth]{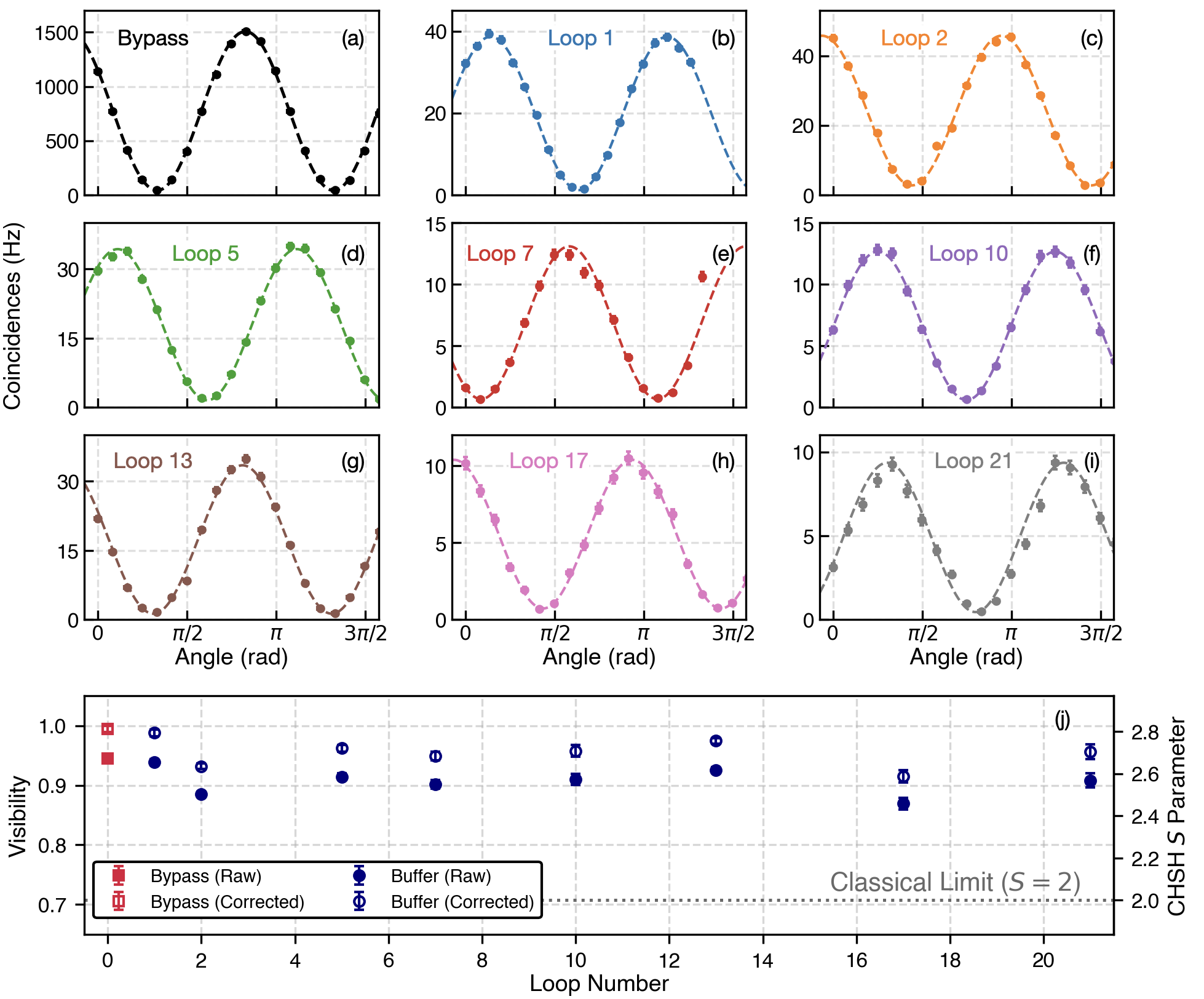}
    \caption{\textbf{Time-bin entanglement.} Measured quantum interference curves (without accidental subtraction) obtained after sending the signal photon through the fiber bypass (a) and through the quantum memory for several loops (b) - (i). Dashed lines represent best fits to the experimental data. (j) Visibility and S parameter extracted from the best fits of panels (a) to (i): raw data (solid points) and corrected for the visibility of the analyzing interferometer (open points).}
    \label{fig:entanglement}
\end{figure}

The resulting experimental data are illustrated in Fig.~\ref{fig:entanglement}, where highly distinct quantum interference patterns are observed for both the reference bypass and the stored photons. Owing to the exceptionally low insertion loss and high-capacity temporal multiplexing capabilities of our architecture (detailed in the subsequent section), the coincidence rates remain remarkably high, ranging from several Hz to tens of Hz even when the signal photon is actively stored within the memory cavity.

The extracted visibilities and the corresponding Clauser-Horne-Shimony-Holt (CHSH) $S$-parameters are plotted in Fig.~\ref{fig:entanglement}(j) and summarized in Table~\ref{tab:visibilities}. Time-bin entanglement remains preserved for all investigated storage times, with every measured visibility yielding a CHSH parameter significantly above the classical bound, although moderate variations in visibility are observed between different round trips. The extracted visibilities correspond to CHSH parameters above the classical bound with the inequality being violated by more than 20 standard deviations in all cases. The data presented and fitted in Fig.~\ref{fig:entanglement} are entirely raw and uncorrected, with neither background nor accidental-coincidence subtraction applied, and without correcting for the visibility of the analyzing MZI. For completeness, we provide the metrics both in their raw form and after correcting just for the intrinsic visibility of the analyzing MZI (measured to be 95\% using a continuous-wave laser). We attribute the marginally higher $S$-parameter observed in the baseline bypass configuration simply to the higher raw coincidence rate, which naturally yields an elevated signal-to-noise ratio. Representative two-dimensional coincidence histograms and the visibility-extraction procedure are reported in Supplementary Information Sec.~S6.4.

\begin{table}[hbt!]
\centering
\caption{Raw and corrected Visibility, CHSH $S$ parameter, and violation significance extracted from sinusoidal fits.}
\label{tab:visibilities}
\renewcommand{\arraystretch}{1.2}
\begin{tabular}{@{} l c c c c @{}}
\toprule
\textbf{Loop} & \textbf{Raw $V$} & \textbf{Corr. $V$} & \textbf{Corr. $S$ Parameter} & \textbf{Violation (Corr.)} \\
\midrule
0 & $0.945 \pm 0.005$ & $0.994 \pm 0.005$ & $2.812 \pm 0.015$ & $55.4\sigma$ \\
1 & $0.939 \pm 0.005$ & $0.988 \pm 0.005$ & $2.794 \pm 0.016$ & $51.1\sigma$ \\
2 & $0.885 \pm 0.004$ & $0.932 \pm 0.005$ & $2.635 \pm 0.013$ & $48.0\sigma$ \\
5 & $0.914 \pm 0.006$ & $0.962 \pm 0.006$ & $2.722 \pm 0.017$ & $43.2\sigma$ \\
7 & $0.902 \pm 0.007$ & $0.949 \pm 0.008$ & $2.684 \pm 0.021$ & $32.1\sigma$ \\
10 & $0.910 \pm 0.009$ & $0.958 \pm 0.010$ & $2.709 \pm 0.028$ & $25.3\sigma$ \\
13 & $0.926 \pm 0.005$ & $0.974 \pm 0.005$ & $2.756 \pm 0.014$ & $53.2\sigma$ \\
17 & $0.869 \pm 0.010$ & $0.915 \pm 0.010$ & $2.588 \pm 0.029$ & $20.4\sigma$ \\
21 & $0.908 \pm 0.012$ & $0.956 \pm 0.012$ & $2.704 \pm 0.035$ & $20.0\sigma$ \\
\bottomrule
\end{tabular}
\end{table}

\section{Multimodality}
\label{Multimodality}

\begin{figure}
    \centering
    \includegraphics[width=1\textwidth]{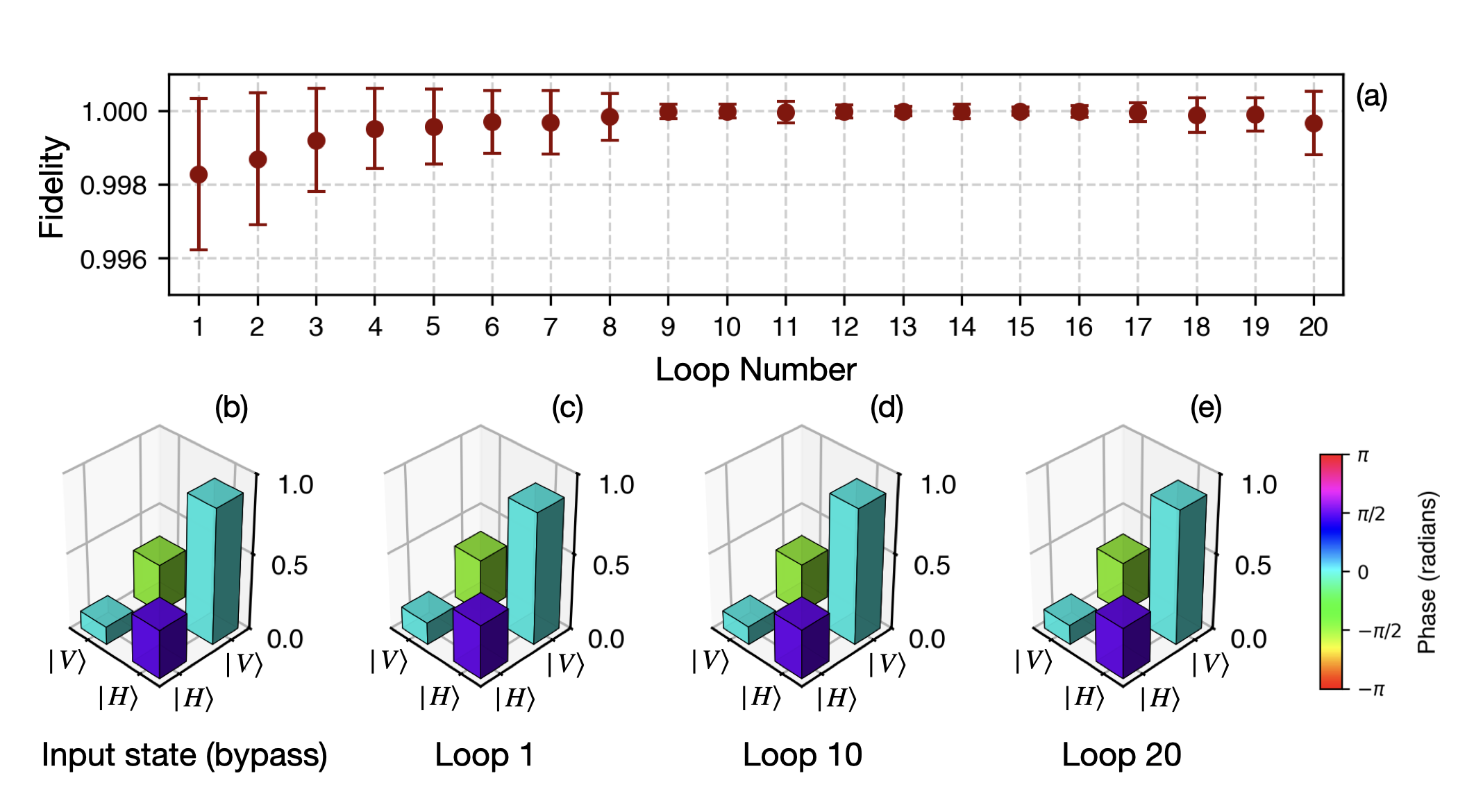}
    \caption{\textbf{Preservation of polarization encoded photons.} (a) Measured fidelity to the initial state as a function of storage loop number (from 1 to 20) for an arbitrary polarization input state. (b)-(e) Reconstructed density matrices obtained via quantum state tomography for the initial input state (b), and the retrieved states from the memory output after 1 (c), 10 (d), and 20 (e) storage loops.}
    \label{fig:fidelity_pol}
\end{figure}

The capability of a quantum storage device to reliably support multi-degree-of-freedom (DoF) encodings is a highly advantageous feature, as it enables the manipulation of high-dimensional quantum states and ensures compatibility with a diverse range of quantum communication protocols. Here, we demonstrate that in addition to time-bin qubits, our quantum memory natively supports both polarization and frequency-bin encodings, thereby encompassing all the degrees of freedom easily routed through standard single-mode fiber-optic infrastructure. These two additional characterizations were conducted using weak coherent states (attenuated laser pulses).

For polarization encoding, a half-wave plate (HWP) and a quarter-wave plate (QWP) were positioned before the memory input to prepare the state $\cos(\vartheta)|H\rangle + e^{i\varphi}\sin(\vartheta)|V\rangle$, where $|H\rangle$ and $|V\rangle$ denote the horizontal and vertical polarization states, respectively. For this benchmark, the state parameters were arbitrarily configured to $\vartheta = 70^\circ$ and $\varphi = -90^\circ$. At the memory output, a state-analysis stage consisting of a QWP, an HWP, and a linear polarizer was employed. Because the residual birefringence inherent to standard optical fibers induces an incremental polarization rotation during each round-trip, an in-line fiber polarization controller was integrated between the two Sagnac interferometers to dynamically compensate for this effect. The insertion of this controller extended the fundamental cavity round-trip time from 1030~ns to 1062~ns. Utilizing this approach, the cumulative polarization rotation could be compensated to an accuracy of better than $2^\circ$ for both $\vartheta$ and $\varphi$. 
The resulting state fidelities and reconstructed states are illustrated in Fig.~\ref{fig:fidelity_pol}(a). Notably, no observable state degradation is detected within the polarization degree of freedom even well beyond a single storage lifetime. The complete datasets for this polarization state and for two additional input-state configurations ($\vartheta = 60^\circ$ and $\varphi = -90^\circ$, $\vartheta = 45^\circ$ and $\varphi = -45^\circ$) are provided in Supplementary Information, Sec.~S6.2, showing fidelities exceeding 99\% in all cases.

\begin{figure}
    \centering
    \includegraphics[width=1\textwidth]{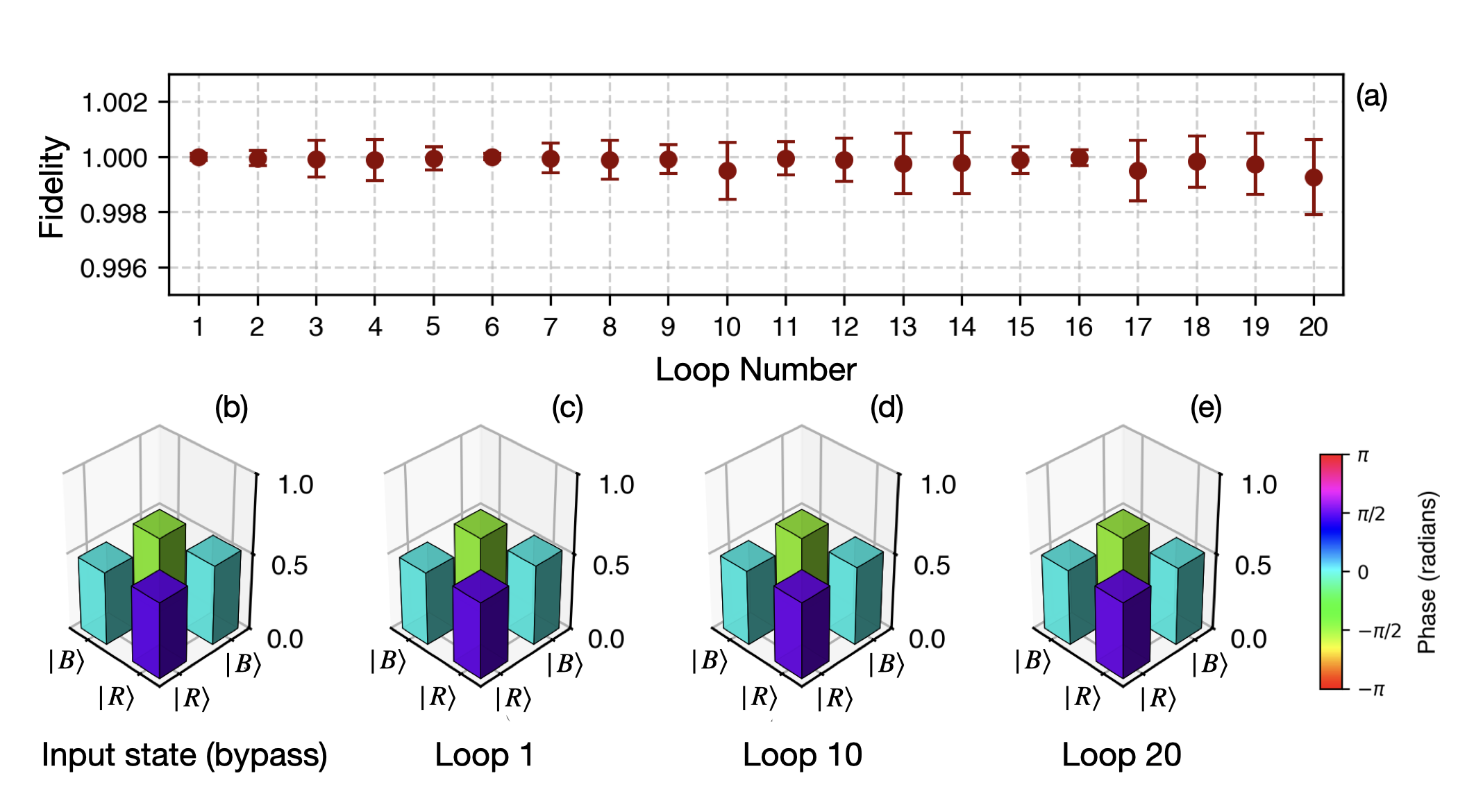}
    \caption{\textbf{Preservation of frequency-bin encoded photons.} (a) Measured fidelity to the initial state as a function of storage loop number (from 1 to 20) for an arbitrary frequency-bin input state. (b)–(e) Reconstructed density matrices obtained via quantum state tomography for the initial input state (b), and the retrieved states from the memory output after 1 (c), 10 (d), and 20 (e) storage loops.}
    \label{fig:fidelity_fbin}
\end{figure}

For the frequency-bin encoding configurations, an electro-optic modulator (EOM) was used to prepare the qubit state $\cos(\vartheta)|R\rangle + e^{i\varphi}\sin(\vartheta)|B\rangle$, where $|R\rangle$ and $|B\rangle$ denote two adjacent optical frequency components separated by \(f_{\mathrm{RF}}=7~\mathrm{GHz}\). The EOM was driven by a microwave tone generator at a modulation frequency of $f_{\mathrm{RF}}$, and the state parameters were chosen to be $\vartheta = 46^\circ$ and $\varphi = -90^\circ$. As with the previous experiments, the stored states were benchmarked against the 2~m reference fiber bypass. The characterization results are summarized in Fig.~\ref{fig:fidelity_fbin}, demonstrating an equally excellent preservation of state fidelity within this third encoding domain. The complete frequency-bin dataset is reported in Supplementary Information Sec.~S6.3.

\section{ Temporal multimode storage capacity}
\label{Multiplexing}

\begin{figure}
    \centering
    \includegraphics[width=1\linewidth]{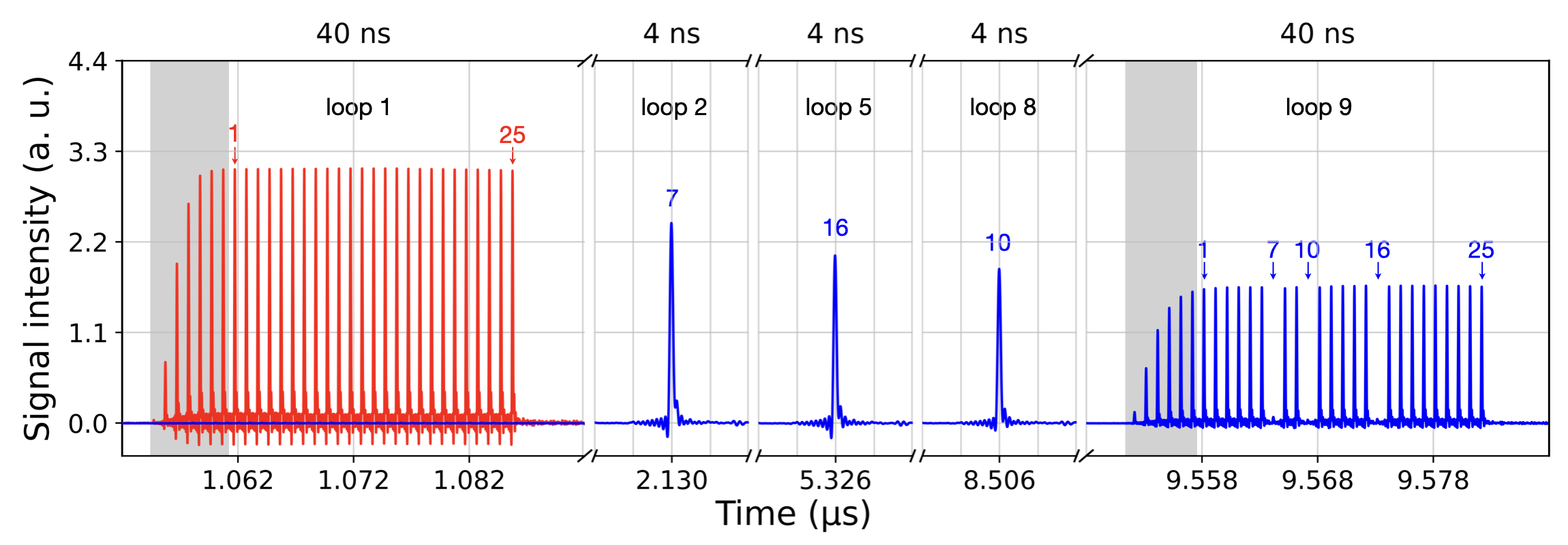}
    \caption{\textbf{Temporal multiplexing:} Dynamic routing of 25 temporal modes at ITU channel 21. The red curve shows the collective retrieval of all 25 modes after 1 storage loop. The blue curve demonstrates on-demand, selective retrieval, where mode 7 is extracted after 2 loops, mode 16 after 5 loops, mode 10 after 8 loops, and all remaining modes after 9 loops. The gray-shaded regions indicate temporal modes experiencing partial switching due to the finite rise time of the 1060 nm control laser.}
    \label{fig:multiplexing}
\end{figure}

Finally, we demonstrate that our quantum memory addresses a pivotal requirement for scalable quantum repeaters and networks: high-capacity temporal multimodality \cite{Afzelius2009}. Temporal multimodality - the capacity to simultaneously store multiple independent qubits within distinct temporal modes - is essential for dynamical entanglement distribution rates without a prohibitive scaling of physical hardware resources. As schematically illustrated in Fig.~\ref{fig:principle}(e), our fiber-integrated architecture natively supports this capability, allowing a large ensemble of temporal modes to be independently injected and selectively retrieved from the fiber cavity using dynamically programmed optical control pulses.

We first evaluate this active temporal multimodal storage capacity using an attenuated pulsed laser emitting 9~ps pulses at ITU CH21 with a 1~GHz repetition rate. As shown in Fig.~\ref{fig:multiplexing}, a sequence comprising 25 distinct temporal modes is injected into the quantum memory and retrieved after a single round-trip loop (red curve), confirming efficient multi-mode loading. In a subsequent dynamic routing experiment (blue curve), the same temporal sequence is injected, but individual modes are selectively extracted at arbitrary, pre-programmed round-trip intervals: mode 7 is retrieved at loop 2, mode 16 at loop 5, and mode 10 at loop 8, while all remaining modes are released at loop 9. 
This demonstration unambiguously proves that temporal modes separated by a time interval as small as 1~ns can be independently addressed on demand. This operation of addressing single pulses is however only available once every loop time. 

Such high-density temporal resolution is notoriously difficult to achieve in matter-based quantum memories due to atomic bandwidth limitations \cite{Jobez2015}, yet it arises naturally within our all-optical platform. Further time-domain measurements demonstrating selective input and delayed retrieval of temporal modes are provided in Supplementary Information Sec.~S7.

Importantly, the information capacity of this platform could be further augmented by incorporating spectral multimodal storage; the 1~GHz mode grid utilized here addresses only approximately $1/125$ of the memory's 12.5~THz 3~dB bandwidth. While frequency-bin encodings in this architecture are processed collectively - meaning all spectral modes are injected and retrieved simultaneously without individual routing - such collectively addressable multi-frequency channels remain highly valuable. They enable global quantum operations to be performed concurrently across an ensemble of co-propagating states, while drastically simplifying the classical control overhead required for dense, highly multimode networks \cite{Collins2007, Sinclair2014}.

This temporal multimodality capability operates seamlessly in the quantum regime. Indeed, we actively leveraged this feature during the entanglement preservation experiments detailed in Section~\ref{Single qubit fidelity and preservation entanglement.} and figure \ref{fig:entanglement} by operating the memory with up to 200 temporal modes circulating simultaneously inside the device. This capability, detailed in Section S7.1 of the Supplementary information, is of fundamental importance in high-dimensional quantum communication, as it allows to maximize secret key rates and overcome link attenuation \cite{Nunn2008}.

\section{Discussion}
The remarkable state fidelities and robust entanglement preservation demonstrated by our device underscore its viability as a high-performance building block for scalable quantum networks. From a physical perspective, the high fidelities observed are a natural consequence of the architecture; once a flying qubit is successfully routed into the cavity, its storage is governed entirely by passive propagation through a standard single-mode optical fiber.

Although standard telecom fibers can introduce polarization rotations through residual birefringence, these transformations remain stable over the microsecond storage times investigated here and can therefore be compensated, so they do not lead to intrinsic decoherence.

Nevertheless, from an experimental and engineering standpoint, validating these high fidelities is of paramount importance. It conclusively demonstrates that the dynamic, high-speed optical switching required to operate the Sagnac-loop mirrors does not introduce appreciable phase noise, polarization scrambling, or measurable state degradation. This firmly validates our all-optical architecture for high-fidelity quantum information routing over all degrees of freedom compatible with fiber networks.

This architectural simplicity translates directly into operational practicality and reliability. In stark contrast to matter-based quantum memories - such as laser-cooled atomic ensembles or rare-earth-ion-doped crystals, which demand complex, time-consuming preparation sequences like magneto-optical trapping, optical pumping, or cryogenic cooling - our all-optical quantum memory requires virtually no preparation time. It operates with a continuous duty cycle and is effectively "always on", ready to accept incoming flying qubits at any arbitrary arrival time. Furthermore, the absence of narrow atomic transitions yields an ultra-wide 3~dB bandwidth exceeding 12.5~THz, inherently unlocking massive multi-channel multiplexing capabilities. Consequently, quantum information encoded across orthogonal degrees of freedom - including polarization, time-bin, and high-density frequency channels handled via conventional wavelength-division multiplexing (WDM) - can be stored concurrently within the same physical loop without requiring structural modifications or active spectral tuning.

These operational advantages are contextualized against the state-of-the-art in Table~\ref{tab:sota_memories}. While matter-based platforms achieve impressive storage lifetimes extending well into the millisecond or second regimes, they are fundamentally constrained by narrow operational bandwidths (typically in the megahertz range), high insertion losses, and probabilistic storage mechanisms that severely throttle network data rates. Conversely, previous all-optical memorys and delay lines have provided broad bandwidths but have historically suffered from prohibitive round-trip insertion losses that rapidly degrade the quantum state after only a few cycles. By achieving an end-to-end input-output loss of just 0.46~dB (90\% transmission) and a characteristic storage lifetime of $\sim 18~\mu$s (equivalent to a mere 0.24~dB loss per round-trip) at room temperature, our device successfully bridges the long-standing dichotomy between low-loss storage and massive optical bandwidth. 

While a storage lifetime of $\sim 18~\mu$s is below the millisecond-scale thresholds required for long-distance entanglement swapping across continental baselines, it provides a massive operational window for localized high-speed networks. For instance, in a system operating at a 1~GHz clock rate - standard for modern telecommunications - this storage duration corresponds to 18,000 clock cycles. This provides more than enough margin for the local synchronization of multiple probabilistic sources within a network node.

An intrinsic feature of this fiber-loop architecture is its discrete temporal operability, which is fundamentally bounded by the cavity round-trip time. Far from being a limitation, this characteristic perfectly aligns with the synchronous demands of scalable quantum computing and communication frameworks. Most advanced quantum networks and linear optical quantum computing (LOQC) circuits are clocked systems that rely on pulsed laser regimes and global clock cycles to synchronize independent channels. By tailoring the physical length of the fiber loop to match the fundamental clock cycle of the surrounding network, the memory can seamlessly integrate into synchronized quantum protocols, naturally releasing stored qubits exactly when the next downstream operation is scheduled to occur.

Consequently, this device offers immediate utility for near-term quantum technologies, particularly for the synchronization of probabilistic quantum events. In multi-photon protocols leveraging spontaneous parametric down-conversion (SPDC) or four-wave mixing sources, the simultaneous generation probability of $N$ independent photons scales exponentially downward ($P \propto p^N$), imposing a severe bottleneck on multi-qubit state generation. Integrating our ultra-low-loss memory allows heralded photons to be deterministically delayed until a full multi-photon coincidence event is achieved. Given our measured per-cycle retention efficiency of $\eta \approx 94.6\%$ (derived from the $0.24$~dB round-trip loss via $10^{-0.24/10}$), the probability of successfully synchronizing a multi-photon state scales highly favorably as $\eta^N$. Quantitatively, attempting to synchronize a 4-photon or 8-photon state using our ultra-low-loss quantum memory yields orders-of-magnitude improvements in multi-photon coincidence rates compared to standard lossy delay lines or low-efficiency atomic memories. In protocols employing identical photons the usable bandwidth of the multiplexed photons would be limited by spectral dispersion effects in the fiber to a few GHz. An estimation of the influence of the fiber dispersion on the visibility of a typical Hong Ou Mandel experiment is reported section S4 of the Supplementary Information.

A key advantage of our system is its reconfigurability: adjusting the fiber length within the loop directly tunes the overall memory lifetime. However, this tunability is subject to physical bounds. At the lower limit, shortening the loop requires a higher peak power and repetition rate from the control laser, both of which increase the average circulating power within the fiber. Assuming an optical damage threshold of approximately 1.5 W of average power for standard SMF-28 fiber, control pulses of about 1 ns, and a repetition rate set to the inverse of the loop duration, the minimum loop time is bounded at roughly 100 ns. 
Conversely, the maximum loop time is constrained to approximately 100 $\mu$s, primarily by the 0.2 dB/km propagation loss of optical fiber. At this upper limit, fiber attenuation reduces the single-round-trip photon detection probability to 1/$e$. 

Importantly, the noise mechanisms discussed in Section~\ref{sec:working_principle} do not limit the loop duration, as Raman scattering and photoluminescence induced by the intense 1060 nm control pulses scale linearly with both control pump power and Sagnac fiber length. Because the required pump power scales inversely with the Sagnac fiber length, the resulting noise figure remains independent of the overall loop time, though it still depends on the specific optical fiber used.

\newgeometry{margin=15mm}
\begin{landscape}
\begin{table}[p]
\centering
\fontsize{8}{8.2}\selectfont
\caption{{\fontsize{10}
{9.2}\selectfont\textbf{State-of-the-art comparison of quantum memory architectures.}
The lifetime $\tau$ is derived from exponential retrieved-signal decay, $\eta(t)=\eta_0\exp(-t/\tau)$, where $\eta_0$ is the extrapolated storage-curve intercept, so that $\eta(\tau)/\eta_0=1/e$.
The equivalent round-trip count is $N_{1/e}=\tau/T_{\mathrm{rt}}$.
The labels $\tau_G$, $t_{\mathrm{AFC}}$, and $t_{\mathrm{qb}}$ denote a Gaussian $1/e$ decay time, a fixed AFC delay, and a qubit-characterization delay, respectively.
Bandwidth denotes memory acceptance, AFC extent, or input-signal bandwidth, as available; values in Refs. \cite{hsiao2018,england2015} are estimated assuming transform-limited Gaussian pulses. The time--bandwidth product $\mathrm{TBP}=\tau B$ uses the tabulated values.
Efficiencies retain the original experimental normalization; values in Ref. \cite{Lee2024FiberLoop} are estimated from nominal losses, while the I/O efficiency of this work is extrapolated to zero storage time.
For Ref.~\cite{zhong2015}, the wavelength refers to optical spin readout, and $T_{sp}$ denotes the nuclear-spin coherence time extracted from spin-echo amplitude decay under dynamical decoupling.
In the Entang.\ (Fid.) column, ``Yes'' indicates demonstrated entanglement preservation; $F_{\mathrm{cl}}$, $F_s$, $F_{\mathrm{Bell}}$, and $F_\chi$ denote classical-field or spectral-overlap, state, Bell-state, and process fidelities, respectively.
$S_{\mathrm{CHSH}}$ is the Clauser--Horne--Shimony--Holt parameter; $V$ denotes interference visibility, with $V_{\mathrm{HOM}}$ identifying Hong--Ou--Mandel interference.
$g_h^{(2)}(0)$ and $g_{\mathrm{XC}}^{(2)}$ denote herald-conditioned intensity autocorrelation and signal--herald cross-correlation, respectively.
Encoding DoF lists the degrees of freedom used in experimentally characterized coherent encodings in storage-and-retrieval tests. Encoding tests may employ single-photon states or coherent optical fields.
A dash denotes a quantity not reported; --- means not applicable; the superscript $\dagger$ identifies values calculated here from other reported parameters.
Acronyms: EIT (electromagnetically induced transparency); AFC (atomic frequency comb); DoF (degrees of freedom); OAM (orbital angular momentum); Pol.\ (polarization); Freq.\ (frequency).}}
\label{tab:sota_memories}
\smallskip
\setlength{\tabcolsep}{1.9pt}
\renewcommand{\arraystretch}{1.08}
\begin{tabularx}{\linewidth}{@{}
>{\hsize=1.7\hsize\linewidth=\hsize\raggedright\arraybackslash}X
>{\hsize=0.85\hsize\linewidth=\hsize\centering\arraybackslash}X
>{\hsize=0.9\hsize\linewidth=\hsize\centering\arraybackslash}X
>{\hsize=0.75\hsize\linewidth=\hsize\centering\arraybackslash}X
>{\hsize=0.90\hsize\linewidth=\hsize\centering\arraybackslash}X
>{\hsize=0.80\hsize\linewidth=\hsize\centering\arraybackslash}X
>{\hsize=0.90\hsize\linewidth=\hsize\centering\arraybackslash}X
>{\hsize=0.7\hsize\linewidth=\hsize\centering\arraybackslash}X
>{\hsize=1.60\hsize\linewidth=\hsize\centering\arraybackslash}X
>{\hsize=1.00\hsize\linewidth=\hsize\centering\arraybackslash}X
>{\hsize=0.9\hsize\linewidth=\hsize\raggedright\arraybackslash}X@{}}
\toprule
\textbf{Platform/Protocol} &
\textbf{Wavelength } &
\textbf{Lifetime $(\tau)$ } &
\textbf{Round trips} &
\textbf{Bandwidth} &
\textbf{TBP} &
\textbf{I/O Effic.} &
\textbf{Loop Effic.} &
\textbf{Entang. (Fid.)} &
\textbf{Temporal Multimodality} &
\textbf{Encoding DoF} \\
\midrule

\multicolumn{11}{@{}l}{\textbf{Light--Matter Interfaces (Atomic and Solid-State)}} \\
\midrule

Cold Cs (EIT) \cite{hsiao2018}
& 895 nm
& $325~\mu\mathrm{s}$ [$\tau_G$]
& ---
& $\sim 3.7$ MHz$^\dagger$
& $\sim 1200$$^\dagger$
& $92.0\%$ 
& ---
& ---; ($F_{\mathrm{cl}}=98.1\%$)
& ---
& ---
\\

Nd$^{3+}$:YVO$_4$ (AFC) \cite{tang2015}
& 879.7 nm
& $500$ ns [$t_{\mathrm{AFC}}$]
& ---
& $\sim500$ MHz
& $\sim250$$^\dagger$
& $\sim7\%$ 
& ---
& ---; ($F_{\mathrm{s}}=91.3\%$)
& 100 
& 1 (Pol.)
\\

Tm$^{3+}$:YAG (AFC) \cite{Bonarota2011}
& 793 nm
& $1.6~\mu\mathrm{s}$ [$t_{\mathrm{AFC}}$]
& ---
& $0.93$ GHz
& $\sim 1.5 \times 10^3$$^\dagger$
& $\sim1\%$ 
& ---
& ---; (---)
& 1060 
& ---
\\

Nd$^{3+}$:YSO (AFC) \cite{clausen2011}
& 883 nm
& $25$ ns [$t_{\mathrm{AFC}}$]
& ---
& $120$ MHz
& $3$$^\dagger$
& $21\%$
& ---
& Yes ($S_{\mathrm{CHSH}}=2.64$)
& ---
& 1 (Time-bin)
\\

Tm$^{3+}$:LiNbO$_3$ (AFC) \cite{Saglamyurek2011}
& 795.5 nm
& $7$ ns [$t_{\mathrm{AFC}}$]
& ---
& $5$ GHz
& $35$$^\dagger$
& $\sim0.2\%$ 
& ---
& Yes ($S_{\mathrm{CHSH}}=2.25$)
& ---
& 1 (Time-bin)
\\

Bulk diamond (Raman) \cite{england2015}
& 723 nm
& $\sim5$ ps$^\dagger$
& ---
& $\sim40$ THz$^\dagger$
& $\sim 200$$^\dagger$
& $0.9\%$ 
& ---
& ---; ($g^{(2)}_h(0)=0.65$)
& ---
& ---
\\

Cold Cs (EIT/Raman) \cite{parigi2015}
& 852 nm
& $1~\mu\mathrm{s}$ [$t_{\mathrm{qb}}$]
& ---
& ---
& ---
& $26\%$ 
& ---
& ---; ($F_{\mathrm{s}}=96.7\%$) 
& ---
& 2 (Pol., OAM) 
\\

$^{151}$Eu$^{3+}$:Y$_2$SiO$_5$ \cite{zhong2015}
& $580$ nm
& $370$ min [$T_{sp}$]
& ---
& ---
& ---
& ---
& ---
& ---; (---)
& ---
& ---
\\

\midrule
\multicolumn{11}{@{}l}{\textbf{Optical Loop and Cavity Architectures}} \\
\midrule

Free-space loop \cite{kaneda2015}
& 710 nm
& $\sim0.273~\mu\mathrm{s}$$^\dagger$
& $\sim32.8$$^\dagger$
& ---
& ---
& ---
& $97.0\%$
& ---; ($g^{(2)}_h(0)=0.479$)
& 30
& --- 
\\

Free-space loop \cite{kaneda2017}
& 1590 nm
& $\sim0.830~\mu\mathrm{s}$ 
& $\sim83$
& $\sim95$ GHz$^\dagger$
& $\sim7.9\times10^4$$^\dagger$
& ---
& $98.8\%$
& ---; $V_{\mathrm{HOM}}=95.7\%$ 
& 40
& 1 (Time-bin)
\\

Free-space loop \cite{Arnold2022Broadband}
& 705 nm
& $\sim 3.8~\mu\mathrm{s}$$^\dagger$
& $\sim 3$$^\dagger$
& $>1$ THz
& $> 4 \times 10^6$$^\dagger$
& ---
& $97.3\%$
& ---; ($F_{\chi}=97.5\%$)
& ---
& 1 (Time-bin)
\\

Fiber loop \cite{Lee2024FiberLoop}
& Telecom
& $\sim54~\mu\mathrm{s}$$^\dagger$
& $\sim2.1$$^\dagger$
& ---
& ---
& $\sim27.8\%$$^\dagger$
& $\sim61.9\%$$^\dagger$
& Yes; ($F_{\mathrm{s}}=94\%$)
& ---
& 1 (Pol.)
\\

Fiber-coupled hybrid loop \cite{Cheng2025FiberCoupled}
& 780 nm
& $\sim0.623~\mu\mathrm{s}$$^\dagger$
& $\sim 1.2$$^\dagger$
& $\sim10$ THz 
& $\sim6.2\times10^6$$^\dagger$
& $\sim54\%$
& $43\%$ 
& ---; $F_{\mathrm{s}}=89.4\%$ 
& ---
& 1 (Pol.)
\\

Fiber cavity \cite{bustard2022}
& 902.5 nm
& $\sim206$ ns 
& $\sim16$
& $\sim386$ GHz$^\dagger$
& $\sim8.0\times10^4$$^\dagger$
& $73\%$
& $\sim 94 \%$$^\dagger$
& ---; ($F_\mathrm{cl}>99.97\%$)
& ---
& --- 
\\

Fiber cavity \cite{BonsmaFisher2024}
& 1260 nm
& $1.64~\mu\mathrm{s}$ 
& 32.8
& $81$ GHz 
& $1.33\times10^5$$^\dagger$
& $15\%^{\dagger}$
& ---
& ---;($g^{(2)}_\mathrm{XC}>2$ up to $3.5 \mu \mathrm{s}$)
& ---
& --- 
\\

\rowcolor[gray]{0.95}
\textbf{Fiber loop (This work)}
& \textbf{1560 nm}
& $\boldsymbol{18.5~\mu\mathrm{s}}$ 
& $\boldsymbol{18}$
& $\boldsymbol{>12.5}$ \textbf{THz}
& $\boldsymbol{>2.3\times10^8}$$^\dagger$
& $\boldsymbol{\sim90\%}$ 
& $\boldsymbol{94.6\%}$
& \textbf{Yes}; $V=98.8\%$
& \textbf{200}
& \textbf{3} (Pol., Freq.-, Time-bin)
\\

\bottomrule
\end{tabularx}

\end{table}
\end{landscape}
\restoregeometry

\section{Conclusion}

In summary, we have demonstrated a robust, all-optical quantum memory architecture that successfully resolves the long-standing trade-off between storage bandwidth and retrieval efficiency. By exploiting the cross-phase modulation nonlinearity within a dual-Sagnac interferometer configuration, we have eliminated the primary bottlenecks that have historically limited fiber-loop memories. The resulting device operates at room temperature with a record-low input-output insertion loss of 0.46 dB, enabling a characteristic storage lifetime of 18 $\mu$s. This performance is more than sufficient for the synchronization of gigahertz-rate photonic sources and entanglement-swapping operations within quantum-network nodes or urban-scale networks.

Furthermore, the capacity to accommodate hundreds of concurrently circulating modes highlights the potential of this architecture to serve as a high-density multiplexing hub in complex quantum network topologies.

These figures of merit place our device among the lowest-loss all-optical quantum memories reported to date, enabling coherent storage over numerous round-trips. Owing to its purely optical mechanism, the memory is intrinsically broadband and fully preserves both the spectral and temporal structures of the stored photons. Operating at room temperature within the telecommunications C-band ($\sim$1550 nm), the platform enables the deterministic capture and release of single photons without requiring bulk optics, complex cryogenic infrastructure, or narrowband atomic transitions.

The significance of these results, however, extends beyond raw storage metrics. Our experimental validation of high state fidelities (>99\%) across time-bin, frequency-bin, and polarization encodings establishes this architecture as a genuinely universal quantum memory. Unlike matter-based interfaces that mandate specific operational bandwidths or polarization states, our all-optical approach remains completely agnostic to the underlying qubit encoding. This flexibility is critical for future heterogeneous quantum networks, where a routing node must seamlessly memory polarization-encoded qubits from a satellite downlink alongside time-bin qubits from a terrestrial fiber link.

Furthermore, the verified preservation of time-bin entanglement confirms the device's compatibility with advanced quantum information protocols, such as entanglement swapping and quantum teleportation. By uniting ultra-low insertion loss, deterministic optical control, telecom-native operation, multi-degree-of-freedom compatibility, and entanglement preservation within a single fiber-integrated architecture, our platform provides a universal memorying solution tailored for scalable photonic quantum networks. It is immediately compatible with existing fiber infrastructure and naturally suited for advanced multiplexing strategies spanning the time, frequency, and polarization domains.

We emphasize that our architecture is designed to serve as a high-performance quantum cache or memory rather than a long-term storage medium. Analogous to the memory hierarchy of classical computing (e.g., SRAM versus magnetic disks), our device prioritizes high-rate throughput and synchronization. This fills a critical gap in quantum repeater nodes where matter-based memories, despite their longer lifetimes, remain constrained by sub-gigahertz bandwidths and severe spatial coupling penalties.

Looking forward, the combination of low loss and high multi-mode capacity opens a clear avenue toward massively multiplexed quantum memories for high-rate quantum repeaters, network synchronization, and distributed photonic quantum computing. Continuous improvements in storage lifetime and total efficiency, alongside integration with on-chip photonic sources, frequency processors, and active network nodes, will position this architecture as a central building block for the future quantum internet.

\section*{Methods}
\subsection*{All-optical input-output switching}
The input-output mechanism of our quantum memory relies on the principles of a nonlinear optical loop mirror, where all-optical switching is achieved via XPM in an optical fiber. This process imparts a $\pi$ phase shift exclusively to the counter-clockwise propagating signal within the Sagnac interferometer \cite{Asobe1993}.
To achieve high-fidelity switching with near-unity efficiency and minimal temporal distortion, the control pulse must remain temporally overlapped with the signal pulse across the entire interferometer length (100 m, in the present implementation). Furthermore, it must provide a constant optical power throughout this interaction length. We accomplish this by shaping the 1060 nm control pulses into a flat-top profile with durations ranging from 5 to 50 ns. These durations are significantly longer than both the 1550 nm signal pulse (10 ps in our experiments) and the signal-control walk-off, which we measure to be $\sim$ 130 ps for each Sagnac interferometer.
For the simultaneous injection or retrieval of multiple pulses, we implement a 5 ns "dead time" to allow the control pulse to reach a steady-state intensity, as indicated by the shaded region in Fig. 7. Alternatively, when injecting or retrieving a single signal pulse from a 1 GHz pulse train, the control laser is limited to producing Gaussian-like pulses rather than flat-top profiles. Nevertheless, because the signal pulse duration remains substantially shorter than the FWHM of the control pulse, the switching mechanism still maintains near-unity efficiency.

\subsection*{Heralded single-photon generation and time-bin encoding}
Heralded single photons were generated using a 2 cm long silicon rib waveguide. The waveguide has a 3 $\mu$m wide base and a 450 nm wide cap etched to a depth of 150 nm over a total thickness of 310 nm. The waveguide is embedded in silica. The guide is pumped by a pulsed laser with 9 ps pulses and 1 GHz repetition rate tuned to ITU channel 23. DWDM filters were used to clean ASE noise from the laser. Signal and idler photon pairs generated by SFWM in the waveguide, whose source characterization is reported in Supplementary Sec.~S3, are sent to 100 GHz DWDM filters to separate idler photons (ITU channel 25) from signal photons (ITU channel 21). The choice of frequencies comes from the fact that channel 21 is the one closer to the maximum of efficiency for which low loss DWDM filters were commercially available at the time of experiments. Idler photons were then sent to superconducting single photon detectors to herald single photon states on signal photons. We encoded time-bin qubits by sending the heralded single photons to a Mach-Zehnder interferometer with an unbalance of 300 ps between the arms (experimental details are described in  the Supplementary Information Sec.~S5). We used a second Mach-Zehnder interferometer with the same unbalance to perform quantum state tomography on the state at the output of the quantum memory. The pumping power for the source in this experiments was chosen to work at a CAR level of approximately 45.

\subsection*{Time bin entanglement measurements}
The same source of photon pairs described in the previous section was employed along with the same interferometers. Time bins were defined on the pump laser using the first interferometer. DWDM filters were used to separate idler photons (ITU channel 25) from signal photons (ITU channel 21) and the latter were input to the quantum memory (the fiber bypass was also used in this case as a reference for comparison). 
Then  both signal and idler photons were sent to the second interferometer followed by superconducting single photon detectors.
We notice that additional attenuation was necessary on the idler channel to avoid saturation of the single photon detectors. The same attenuation was needed for the signal channel when using the bypass. These measurements were carried out exploting temporal multimodality as outlined in section \ref{Multiplexing} and section S7.1 of the Supplementary Information file to partially compensate for the duty cycle of the memory.

\section*{Acknowledgements}

D.C., S.C. and A.B. acknowledge European Union funding from the STARLight project (project ID: 101194170). M.G. and D.B. and M.L. acknowledge the PNRR MUR project PE0000023-NQSTI. The authors acknowledge CEA-Leti, Grenoble, for providing the silicon photonic chip used as a source of photon pairs.

\section*{Competing Interests}
The authors declare the following competing financial interests: a patent application related to the all-optical quantum memory device described in this paper has been filed by the University of Pavia, listing N.T., M.L., M.G., and D.B. as inventors.

\section*{Author contributions}

D.B. and M.G. conceived the original idea. D.C. and N.T. performed all the experiments with contributions from S.C. and A.B., under the supervision of M.G. and D.B.. M.L. supervised the theoretical aspects. M.L., M.G., and D.B. coordinated and supervised the project. All authors contributed to the preparation of the manuscript.

\section{Data availability}
The data that support the findings of this study are available from the corresponding
author upon reasonable request.

\bibliographystyle{unsrt}
\bibliography{bibmem}

\section*{Supplementary Information for ``A Universal Telecom-Band Quantum Memory Enabled by Cross-Phase Modulation''}
\addcontentsline{toc}{section}{Supplementary Information}

\setcounter{section}{0}
\setcounter{subsection}{0}
\setcounter{figure}{0}
\setcounter{table}{0}
\setcounter{equation}{0}

\renewcommand{\thesection}{S\arabic{section}}
\renewcommand{\thesubsection}{S\arabic{section}.\arabic{subsection}}
\renewcommand{\thefigure}{S\arabic{figure}}
\renewcommand{\thetable}{S\arabic{table}}
\renewcommand{\theequation}{S\arabic{equation}}

\newcommand{\BSFWMrefs}{
N. Maring \textit{et al.}, Nat. Commun. 8, 1 (2017);
C. Reimer \textit{et al.}, Optica 3, 476 (2016);
R. Ikuta \textit{et al.}, Phys. Rev. Lett. 117, 110503 (2016);
M. Kues \textit{et al.}, Nature 546, 622 (2017)
}

\newcommand{\XPMpoldep}{%
Agrawal, G. in Nonlinear fiber optics chap. 7 (Academic Press, 2013)%
}
\newcommand{\XPMyarivBOOK}{%
Yariv, A. \& Yeh, P. Photonics. Optical Electronics in Modern Communications
10th Anniversary Edition (Oxford University Press, Oxford, 2006)%
}

\newcommand{\XPMstimaNdue}{%
T. Kato, Y. Suetsugu, M. Takagi, E. Sasaoka, and M. Nishimura, ``Measurement of the nonlinear refractive index in optical fiber by the cross-phase-modulation method with depolarized pump light,'' \textit{Optics Letters} \textbf{20}, 988--990 (1995)%
}

\newcommand{\XPMinsensitive}{%
Bülow, H. \& Veith, G. Polarisation-independent switching in a nonlinear
optical loop mirror by a dual-wavelength switching pulse. Electronics
Letters \textbf{29}, 588–589 (7 1993)%
}

\newcommand{\XPMrefskum}{%
M.~A.~Hall, J.~B.~Altepeter, and P.~Kumar, Phys. Rev. Lett. \textbf{106}, 053901 (2011)%
}

\newcommand{\XPMrefsPRIORkum}{%
H.~Bülow and G.~Veith, Electron. Lett. \textbf{29}, 588 (1993)%
}

\newcommand{\entangRefs}{%
M. Bacchi \textit{et al.}, arXiv:2505.04598 (2025)%
}

\newcommand{\refSparam}{%
I. Marcikic \textit{et al.}, Phys. Rev. Lett. \textbf{93}, 180502 (2004)%
}

\newcommand{\refCorneringAttenuation}{%
Corning Incorporated. Corning® SMF-28® Ultra Optical Fiber Product
Information Available at: https://www.corning.com/media/worldwide/
coc/documents/Fiber/product-information-sheets/PI-1470-AEN.
pdf. 2020.%
}

\newcommand{\refCorneringSPlice}{%
Corning SMF-28 Ultra Optical Fiber Fusion Splicing Report. Corning application note. https://www.corning.com/media/worldwide/coc/documents/Fiber/application-notes/AN2008.pdf (2017)%
}

\renewcommand{\thesection}{S\arabic{section}}

\maketitle

\section{Cross-phase modulation in optical fibers}
\label{sec:switching}

Our fiber-loop quantum memory exploits cross-phase modulation (XPM) to implement all-optical switching. XPM is a third-order Kerr nonlinear effect in which the intensity of a control field modifies the refractive index experienced by a signal field, thereby inducing a nonlinear phase shift. The resulting phase shift, $\Delta \phi^{\mathrm{XPM}}$, is proportional to the control-pulse peak power and depends on the relative polarization of the interacting fields through a coefficient $k=2$ for copolarized fields and $k=2/3$ for orthogonally polarized fields. It can be written as:\footnote{\label{fn:yariv}\XPMyarivBOOK}

\begin{equation}
\Delta \phi^{\mathrm{XPM}}_{\mathrm{signal}}=k\frac{n_2 \omega_{\mathrm{signal}}}{c A_{\mathrm{eff}}}
L_{\mathrm{int}} P_{\mathrm{control}}
\label{eq_deltaphi_xpm_no_pol}
\end{equation}

where $n_2$ is the nonlinear Kerr coefficient, $\omega_{\mathrm{signal}}$ is the signal angular frequency, $c$ is the speed of light in vacuum, $L_{\mathrm{int}}$ is the interaction length, $P_{\mathrm{control}}$ is the control-pulse peak power and $A_{\mathrm{eff}}$ is the effective signal-control modal overlap area calculated as:
\begin{equation}
    A_{\mathrm{eff}}=\frac{\pi}{8}(\mathrm{MFD}_{\mathrm{signal}}^2+\mathrm{MFD}_{\mathrm{control}}^2)
    \label{eqAeff}
\end{equation}
where \(\mathrm{MFD}_{\mathrm{signal}}=(10.4 \pm 0.5)~\mu\mathrm{m}\) is the Gaussian mode-field diameter of the signal field at \(1550~\mathrm{nm}\), \footnote{\refCorneringAttenuation} while \(\mathrm{MFD}_{\mathrm{control}}=(7.8 \pm 0.2)~\mu\mathrm{m}\) is the Gaussian mode-field diameter of the control field at \(1064~\mathrm{nm}\), obtained from TE-polarized mode simulations performed with the finite-difference eigenmode (FDE) solver in Lumerical MODE.

Since the XPM-induced phase shift depends on the relative signal-control polarization, polarization fluctuations during propagation lead to fluctuations in the nonlinear phase shift and, consequently, in the switching efficiency. To remove this sensitivity, we use a polarization-insensitive XPM scheme based on two temporally overlapped control pulses with orthogonal polarizations, equal peak powers, and slightly different wavelengths.\footnote{\XPMinsensitive} In this configuration, any signal polarization can be expressed in the basis defined by the two orthogonal control polarizations, so that each component experiences one copolarized and one orthogonally polarized control contribution. The two contributions in Eq.~\ref{eq_deltaphi_xpm_no_pol} therefore add up, giving:

\begin{equation}
\Delta \phi^{\mathrm{XPM}}_{\mathrm{signal}}=\left(2+\frac{2}{3}\right)
\frac{n_2 \omega_{\mathrm{signal}}}{c A_{\mathrm{eff}}}
L_{\mathrm{int}} P_{\mathrm{control}}
\label{eq_deltaphi_xpm_pol_insensitive_single}
\end{equation}

where $P_{\mathrm{control}}$ is now the peak power of each control pulse. Since the two orthogonally polarized control pulses are set to have equal peak power, the total control power is $P_{\mathrm{tot}}=2P_{\mathrm{control}}$. The polarization-insensitive XPM phase shift can therefore be expressed in terms of $P_{\mathrm{tot}}$ as:

\begin{equation}
\Delta \phi^{\mathrm{XPM}}_{\mathrm{signal}}= \frac{4}{3}
\frac{n_2 \omega_{\mathrm{signal}}}{c A_{\mathrm{eff}}}
L_{\mathrm{int}} P_{\mathrm{tot}} .
\label{eq_deltaphi_xpm_pol_insensitive_total}
\end{equation}

Although the effective nonlinear coefficient $4/3$ is lower than the value obtained for perfectly copolarized signal and control fields, this configuration makes the induced phase shift independent of the input signal polarization, provided that the two control pulses are balanced in power and temporally overlapped. 

\section{Switching efficiency}

We performed a switching-efficiency characterization to identify the control-pulse conditions that maximize the XPM-induced transmission of the Sagnac interferometers. We report the measurement performed on the input Sagnac interferometer, S1. The Sagnac interferometer under test was initially adjusted to its reflective state using the intraloop fiber polarization controller, while the other Sagnac interferometer was configured to be transmissive and therefore acted only as a delay line. The two orthogonally polarized control pulses injected into the memory had a duration of $\tau_{\mathrm{control}}=30~\mathrm{ns}$ and a repetition rate of $R_{\mathrm{control}}=100~\mathrm{kHz}$.

In this configuration, the signal transmitted through the memory was recorded on a photodetector while varying the total peak power $P_{\mathrm{peak,tot}}$ of the combined control field. For each setting, the total average power $P_{\mathrm{avg,tot}}$ was measured with a power meter, and the corresponding total peak power was obtained as:
\begin{equation}
P_{\mathrm{peak,tot}}= \frac{P_{\mathrm{avg,tot}}}{\mathrm{Duty\ cycle}}=\frac{P_{\mathrm{avg,tot}}}
{R_{\mathrm{control}} \cdot \tau_{\mathrm{control}}}.
\end{equation}

The acquired signal was compared with a reference signal obtained when both Sagnac interferometers were configured in the transmissive state. This reference includes the passive losses of the open-memory configuration, but does not include the XPM-induced losses. The switching efficiency $\eta_{S1}$ was calculated as the ratio between the integrated area of the switched signal, $A_{\mathrm{Switched}}$, and that of the reference signal, $A_{\mathrm{Reference}}$:
\begin{equation}
\eta_{S1}=\frac{A_{\mathrm{Switched}}}{A_{\mathrm{Reference}}}.
\end{equation}

\begin{figure}[h!]
\centering
\begin{subfigure}{0.68\textwidth}
\centering
\includegraphics[width=\linewidth]{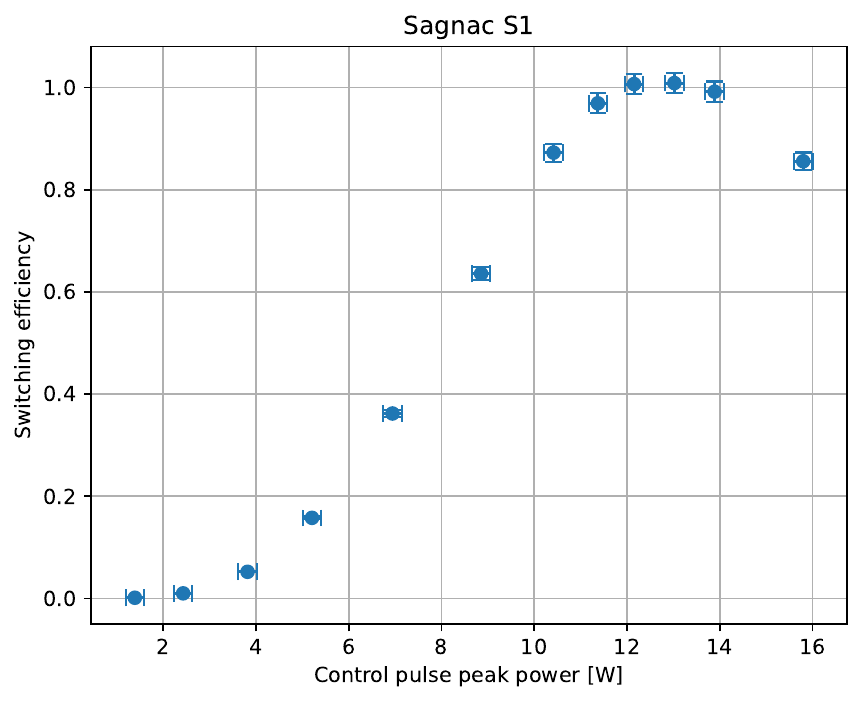}
\end{subfigure}

\caption{
\textbf{Switching efficiency of S1 as a function of the total control-pulse peak power.}
The measurement was performed on the input Sagnac interferometer, S1, using two temporally overlapped control pulses with orthogonal polarizations and equalized powers. The horizontal axis reports the total peak power $P_{\mathrm{peak,tot}}$ of the combined control field. The switching efficiency was obtained by normalizing the integrated area of the switched signal to that of a reference signal acquired with both Sagnac interferometers set to the transmissive state. The maximum switching efficiency and the corresponding control-pulse peak power are reported in Eq.~\ref{eq_max_switch_s2}.}
\label{fig:switching_efficiency}
\end{figure}

The result for S1 is shown in Fig.~\ref{fig:switching_efficiency}; similar performance was verified for S2. The XPM-induced phase shift increases with $P_{\mathrm{peak,tot}}$, leading to a larger transmitted fraction until the optimal switching condition is reached. At higher powers, the phase shift exceeds the optimum value and the switching efficiency decreases. The maximum switching efficiency measured for S1 is:
\begin{equation}
\label{eq_max_switch_s2}
\eta_{S1}=(1.009\pm0.014)
\quad \mathrm{at} \quad
P_{\mathrm{peak,tot}}=(13.0\pm 0.2)~\mathrm{W}.
\end{equation}

We compare this value with the control power expected from the XPM model introduced in Sec.~\ref{sec:switching}. Substituting $\Delta \phi^{\mathrm{XPM}}_{\mathrm{signal}}=\pi$, $n_2=(2.96 \pm 0.15)\times 10^{-20}~\mathrm{m^2/W}$, $A_{\mathrm{eff}}=(66.4 \pm 0.5)~\mu\mathrm{m}^2$ (see Eq.\ref{eqAeff}), and $L_{\mathrm{int}}=(105 \pm 1)~\mathrm{m}$ into Eq.~\ref{eq_deltaphi_xpm_pol_insensitive_total} gives an expected total control peak power of $P_{\mathrm{tot}}=(12.4\pm1.4)~\mathrm{W}$ for the polarization-insensitive configuration. This value is in good agreement with the measured optimum reported in Eq.~\ref{eq_max_switch_s2}.

\clearpage

\section{Photon-pair source characterization}
\label{sec:source}

To select the appropriate operating regime for our integrated source in the
final experiments, we first characterized its performance. In particular, we
investigated the generation of photon pairs through Spontaneous Four Wave Mixing (SFWM) in a \SI{16}{mm}-long low-loss silicon-on-insulator (SOI) rib waveguide. The waveguide was fabricated on a silicon device layer with a thickness of \SI{310}{nm}. The rib structure features a \SI{3}{\micro m} wide base and a
\SI{450}{nm} wide rib, etched to a depth of \SI{150}{nm}, leaving a
\SI{160}{nm} slab. This geometry defines a single-mode guiding structure,
which we used as an integrated source of entangled photon pairs. In the final
experiments, a \SI{20}{mm}-long SOI rib waveguide with the same geometry
and nearly identical performance was employed. Both SOI platforms were
realized in collaboration with the CEA-Leti research center in Grenoble,
France, by deep ultraviolet lithography.
The characterization consisted of identifying the pump-pulse peak power that simultaneously maximizes the photon-pair generation rate while ensuring operation in the low-squeezing regime, where multi-pair generation is negligible, and maintaining a high signal-to-noise ratio. We tested the source using a pump configuration that reproduced the early--late time-bin structure used in the final experiment, although with a different repetition rate and pulse duration. In particular, pairs of early and late pump pulses were sent to the waveguide to generate entangled photon pairs in either the early or late time bin. After the generation stage, signal and idler photons were separated by dense wavelength-division multiplexing (DWDM) filters and detected with two superconducting nanowire single-photon detectors (SNSPDs).

\begin{figure}[h!]
\centering

\begin{subfigure}{0.49\textwidth}
    \centering
    \includegraphics[width=\linewidth]{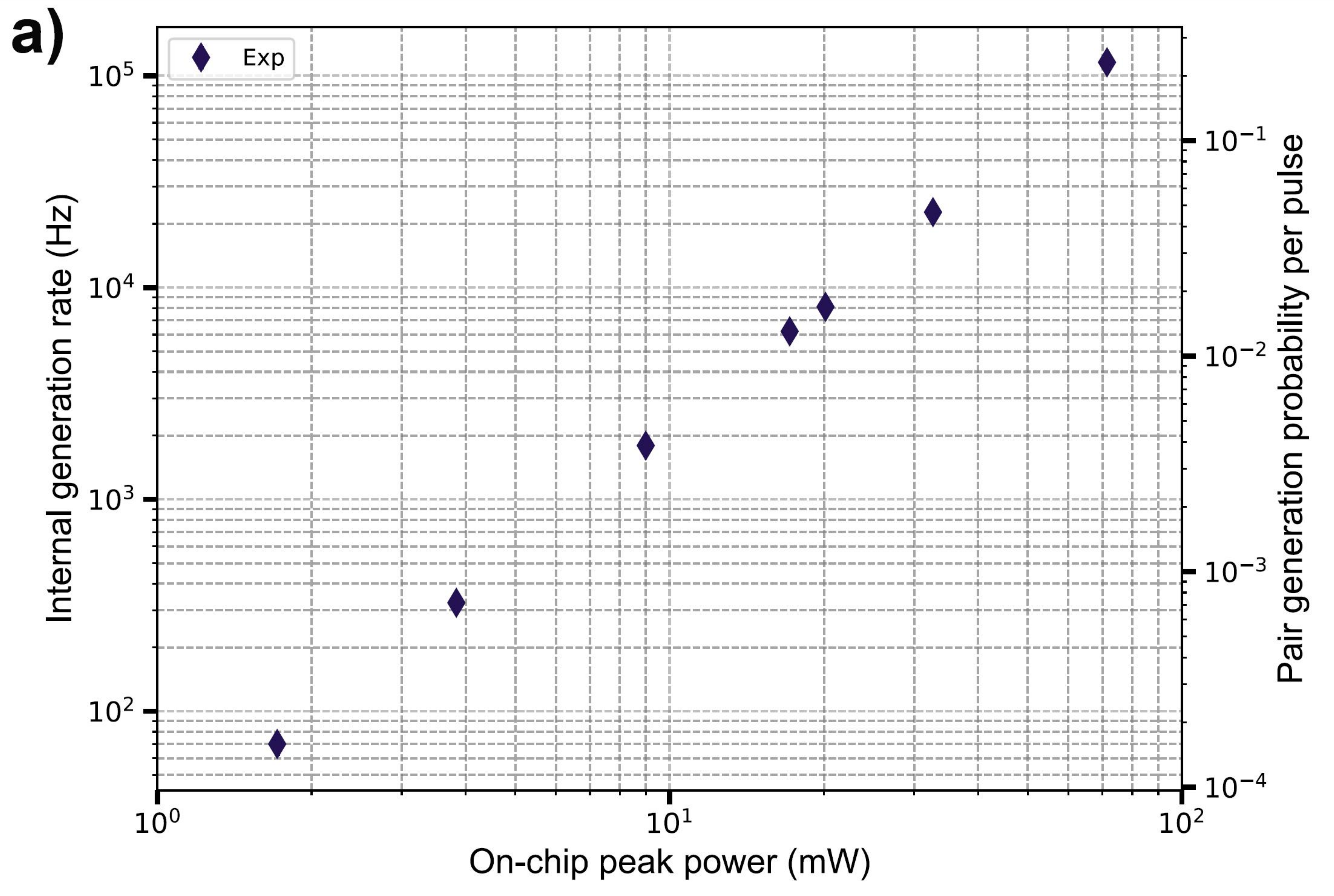}
\end{subfigure}
\hfill
\begin{subfigure}{0.47\textwidth}
    \centering
    \includegraphics[width=\linewidth]{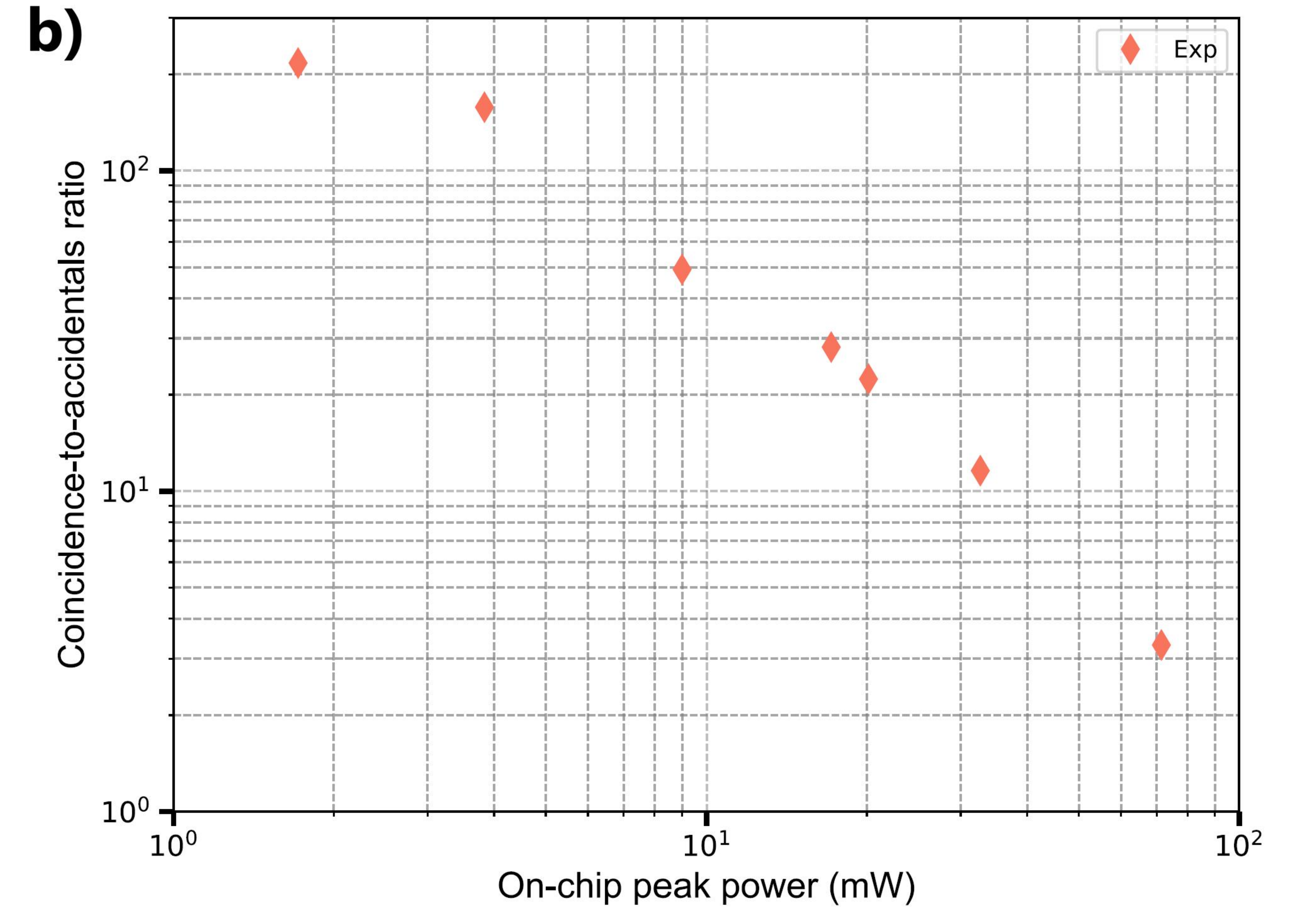}
\end{subfigure}

\caption{
\textbf{Integrated waveguide source characterization:} generation rate, pair probability, and coincidence-to-accidentals ratio (CAR) as a function
of coupled pump-pulse peak power. 
(a) Internal generation rate (left axis) and pair generation probability per pulse (right axis). 
(b) Coincidence-to-accidentals ratio (CAR), quantifying the signal-to-noise ratio of the generated photon pairs.}

\label{fig:source_characterization}
\end{figure}

To analyze the SFWM process, we collected coincidence counts while varying the peak power of the pump pulses injected into the waveguide by using a variable optical attenuator (VOA). Coincidence counts were obtained by
recording signal and idler arrival times on the SNSPDs with respect to the pump-laser reference clock, using time-tagging electronics. From the coincidence data, we extracted the internal generation rate, the pair generation probability per pulse, and the coincidence-to-accidentals ratio (CAR), which quantifies the contrast between true photon-pair coincidences and accidental background events. Results are shown in log-log scale in Fig. \ref{fig:source_characterization}.

From these results, we chose an on-chip pump-pulse peak power of approximately
20~mW. In the 100~kHz characterization measurement (1 ns pulse width), this operating point yielded an
internal generation rate of approximately 5~kHz, a pair generation probability per pulse
of approximately $10^{-2}$, and a CAR of approximately 20.
Because the final experiments used a different repetition rate and pulse duration (1 GHz and 9 ps), the absolute internal generation rate of the final setup is not inferred from the 100 kHz data. Instead, this characterization was used to identify a conservative on-chip peak-power range for operating the source in a low-squeezing regime.

\clearpage

\section{Effect of Chromatic Dispersion on Hong Ou Mandel
Interference of Lorentzian Single Photons in Optical Fiber}

In this section we investigate the reduction of Hong Ou Mandel (HOM)
interference visibility between two initially identical single photons when one photon propagates through a dispersive optical fiber. The scope of this section is to investigate the spectral characteristic single photons should have to be stored in our quantum memory for a certain time and then succesfully undergo HOM interference with an identical photon that has not been stored in teh memory

Both photons are assumed to be Fourier-transform limited, with identical Lorentzian spectral intensity distributions centered at $\lambda_0=1550$~nm. One photon propagates through a length $L$ of standard SMF-28 optical fiber, corresponding to a propagation time
$T$, while the other photon propagates through a dispersion-free reference path.
We assume that the relative group delay accumulated during propagation is compensated before the photons interfere at the beamsplitter.
Consequently, the only source of distinguishability considered here is the frequency-dependent spectral phase
introduced by chromatic dispersion.
We neglect polarization effects, nonlinear optical effects, third-order dispersion, and frequency-dependent losses.

For SMF-28 optical fiber at 1550~nm, we assume

\begin{align}
\lambda_0 &= 1550~\mathrm{nm},\\
n_g &= 1.4682,\\
D &= 17~\mathrm{ps/(nm\,km)}.
\end{align}

The group velocity dispersion coefficient is related to the dispersion parameter $D$ by

\begin{equation}
\beta_2 =
-\frac{\lambda_0^2}{2\pi c}D
\end{equation}

where $c$ is the speed of light in vacuum.

Substituting the parameters gives

\begin{equation}
\beta_2 \simeq -21.68~\mathrm{ps^2/km}.
\end{equation}

The propagation length associated with a propagation time $T$ is

\begin{equation}
L=\frac{cT}{n_g}.
\end{equation}

The corresponding lengths are reported in
Table~\ref{tab:lengths}.

\begin{table}[htbp]
\centering
\caption{Fiber length corresponding to each propagation time.}
\label{tab:lengths}
\begin{tabular}{cc}
\toprule
Propagation time $T$ ($\mu$s) & Fiber length $L$ (km)\\
\midrule
1  & 0.204\\
2  & 0.408\\
5  & 1.021\\
10 & 2.042\\
20 & 4.084\\
\bottomrule
\end{tabular}
\end{table}

For two pure single-photon states $|\psi_1\rangle$
and $|\psi_2\rangle$, the ideal HOM visibility is

\begin{equation}
V_{\mathrm{HOM}} =
\left|\langle\psi_1|\psi_2\rangle\right|^2.
\label{eq:hom}
\end{equation}

Here, the visibility is defined through the coincidence probability at a balanced beamsplitter,

\begin{equation}
P_{\mathrm{coin}} =
\frac{1}{2}\left(1-V_{\mathrm{HOM}}\right).
\end{equation}

Perfectly indistinguishable photons have
$V_{\mathrm{HOM}}=1$, whereas completely distinguishable
photons have $V_{\mathrm{HOM}}=0$.

Let $\omega_0$ be the central angular frequency and define

\begin{equation}
\Omega=\omega-\omega_0.
\end{equation}

Expanding the propagation constant around $\omega_0$,

\begin{equation}
\beta(\omega)=
\beta_0+\beta_1\Omega+
\frac{\beta_2}{2}\Omega^2+\cdots.
\end{equation}

The constant phase $\beta_0L$ does not affect the
interference visibility.
The linear term $\beta_1L\Omega$ corresponds to the
group delay and is assumed to be compensated.

The remaining dispersive phase is therefore

\begin{equation}
\phi_D(\Omega)=
\frac{\beta_2L}{2}\Omega^2.
\label{eq:dispersion_phase}
\end{equation}

If the input spectral amplitude is $\psi(\Omega)$,
the photon after propagation is described by

\begin{equation}
\psi_L(\Omega)=
\psi(\Omega)
\exp\left(
i\frac{\beta_2L}{2}\Omega^2
\right).
\end{equation}

The overlap between the reference photon and the
propagated photon is

\begin{equation}
\langle\psi_0|\psi_L\rangle =
\int_{-\infty}^{+\infty}
|\psi(\Omega)|^2
\exp\left(
i\frac{\beta_2L}{2}\Omega^2
\right)d\Omega.
\end{equation}

Consequently,

\begin{equation}
V_{\mathrm{HOM}}(L)=
\left|
\int_{-\infty}^{+\infty}
|\psi(\Omega)|^2
\exp\left(
i\frac{\beta_2L}{2}\Omega^2
\right)d\Omega
\right|^2.
\label{eq:visibility_integral}
\end{equation}

This expression describes the reduction of HOM visibility
caused exclusively by quadratic spectral phase.
We assume that the spectral intensity distribution of
each photon is Lorentzian, with a full width at half maximum (FWHM) $\Delta\nu$ expressed in Hz.
In terms of angular-frequency detuning $\Omega$,
the normalized spectrum is

\begin{equation}
|\psi(\Omega)|^2=
\frac{\gamma}{\pi}
\frac{1}{\Omega^2+\gamma^2},
\label{eq:lorentzian}
\end{equation}

where

\begin{equation}
\gamma=\pi\Delta\nu
\end{equation}

is the half width at half maximum expressed in
angular-frequency units.

The normalization is

\begin{equation}
\int_{-\infty}^{+\infty}
|\psi(\Omega)|^2d\Omega=1.
\end{equation}

Substituting Eq.~\eqref{eq:lorentzian} into
Eq.~\eqref{eq:visibility_integral} gives

\begin{equation}
V_{\mathrm{HOM}} =
\left|
\frac{\gamma}{\pi}
\int_{-\infty}^{+\infty}
\frac{
\exp\left(i\beta_2L\Omega^2/2\right)
}{
\Omega^2+\gamma^2
}
d\Omega
\right|^2.
\label{eq:lorentzian_integral}
\end{equation}

The integral can be evaluated analytically using
the complementary error function.

Defining the dimensionless parameter

\begin{equation}
q=
\frac{\beta_2L}{2}
\left(\pi\Delta\nu\right)^2,
\label{eq:q}
\end{equation}

the spectral overlap becomes

\begin{equation}
\langle\psi_0|\psi_L\rangle
=
e^{-iq}
\operatorname{erfc}
\left(\sqrt{-iq}\right).
\end{equation}

Therefore, the HOM visibility is

\begin{equation}
V_{\mathrm{HOM}}(L,\Delta\nu)
=
\left|
e^{-iq}
\operatorname{erfc}
\left(\sqrt{-iq}\right)
\right|^2,
\label{eq:final}
\end{equation}

where $q$ is given by Eq.~\eqref{eq:q}.
Since $L=cT/n_g$, the visibility can also be expressed
directly in terms of the propagation time:

\begin{equation}
q=
\frac{\beta_2cT}{2n_g}
\left(\pi\Delta\nu\right)^2.
\end{equation}

Equation~\eqref{eq:final} is the analytical expression
used in the numerical calculations below.

We calculate the HOM visibility for Lorentzian
linewidths between 1 and 100~GHz, considering
propagation times

\begin{equation}
T=1,\;2,\;5,\;10,\;20~\mu\mathrm{s}.
\end{equation}

\begin{figure}[htbp]
\centering
\includegraphics[width=0.95\linewidth]
    {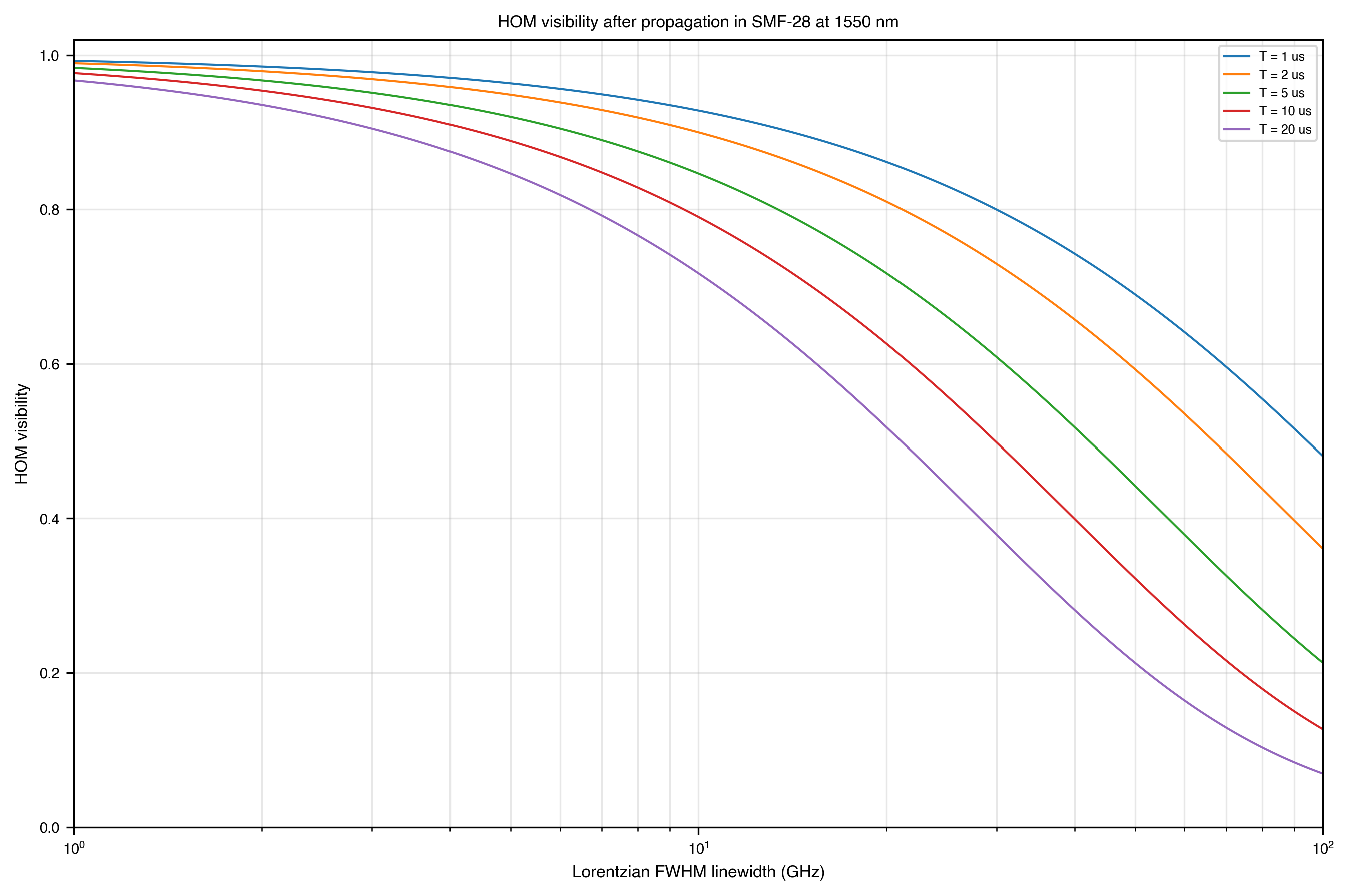}
\caption{
Calculated HOM interference visibility as a function of the FWHM Lorentzian spectral linewidth for two
initially identical, Fourier-transform-limited photons at 1550~nm. One photon propagates through SMF-28 optical fiber,
while the other remains dispersion-free.
The curves correspond to propagation times
$T=1,2,5,10,$ and $20~\mu$s. The relative group delay is assumed to be compensated,
so that the visibility reduction is due exclusively to chromatic dispersion.
}
\label{fig:visibility}
\end{figure}

The calculated visibility curves are shown in
Fig.~\ref{fig:visibility}.
Numerical values at selected linewidths are reported
in Table~\ref{tab:visibility}.
The results demonstrate that chromatic dispersion can significantly reduce the HOM visibility of broadband photons after propagation through several kilometres of standard telecom fiber.

\begin{table}[htbp]
\centering
\caption{Calculated HOM visibility for different
linewidths and propagation times.}
\label{tab:visibility}
\begin{tabular}{cccccc}
\toprule
$\Delta\nu$ & \multicolumn{5}{c}{$T$ ($\mu$s)}\\
(GHz) & 1 & 2 & 5 & 10 & 20\\
\midrule
1   & 0.993 & 0.990 & 0.984 & 0.977 & 0.967\\
2   & 0.985 & 0.979 & 0.967 & 0.954 & 0.936\\
5   & 0.963 & 0.949 & 0.920 & 0.889 & 0.847\\
10  & 0.928 & 0.900 & 0.847 & 0.790 & 0.717\\
20  & 0.862 & 0.810 & 0.717 & 0.626 & 0.518\\
50  & 0.690 & 0.593 & 0.442 & 0.322 & 0.213\\
100 & 0.481 & 0.361 & 0.213 & 0.127 & 0.069\\
\bottomrule
\end{tabular}
\end{table}

For example, after $20~\mu$s of propagation,
corresponding to approximately 4.08~km of SMF-28,
the predicted visibilities are

\begin{align}
V_{\mathrm{HOM}}(10~\mathrm{GHz}) &\simeq 0.717,\\
V_{\mathrm{HOM}}(20~\mathrm{GHz}) &\simeq 0.518,\\
V_{\mathrm{HOM}}(100~\mathrm{GHz}) &\simeq 0.069.
\end{align}

Conversely, for linewidths of approximately 1~GHz, the predicted visibility remains above 96\% even after $20~\mu$s.

The reduction is particularly pronounced for Lorentzian spectra because their long spectral tails make them sensitive to the quadratic spectral phase accumulated during propagation.
For an otherwise identical Gaussian spectral intensity, the degradation would generally be smaller at the same FWHM linewidth.

An important implication for fiber-loop quantum memories is that their broad operational spectral acceptance does not automatically imply that photons spanning
the entire bandwidth can be stored without
temporal-mode distortion.

The relevant figure of merit for interference-based applications is the spectral overlap after propagation,
including the dispersive spectral phase.

In the present model, attenuation reduces the probability of successful photon retrieval but does not directly reduce HOM visibility conditional on the detection of both photons.
The calculated visibility reduction arises exclusively from dispersive changes to the temporal and spectral
mode of the propagated photon.

For Fourier-transform-limited photons with Lorentzian spectral intensity, chromatic dispersion in standard
SMF-28 fiber can significantly reduce HOM interference visibility at linewidths in the tens-of-GHz range.

The effect becomes increasingly important with propagation time and may therefore represent a limitation for interference-based applications of broadband fiber-loop quantum memories.

Dispersion compensation or spectral narrowing can mitigate this effect, although each approach introduces different practical trade-offs.

\section{Experimental setup}

\begin{figure}
    \centering
    \includegraphics[width=1\linewidth]{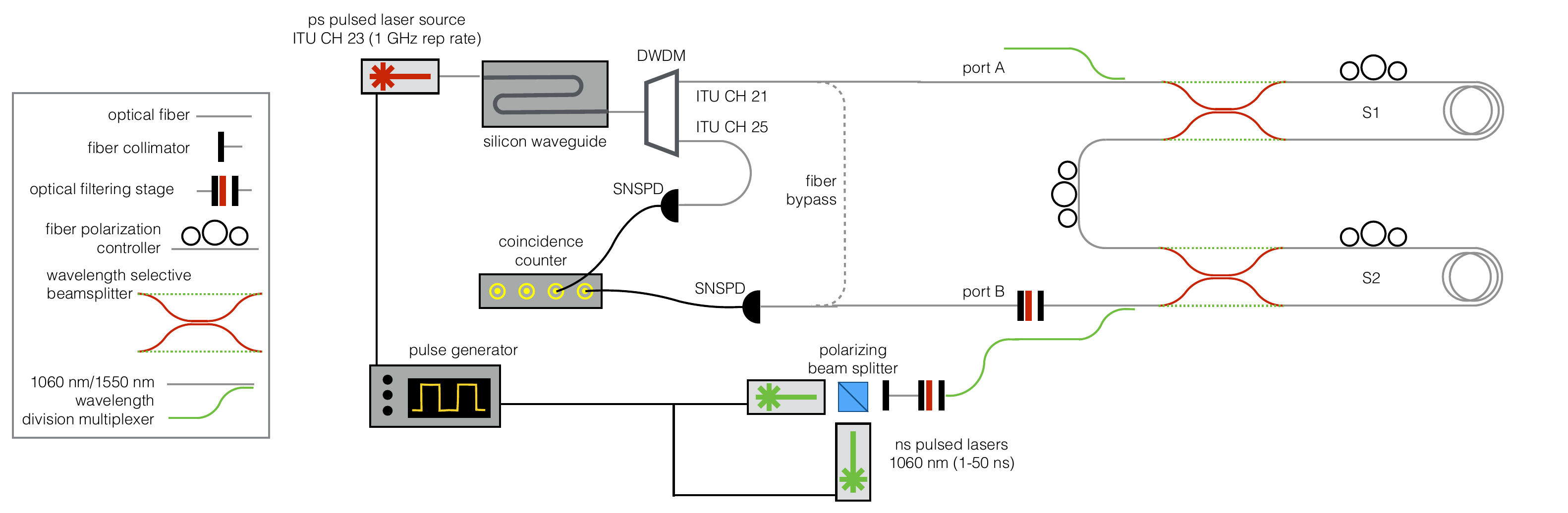}
    \caption{\textbf{Experimental setup used to characterize the fiber-loop quantum memory.}
A 9 ps pulsed laser centered on ITU channel~23 and operating at 1~GHz pumps a silicon waveguide, where photon pairs are generated by SFWM.
The generated photons are separated by 100~GHz DWDM filters, with the signal photon selected in ITU channel~21 and injected into the memory through port~A, while the idler photon is selected in ITU channel~25 and sent directly to the SNSPD.
Optical switching of the memory is driven by 1060~nm control pulses.
To make the switching operation insensitive to the polarization state of the stored photons, two 1060~nm control lasers with orthogonal polarizations are combined on a polarization beam splitter.
Before entering the memory from port~B, the combined control radiation is spectrally cleaned by a band-pass filter centered at 1060~nm and then injected into the memory counterpropagating with respect to the signal.
The control lasers are driven by programmable pulse generators synchronized to the 1~GHz pump-laser clock, enabling selective loading and retrieval of chosen temporal modes.
After retrieval from port~B, the signal photon passes through a 1500~nm long-pass filter to suppress residual 1060~nm control-field components.
The signal is then detected by an SNSPD and correlated with the idler detection for heralded and coincidence measurements.
A short fiber bypass connecting port~A directly to port~B provides a reference measurement that removes common setup contributions from the measured quantities and therefore isolates the losses and noise introduced by the double-Sagnac memory.
}
\label{fig:experimental_setup_generic}
\end{figure}

The experimental platform used throughout this work is schematically shown in Fig.~\ref{fig:experimental_setup_generic}. 
All experiments relied on the same fiber-loop quantum memory and differed only in the state-preparation and analysis stages inserted before and after the memory. 

Photon pairs were generated by SFWM in a silicon waveguide pumped by a pulsed laser centered on ITU channel~23 and operating at a repetition rate of 1~GHz. 
The generated photons were then spectrally separated by DWDM filters with a 100~GHz bandwidth. 
The signal photon was selected in ITU channel~21 and sent to the memory through port A, while the idler photon was selected in ITU channel~25 and sent directly to the detection system, where it was used either as a herald or as the second photon in coincidence measurements.

Optical switching in the memory was based on XPM between the signal field and 1060~nm optical control fields, which were spectrally cleaned by band-pass filters centered at 1060~nm before being injected from port~B (propagating in the opposite direction with respect to the signal). To make the switching operation insensitive to the polarization state of the stored photons, the control fields were implemented using two 1060~nm control lasers with orthogonal polarizations, combined on a polarization beam splitter before injection into the memory.\footnote{\XPMrefskum} \footnote{\XPMrefsPRIORkum}
The control pulses were driven by electronic pulse generators, whose clocks were derived from the 1~GHz pump laser. By adjusting their repetition rate and relative delays, the control pulses were made to temporally overlap with the target signal pulses inside the appropriate Sagnac interferometer, namely S1 for input and S2 for retrieval operations.
The precise temporal synchronization between the generated signal photons and the optical control pulses, together with the tunable control-pulse duration set between 1 and 50~ns, defined the temporal switching window of the memory. This enabled selective input, storage, and retrieval of the desired temporal modes.

\begin{figure}
    \centering
    \includegraphics[width=1\linewidth]{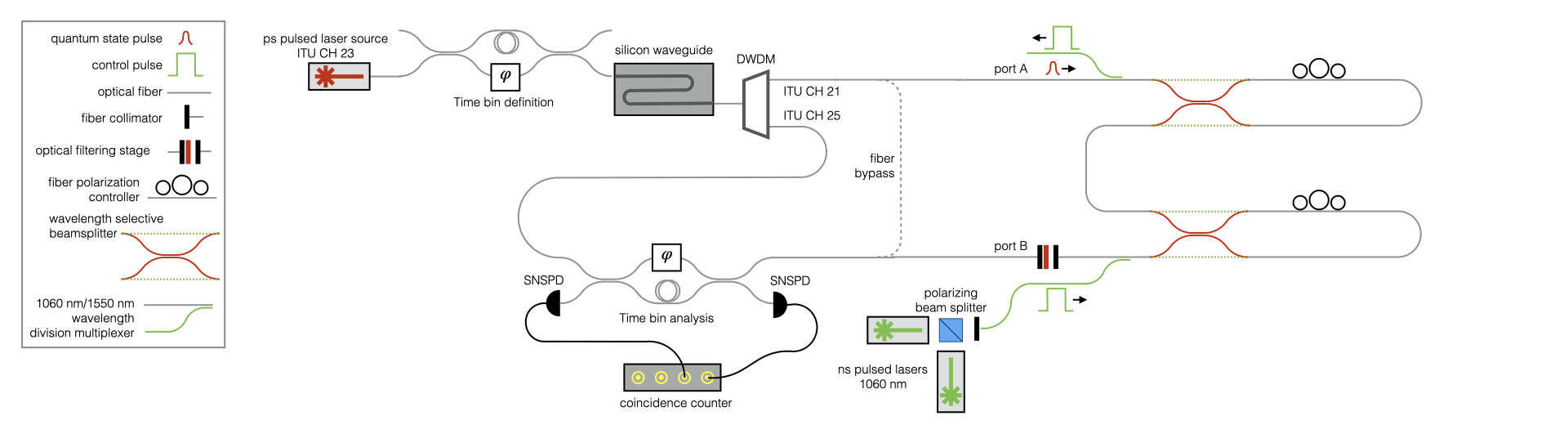}
    \caption{\textbf{Experimental setup for time-bin entanglement preservation during storage.}
An unbalanced Mach--Zehnder interferometer prepares coherent early and late pump pulses, which generate time-bin-entangled photon pairs through spontaneous four-wave mixing in a silicon waveguide.
After wavelength separation, the signal photon is injected into the fiber-loop quantum memory through port A and retrieved through port B, while the idler is sent directly to the analysis stage.
The retrieved signal and the idler are analyzed using a second unbalanced Mach--Zehnder interferometer, common to both photons, with a temporal delay matched to that of the preparation interferometer.
Phase-dependent two-photon coincidences are recorded using superconducting nanowire single-photon detectors (SNSPDs) to assess entanglement preservation after storage.
The dashed fiber bypass provides a reference measurement without storage.
}
\label{fig:experimental_setup_entanglement}
\end{figure}

After being retrieved from port~B, the signal pulse passed through a 1500 nm long-pass filter to suppress residual 1060~nm control-field components before reaching the detection stage.
For heralded single-photon and time-bin entanglement experiments, the retrieved signal was also detected by an SNSPD and correlated with the idler using a coincidence counter. 
For the time-bin entanglement measurements, the preparation and analysis stages were configured as shown in Fig.~\ref{fig:experimental_setup_entanglement}.
An unbalanced Mach--Zehnder interferometer placed before the silicon
waveguide prepared coherent early and late pump pulses, enabling
the generation of time-bin-entangled photon pairs by SFWM.
After retrieval, the signal and the idler were analyzed using a
second unbalanced Mach--Zehnder interferometer, common to both
photons, with a temporal delay matched to that of the pump
interferometer.

For the measurements performed with attenuated classical pulses, the same input and output paths of the memory were used, but the photon-pair source was removed and the 1 GHz pump laser was sent directly to the memory. Moreover, the detection stage consisted of either a photodetector for time-resolved analysis or a CCD camera for spectral analysis.

To isolate the contribution of the memory, reference measurements were performed by replacing it with a short fiber bypass connecting port A directly to port B. In this configuration, the signal followed the default input and output paths, but did not propagate through the loop. 
Normalization to this bypass reference removed only the contributions common to the bypass and memory configurations, including source, coupling, shared filtering, and detection losses. Therefore, the extracted quantities reflected the additional losses and noise introduced by the double-Sagnac memory module, including the Sagnac interferometers, the wavelength selective beamsplitters and the output filter.

\clearpage

\subsection{Optical losses}
\label{sec:optical_losses}

The lifetime extracted from the linear characterization corresponds to a fiber-loop round-trip loss of $0.24~\mathrm{dB}$. This value can be independently estimated from a passive-loss budget of the cavity, including the two Sagnac interferometers and the central connections between them.

The loss contribution of each Sagnac interferometer was estimated from the measured splitting coefficients of its fiber coupler together with the loop propagation and splice losses. For a Sagnac interferometer set in the reflective state, the reflected power is given by:
\begin{equation}
|r_{\mathrm{SI}}|^2 =
4 k^2 \xi^2 e^{-\alpha L}
\label{eq:sagnac_reflectivity}
\end{equation}
where \(k^2\) and \(\xi^2\) are the straight-through and cross power coefficients
of the fiber coupler, respectively, and \(e^{-\alpha L}\) is the loop attenuation factor accounting for propagation and splice losses inside the Sagnac interferometer.

The parameters used for the estimate are reported in
Table~\ref{tab:optical_losses}. The coupler coefficients were independently measured and are reported in dB. They were converted to linear power coefficients as:
\begin{equation}
k^2 = 10^{k^2_{\mathrm{dB}}/10},
\qquad
\xi^2 = 10^{\xi^2_{\mathrm{dB}}/10}
\end{equation}
For each Sagnac, the loop attenuation factor associated with propagation and splice losses was calculated as:
\begin{equation}
e^{-\alpha L_i}
=
10^{-2\left(2\ell_{\mathrm{sp}}+\alpha_{\mathrm{f}}L_i\right)/10}
\label{eq:loop_transmission}
\end{equation}
where \(i=1,2\) labels the two Sagnac interferometers,
\(\ell_{\mathrm{sp}}\) is the loss of a single splice,
\(\alpha_{\mathrm{f}}\) is the fiber propagation loss in \(\mathrm{dB/km}\),
and \(L_i\) is the loop length. The term \(2\ell_{\mathrm{sp}}\) accounts for
the two splices inside each Sagnac loop, while the prefactor 2 accounts for the symmetric contribution of the clockwise (CW) and counter-clockwise (CCW) fields.

Using \(L_1=L_2\sim105~\mathrm{m}\),
\(\alpha_{\mathrm{f}}=0.18~\mathrm{dB/km}\), \footnote{\refCorneringAttenuation} and
\(\ell_{\mathrm{sp}}=0.01~\mathrm{dB}\), \footnote{\refCorneringSPlice} we obtain
$ e^{-\alpha L_i}
\sim
0.9822 $.

For the first Sagnac interferometer, the measured coupler coefficients are
\(k^2_{\mathrm{dB}}=-3.03~\mathrm{dB}\) and
\(\xi^2_{\mathrm{dB}}=-3.02~\mathrm{dB}\), corresponding to
\(k^2=0.4977\) and \(\xi^2=0.4989\). The resulting reflectivity of the first Sagnac interferometer, calculated from Eq.~\eqref{eq:sagnac_reflectivity}, is $ |r_{\mathrm{S1}}|^2
\sim
0.9755$. Expressed as an equivalent mirror loss $\ell_{\mathrm{S1}}$, this gives:
\begin{equation}
\ell_{\mathrm{S1}}
=
-10\log_{10}\left(|r_{\mathrm{S1}}|^2\right)
\sim
0.107~\mathrm{dB}
\end{equation}
Analogously, for the second Sagnac interferometer, using
\(k^2_{\mathrm{dB}}=-3.13~\mathrm{dB}\) and
\(\xi^2_{\mathrm{dB}}=-2.93~\mathrm{dB}\), we obtain $|r_{\mathrm{S2}}|^2
\sim
0.9733$
corresponding to an equivalent mirror loss $\ell_{\mathrm{S2}}$ of:
\begin{equation}
\ell_{\mathrm{S2}}
=
-10\log_{10}\left(|r_{\mathrm{S2}}|^2\right)
\sim
0.117~\mathrm{dB}
\end{equation}

The total optical loss per cavity round trip, \(\ell_{\mathrm{b}}\), is then obtained by summing the two Sagnac mirror losses and the two central splice losses connecting the interferometers:
\begin{equation}
\ell_{\mathrm{b}}
=
\ell_{\mathrm{S1}}
+
\ell_{\mathrm{S2}}
+
2\ell_{\mathrm{sp}} 
\end{equation}
Substituting the estimated values $\ell_{\mathrm{b}}
\sim
0.244~\mathrm{dB}$.
This corresponds to a memory input/output efficiency $\eta_{\mathrm{b}}$:
\begin{equation}
\eta_{\mathrm{b}}
=
10^{-\ell_{\mathrm{b}}/10}
\sim
0.945 
\end{equation}
This independent estimate is in good agreement with the round-trip loss extracted from the memory lifetime measurement.

\begin{table}[t]
\centering
\caption{
Parameters used to estimate the optical round-trip loss of the fiber-loop
memory.
}
\label{tab:optical_losses}
\begin{tabular}{lcc}
\hline
Parameter & Sagnac 1 (S1) & Sagnac 2 (S2) \\
\hline
Loop length \(L_i\) & \(105~\mathrm{m}\) & \(105~\mathrm{m}\) \\
Fiber propagation loss \(\alpha_{\mathrm{f}}\) & \(0.18~\mathrm{dB/km}\) & \(0.18~\mathrm{dB/km}\) \\
Splice loss \(\ell_{\mathrm{sp}}\) & \(0.01~\mathrm{dB}\) & \(0.01~\mathrm{dB}\) \\
Straight-through coefficient \(k^2_\mathrm{dB}\) & \(-3.03~\mathrm{dB}\) & \(-3.13~\mathrm{dB}\) \\
Cross coefficient \(\xi_\mathrm{dB}^2\) & \(-3.02~\mathrm{dB}\) & \(-2.93~\mathrm{dB}\) \\
\hline
\end{tabular}
\end{table}

\section{Experimental datasets}
In this section, we report representative datasets used to evaluate the state-preservation performance of the memory across different encoding schemes, including time-bin, polarization-bin, and frequency-bin states.

Time-bin fidelities were measured using heralded single photons generated by SFWM, as described in Sec.~3.1 and in the Methods section of the main text. Time-bin entanglement measurements were performed using the configuration described in Sec.~3.2 of the main text. In contrast, polarization-bin and frequency-bin fidelities were measured using attenuated classical laser pulses, as described in Sec.~4 of the main text. Regardless of the encoding scheme, each input state can be written in the following general form:
\begin{equation}
    \ket{\psi} = \cos\theta \ket{0} + e^{i\phi}\sin\theta \ket{1}
    \label{eq_stato_generica}
\end{equation}
where the basis states $\ket{0}$ and $\ket{1}$ correspond to the two modes of the specific encoding: early $\ket{E}$ and late $\ket{L}$ temporal modes for time-bin qubits, horizontal $\ket{H}$ and vertical $\ket{V}$ polarizations for polarization-bin qubits, and two optical frequency components red $\ket{R}$ and blue $\ket{B}$, separated by the modulation frequency $f_{RF}$, for frequency-bin qubits.

The output state retrieved from the memory was reconstructed after different numbers of round trips and compared with a reference input state measured by replacing the memory with a short fiber bypass. This reference measurement includes the nonidealities of the state-preparation and analysis stages. As a result, the reported fidelities quantify the degradation introduced by the memory with respect to the experimentally prepared input state, rather than the distance from the nominal ideal state.

Given a reconstructed output density matrix $\rho_{\mathrm{out}}$, the fidelity was calculated with respect to the corresponding bypass reference state. When the reference is described by the pure state $\ket{\psi_{\mathrm{in}}}$, the fidelity $F$ is given by:
\begin{equation}
    F = \bra{\psi_{\mathrm{in}}}\rho_{\mathrm{out}}\ket{\psi_{\mathrm{in}}}
\end{equation}
The datasets reported below show the reconstructed values of $\theta$, $\phi$, $F$, and the corresponding density matrices as a function of the number of memory round trips.

\subsection{Time-bin heralded single photons}
The time-bin heralded single-photon dataset was acquired using the heralded SFWM source and the interferometric preparation-and-analysis configuration described in Sec.~3.1 and in the Methods section of the main text. More specifically, signal photons generated by SFWM in the Si waveguide were selected in ITU channel 21 and encoded in the time-bin basis defined by the early $\ket{E}$ and late $\ket{L}$ temporal modes, separated by 300~ps. Idler photons in ITU channel 25 were detected by an SNSPD and used as heralds. The reconstructed states were obtained from signal-idler coincidence measurements. 

Two configurations were tested: an independent preparation-and-analysis configuration, using two separate unbalanced Mach--Zehnder interferometers, and a self-referenced configuration, using the same interferometer for both preparation and projection.

\subsubsection{Independent state generation and projection}

\begin{figure}[h!]
\centering

\includegraphics[width=0.55\textwidth]{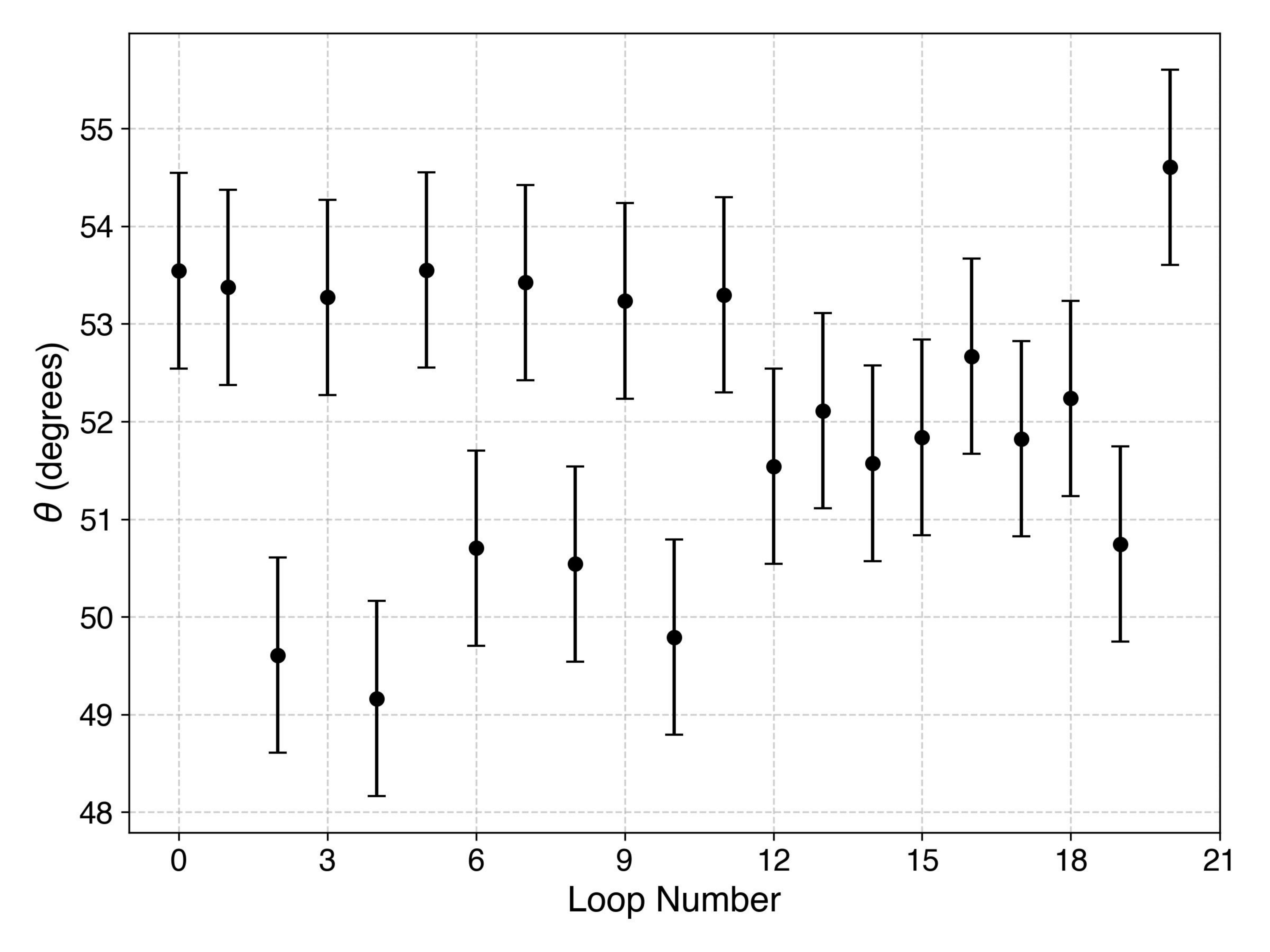}

\includegraphics[width=0.55\textwidth]{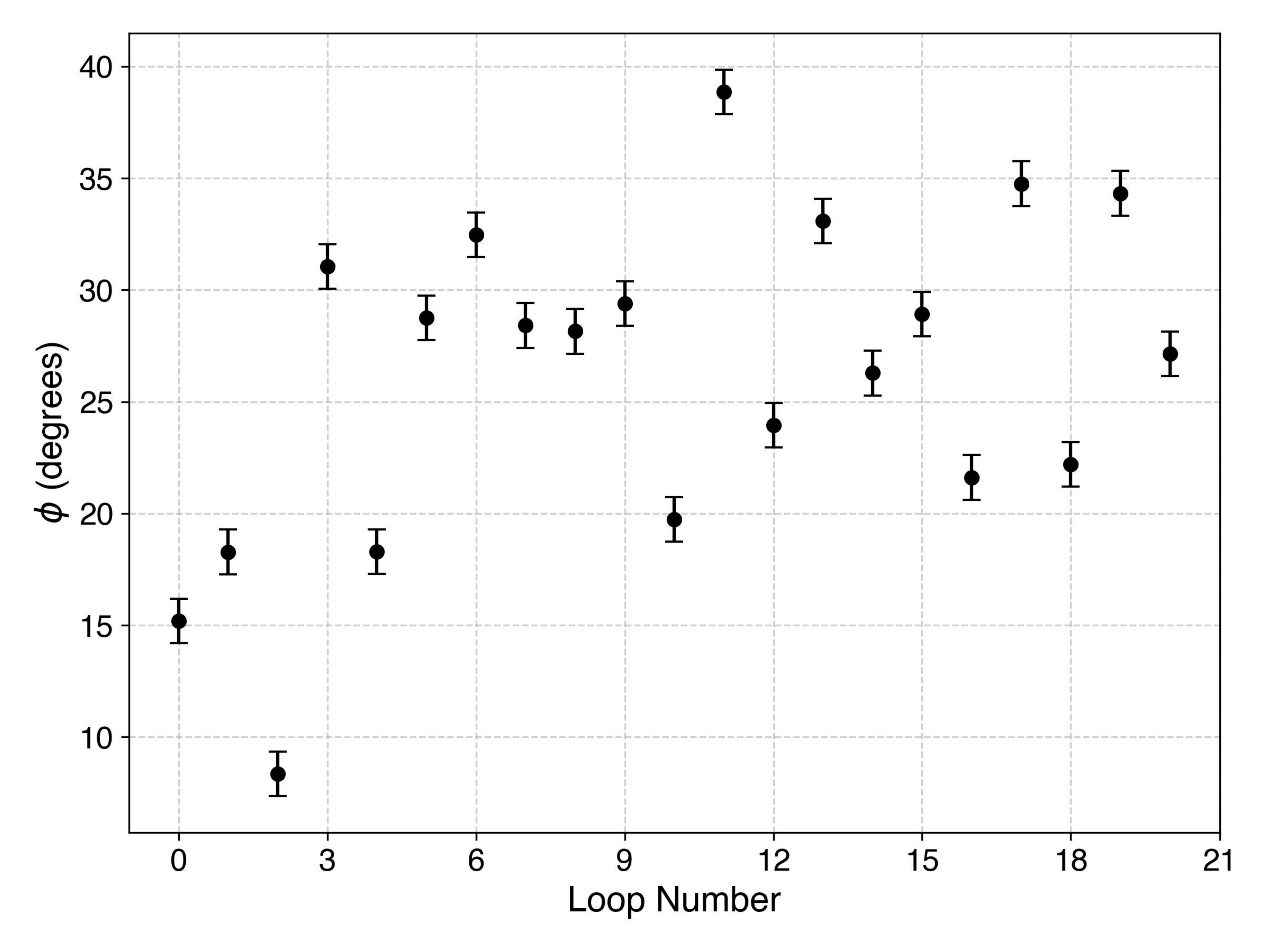}

\includegraphics[width=0.55\textwidth]{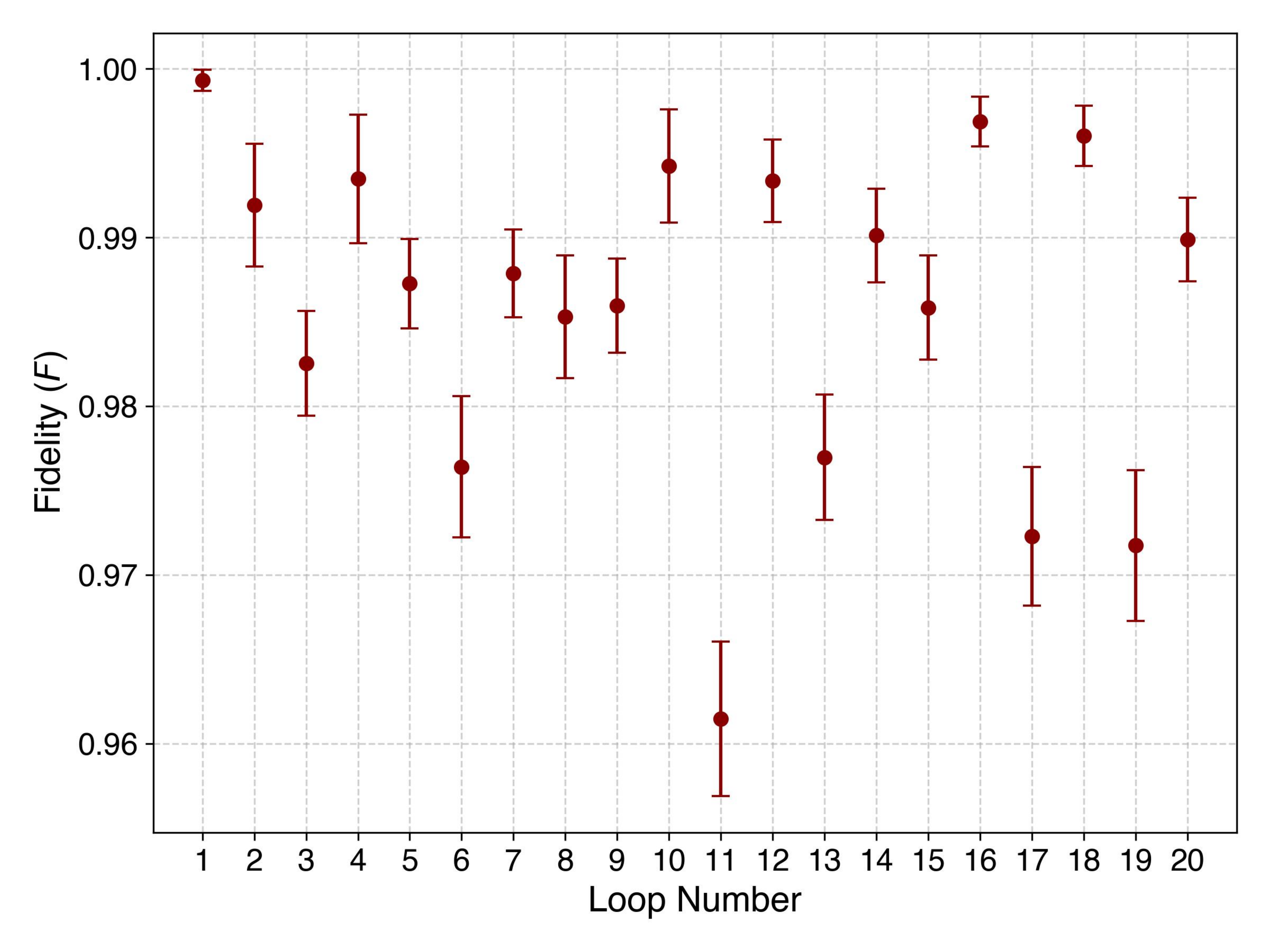}

\caption{\textbf{Heralded time-bin single-qubit storage with independent state preparation and analysis, nominally $\theta=53^\circ$ and $\phi=15^\circ$.}
The panels show the reconstructed values of $\theta$, $\phi$, and $F$ with respect to the bypass reference state as a function of the memory loop number. The input time-bin qubit was prepared and analyzed using two independent unbalanced Mach--Zehnder interferometers with 300~ps path imbalance each.}

\label{fig:time_bin_params}

\end{figure}

\begin{figure}[h!]
\centering

\includegraphics[width=1\textwidth]{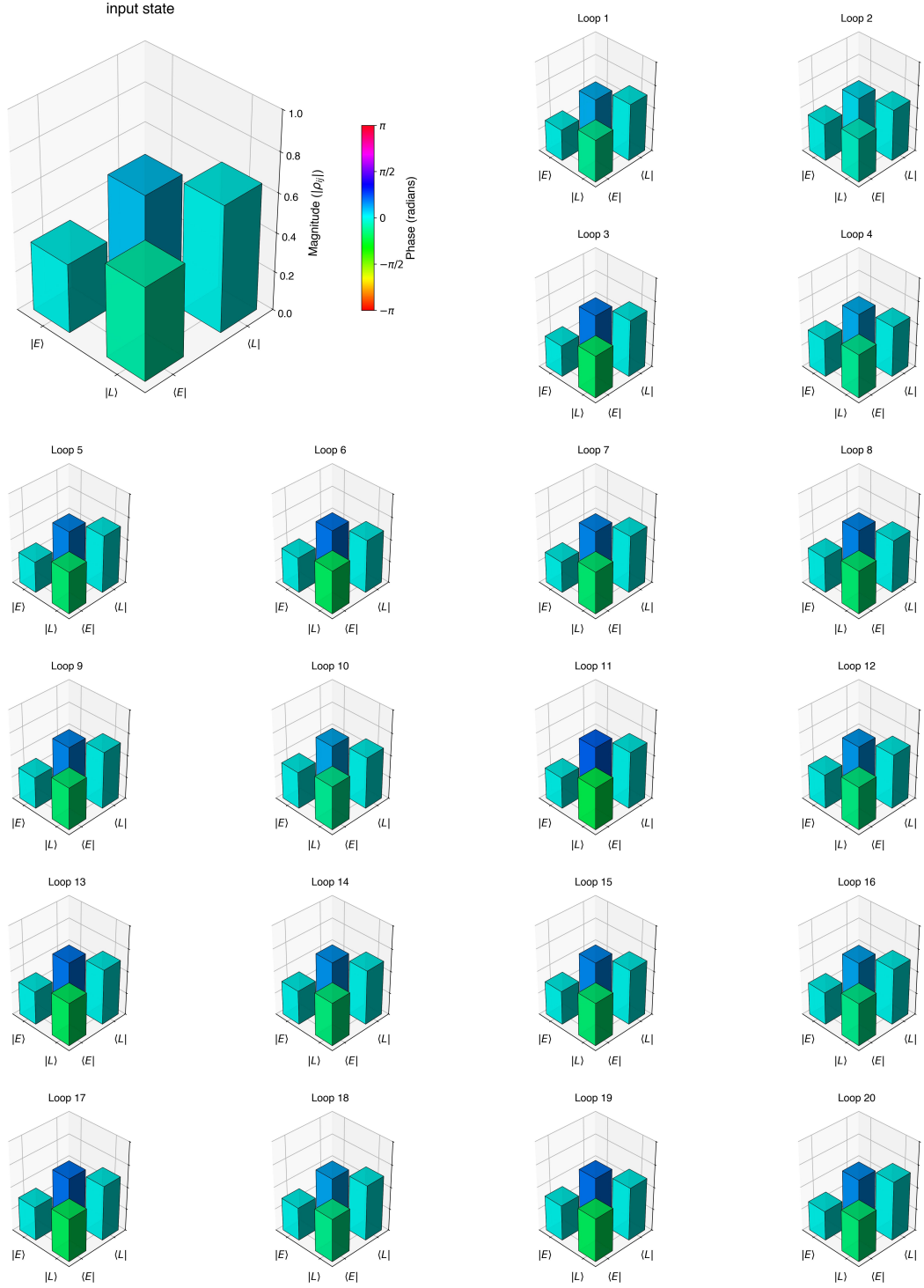}

\caption{\textbf{Density matrices for heralded time-bin single-qubit storage with independent preparation and analysis, nominally $\theta=53^\circ$ and $\phi=15^\circ$.}
Density matrices reconstructed from signal-idler coincidence measurements are shown for the bypass reference state and for the states retrieved after different memory loop numbers. The time-bin basis is defined by the early and late temporal modes, $\ket{E}$ and $\ket{L}$.}
\label{fig:time_bin_density}

\end{figure}

In the independent preparation-and-analysis configuration, the time-bin qubit was prepared using an unbalanced Mach--Zehnder interferometer and analyzed using an equally unbalanced but independently stabilized Mach--Zehnder interferometer. The nominal input state was prepared as:
\begin{equation}
    \ket{\psi_\mathrm{in}} =
    \cos(53^\circ)\ket{E}
    + e^{i15^\circ}\sin(53^\circ)\ket{L}
\end{equation}
The memory output was reconstructed after different numbers of round trips and compared with the reference input state measured by replacing the memory with a short fiber bypass. The corresponding data are shown in Figs. \ref{fig:time_bin_params} and \ref{fig:time_bin_density}. The larger fluctuations observed in this independent-interferometer configuration are attributed to residual relative phase noise between the preparation and analysis interferometers, rather than to memory-induced decoherence, as confirmed by the self-referenced measurement in \ref{subsec_self_referenced_single_qb_data}.

\clearpage

\subsubsection{Self-referenced state generation and projection}
\label{subsec_self_referenced_single_qb_data}
\begin{figure}[h!]
\centering

\includegraphics[width=0.55\textwidth]{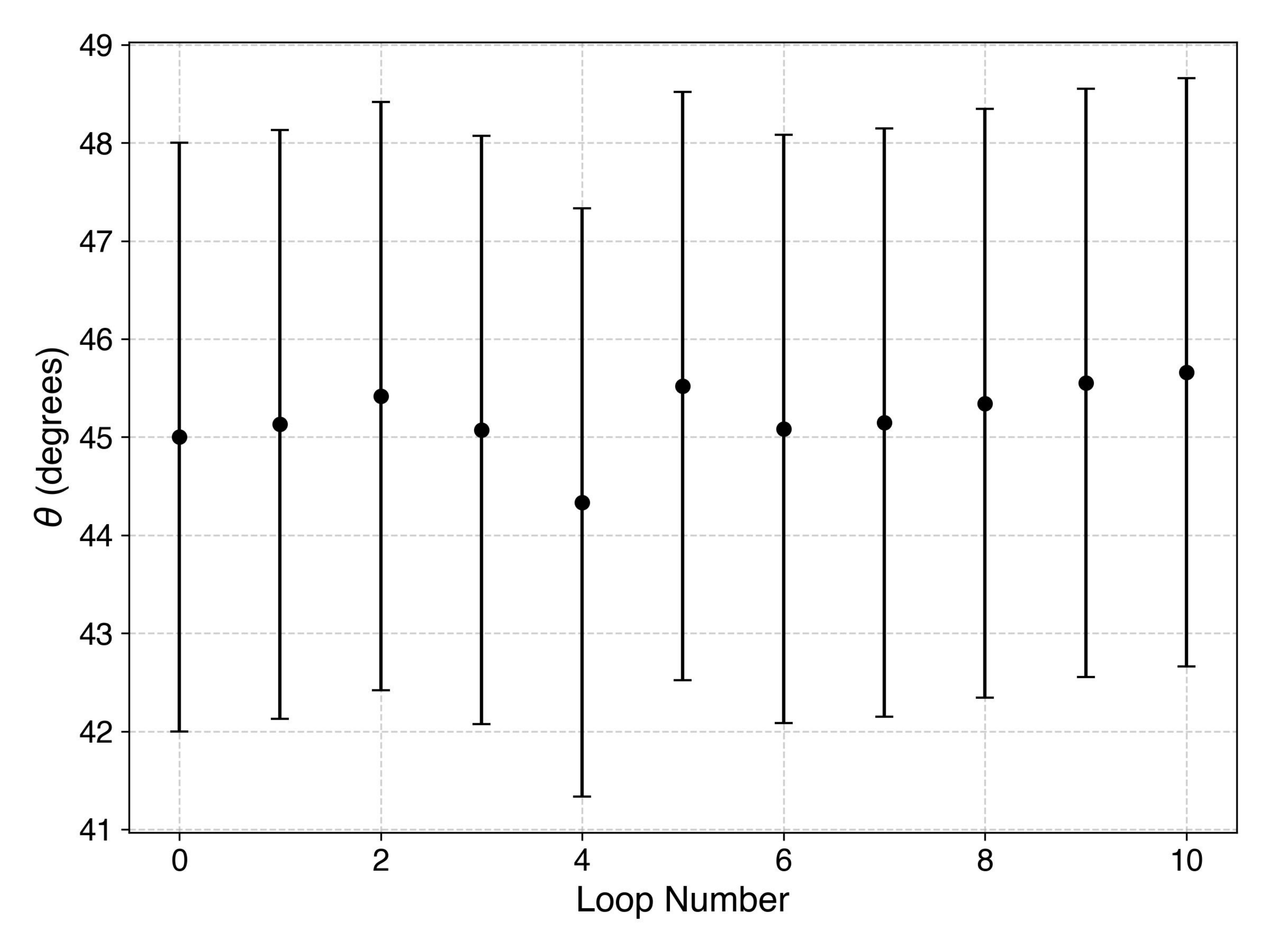}

\includegraphics[width=0.55\textwidth]{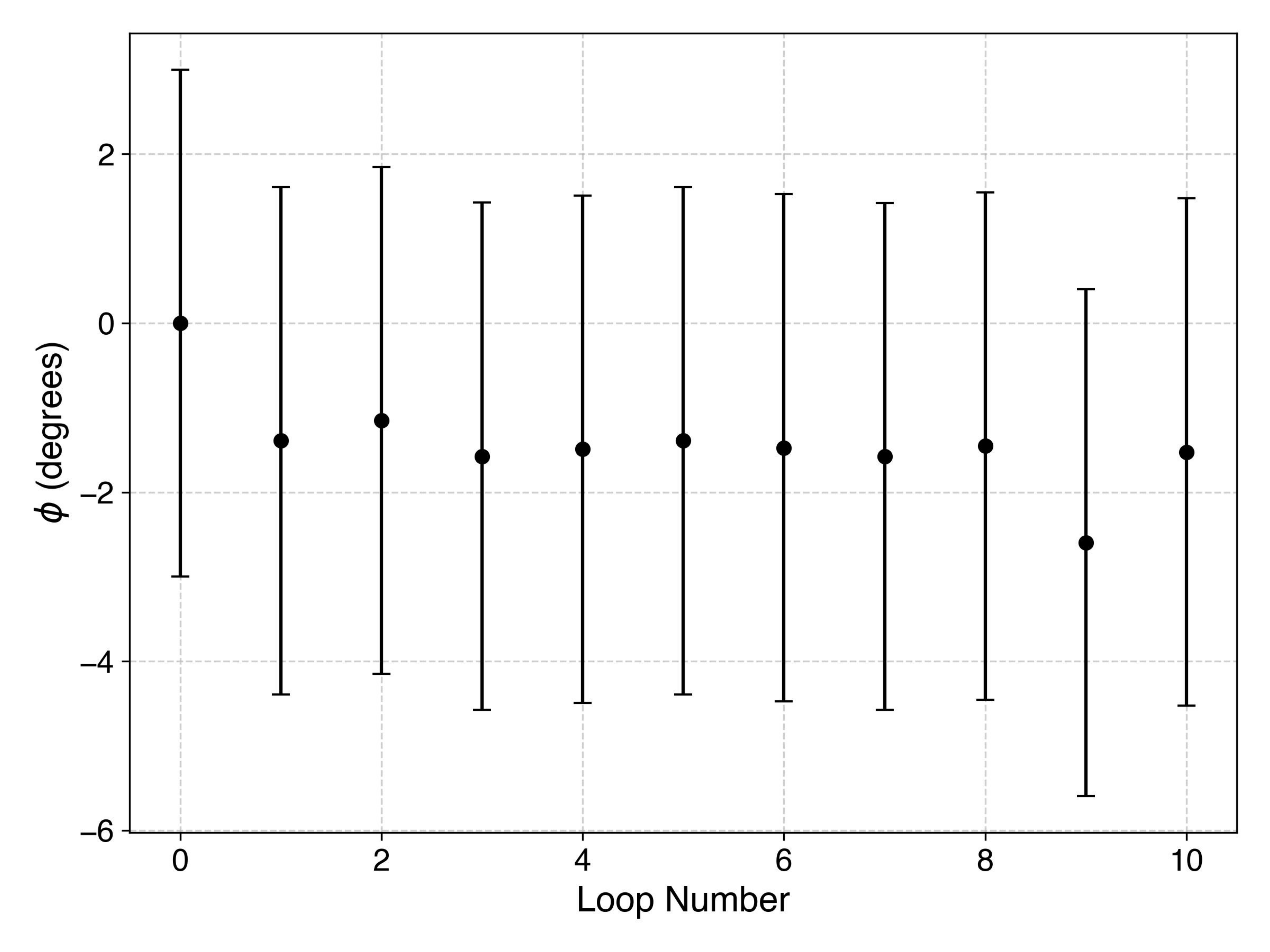}

\includegraphics[width=0.55\textwidth]{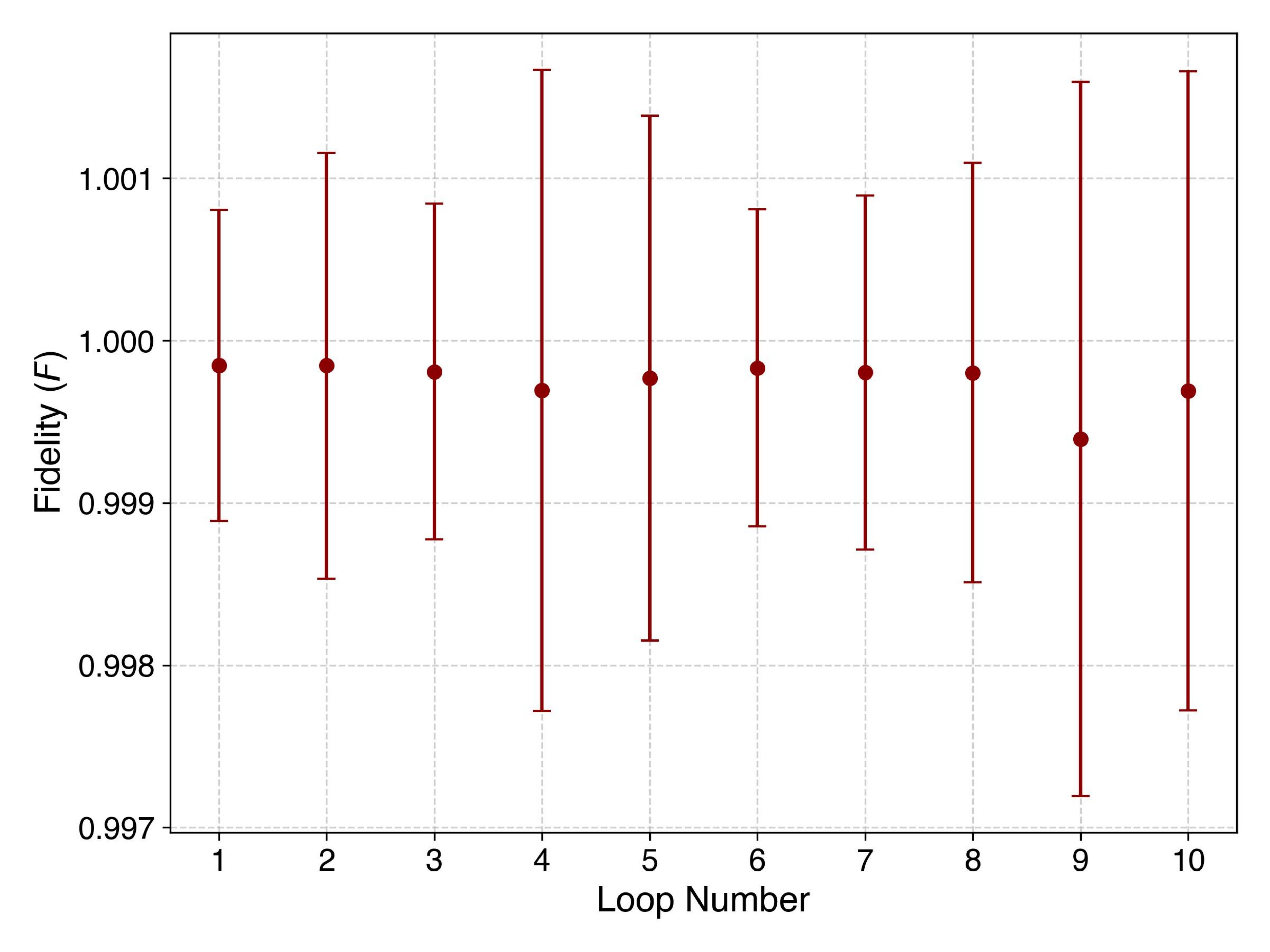}

\caption{\textbf{Heralded time-bin single-qubit storage in the self-referenced configuration.}
The panels show the reconstructed values of $\theta$, $\phi$, and $F$ as a function of the memory loop number. In this measurement, the same unbalanced Mach--Zehnder interferometer was used for both state preparation and projection, so that interferometric phase drifts are common-mode and the intrinsic effect of the memory on the time-bin coherence can be isolated.}
\label{fig:time_bin_params_si}

\end{figure}

A second time-bin dataset was acquired in a self-referenced configuration, where the same unbalanced Mach--Zehnder interferometer was used for both state preparation and projection. In this configuration, the interferometric phase used to define the prepared state is the same as that used for the analysis. This configuration prepares and analyzes the state:
\begin{equation}
    \ket{+} = \frac{1}{\sqrt{2}}\left(\ket{E}+\ket{L}\right)
    \label{eq_state_autoref}
\end{equation}
which corresponds to $\theta=45^\circ$ and $\phi=0^\circ$ in the parametrization of Eq. \ref{eq_stato_generica}.
The retrieved state was measured after different numbers of memory round trips and compared with the corresponding self-referenced input state. The dataset shown in Fig. \ref{fig:time_bin_params_si} provides a direct test of the intrinsic preservation of time-bin coherence by the memory.

\subsubsection{Time bin encoding with broadband photons.}

\begin{figure}[h!]
\centering

\includegraphics[width=0.55\textwidth]{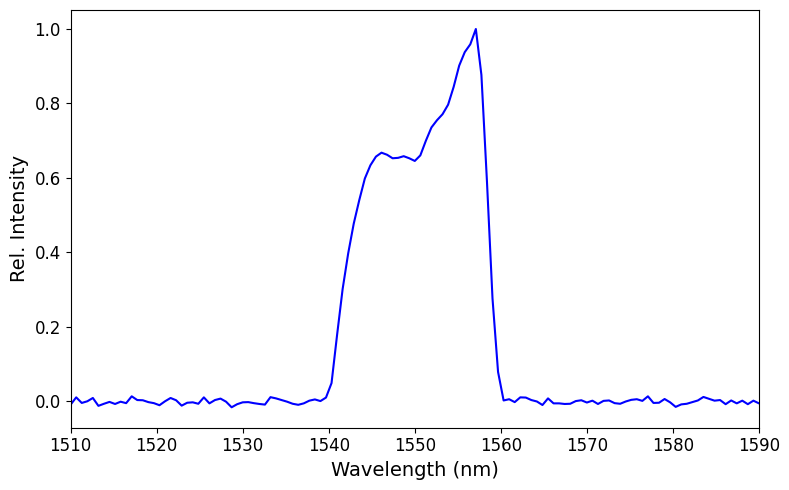}

\caption{\textbf{Heralded time-bin single-qubit storage in the self-referenced configuration using broadband photons.}
(a) Spectrum of the optical pulses used in the experiment.}
\label{fig:broadband_time_bins_spectrum}

\end{figure}

\begin{figure}[h!]
\centering

\includegraphics[width=0.55\textwidth]{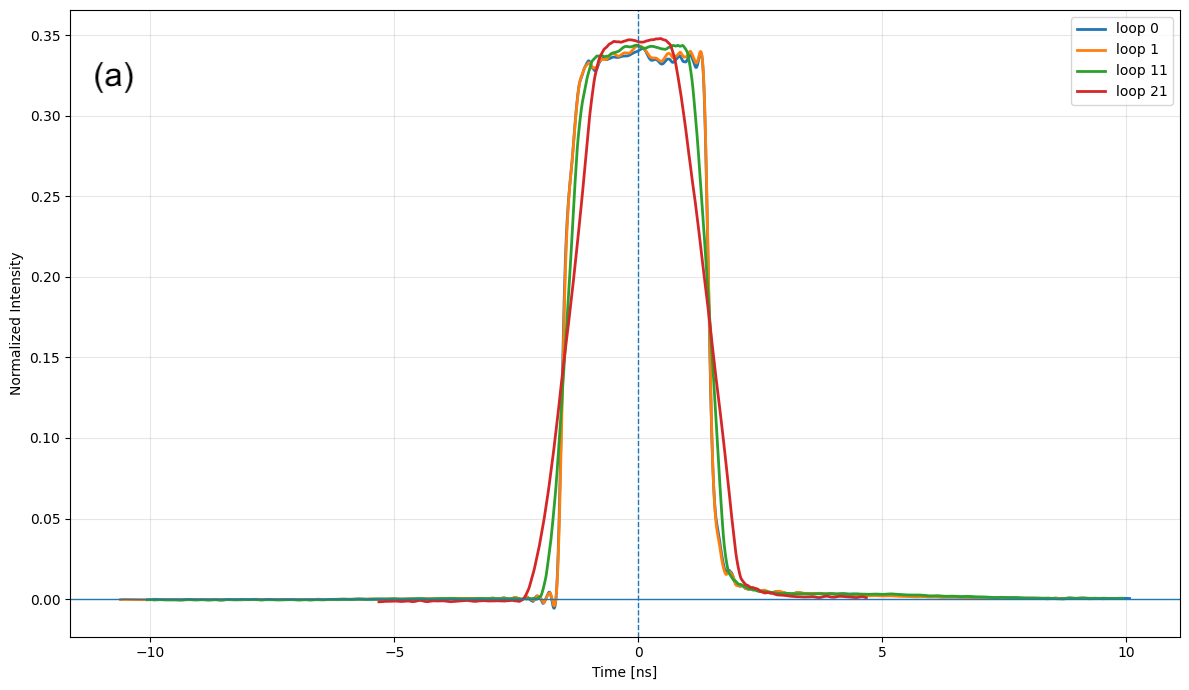}

\includegraphics[width=0.55\textwidth]{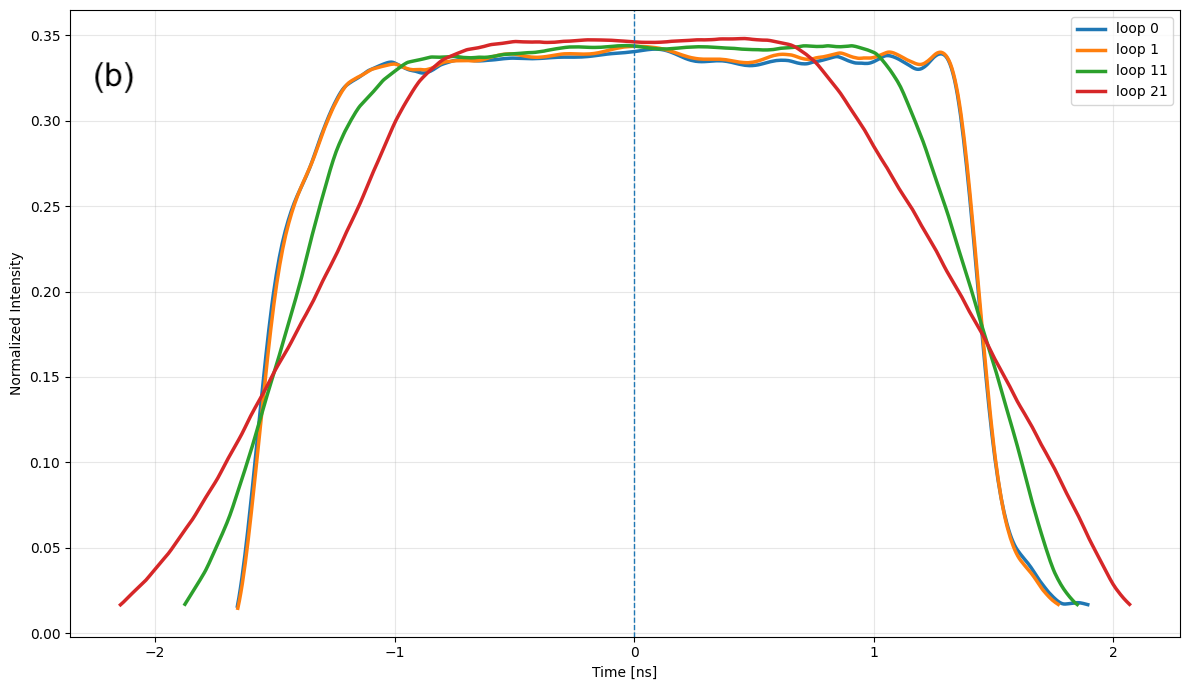}

\includegraphics[width=0.55\textwidth]{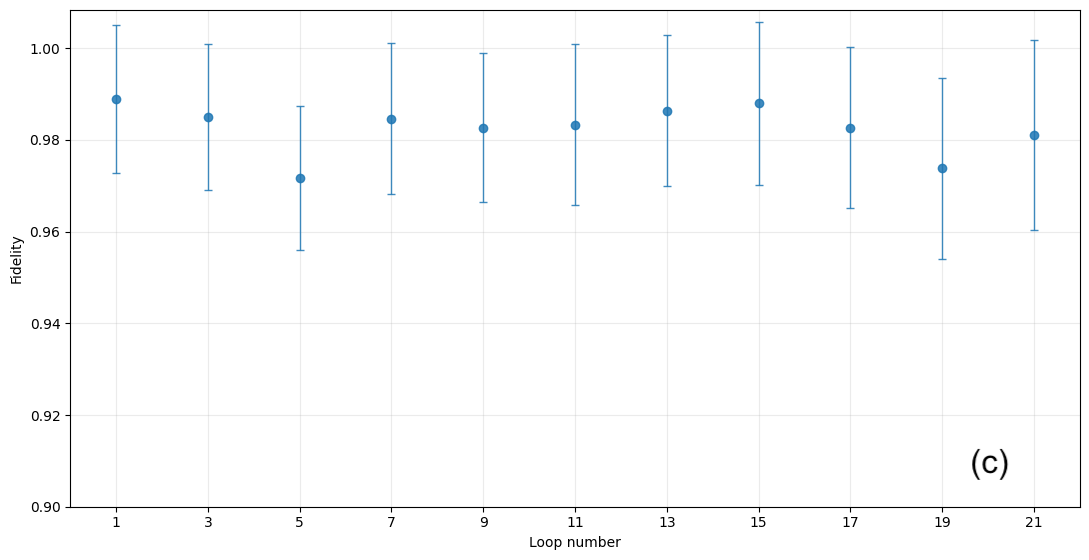}

\caption{\textbf{Heralded time-bin single-qubit storage in the self-referenced configuration using broadband photons.}
(a) Temporal profiles of the broadband photons used in the experiment after the bypass (loop 0) and after spending 1,11 and 21 loops inside the memory. Thje profile (b) Zoom of panel (a) to magnify the effect of fiber dispersion on the time profile. (c) Fidelity of the broadband photons when time-bin encoded.}
\label{fig:broadband_time_bins}

\end{figure}

Fiber based platforms suffer from the intrinsic dispersion of optical fibers, whose value is $\sim$18 ps/nm/km around 1550 nm. This means that broad (>5 nm) quantum state pulses would experience significant chirping due to the fiber dispersion while being stored in the memory, and the chirping would depend on the storage time. Nevertheless, this does not compromise the visibility of quantum interference when qubits are encoded on some degree of freedom. If qubits are encoded for instance in polarization or time bin, each frequency component for both polarizations or both time bins would experience the same dispersion induced delay, thus fully preserving the interference. The only caveat is that the control pulses used for the output operation when outputting the stored photons the control pulses should be long enough to accomodate the temporal broadening induced by dispersion. This is absolutely possible in our system. Specifically, a full bandwidth pulse of 12.5 THz would experience a temporal broadening of about 7 ns due to dispersion after spending 20 loops inside the memory.  This broadening is much shorter than the typical control pulse duration used in the various experiments reported in this work (10 ns to 30 ns). In the case of frequency bin encoding, we notice that in most of the literature the spacing between the bins is typically <20  GHz and hence the band much less than such value.

To practically prove this point we have performed performed a experiment time-bin fidelity in which 20 nm (about 2.5 THz) broad, 1 ns long pulses are time-bin encoded. The spectrum of the encoded pulses is shown in Figure \ref{fig:broadband_time_bins_spectrum}. The experiment is performed in the same self-referencing configuration described in section \ref{subsec_self_referenced_single_qb_data} above using a Mach-Zehnder interferometer with 10 ns unbalance. The pulses were created by passing attenuated light emitted from a superluminescent diode throrugh a standard CWDM bandpass filter centered at 1550 nm and an electro-optical amplitude modulator was used to carve 1 ns long rectangular pulses. The pulses, shown in figure \ref{fig:broadband_time_bins} are subject to a clear visible broadening due to the dispersion

The results show fidelities >97\% for all measured loop numbers, seemingly independent of the time spent inside the memory.

\clearpage

\subsection{Polarization-bin states}
Polarization-bin datasets were acquired using the polarization preparation and analysis configuration described in Sec.~4 of the main text. The computational basis was defined by the horizontal $\ket{H}$ and vertical $\ket{V}$ polarization states. The input states were prepared using free-space polarization optics and reconstructed after retrieval from the memory by projecting the output field onto different polarization bases. The fidelity was calculated with respect to the reference input state measured by replacing the memory with a short fiber bypass.

These measurements were performed with attenuated classical pulses rather than heralded single photons. They therefore characterize the ability of the memory to preserve the polarization degree of freedom during propagation and switching, with the retrieved states compared directly with the corresponding bypass reference states. This is nontrivial since the memory is implemented in optical fiber, where residual birefringence can accumulate polarization rotations over multiple round trips.

\subsubsection{Polarization-bin input state with nominal $\theta=70^\circ$}
\begin{figure}[h!]
\centering

\includegraphics[width=0.55\textwidth]{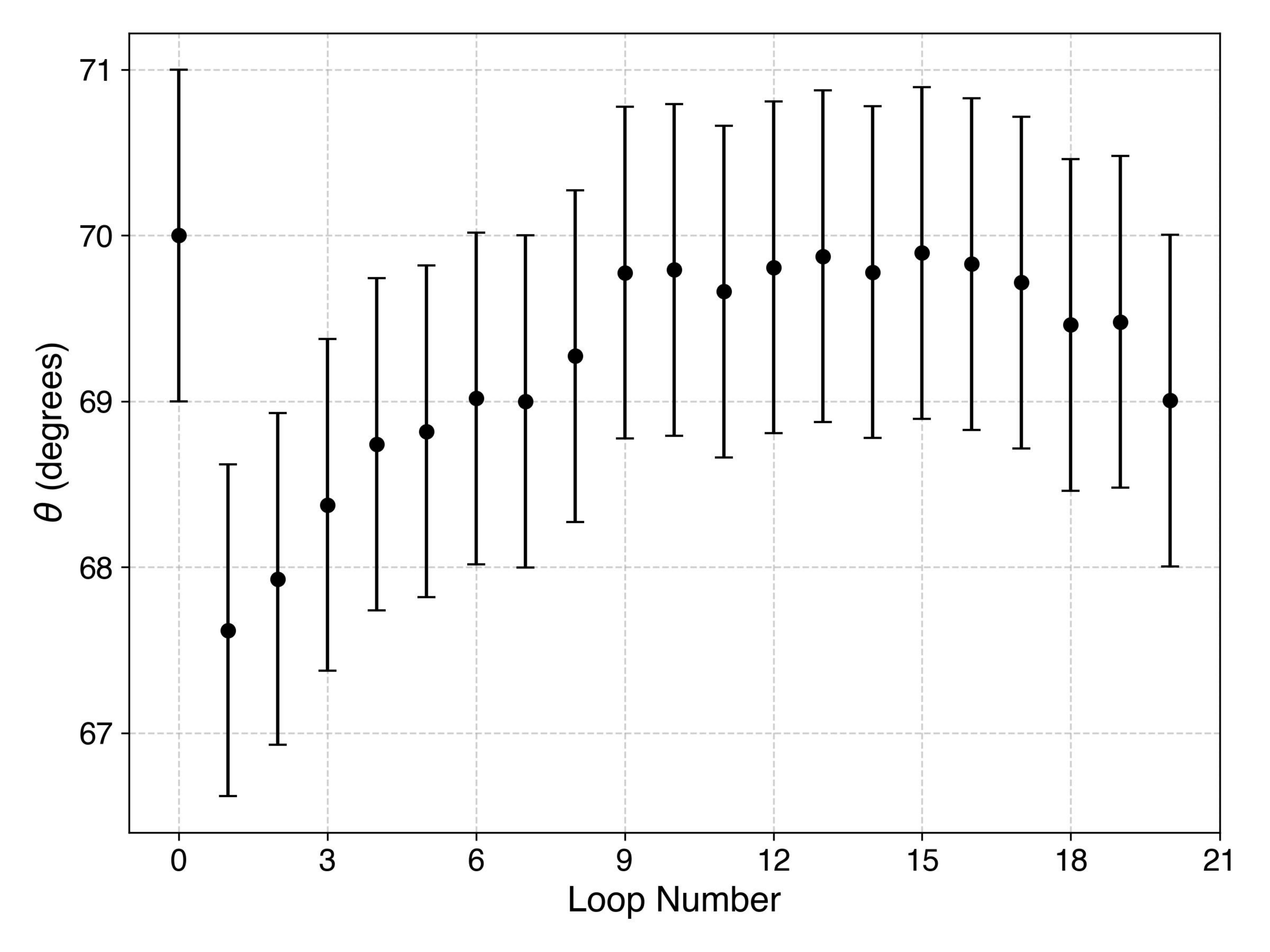}

\includegraphics[width=0.55\textwidth]{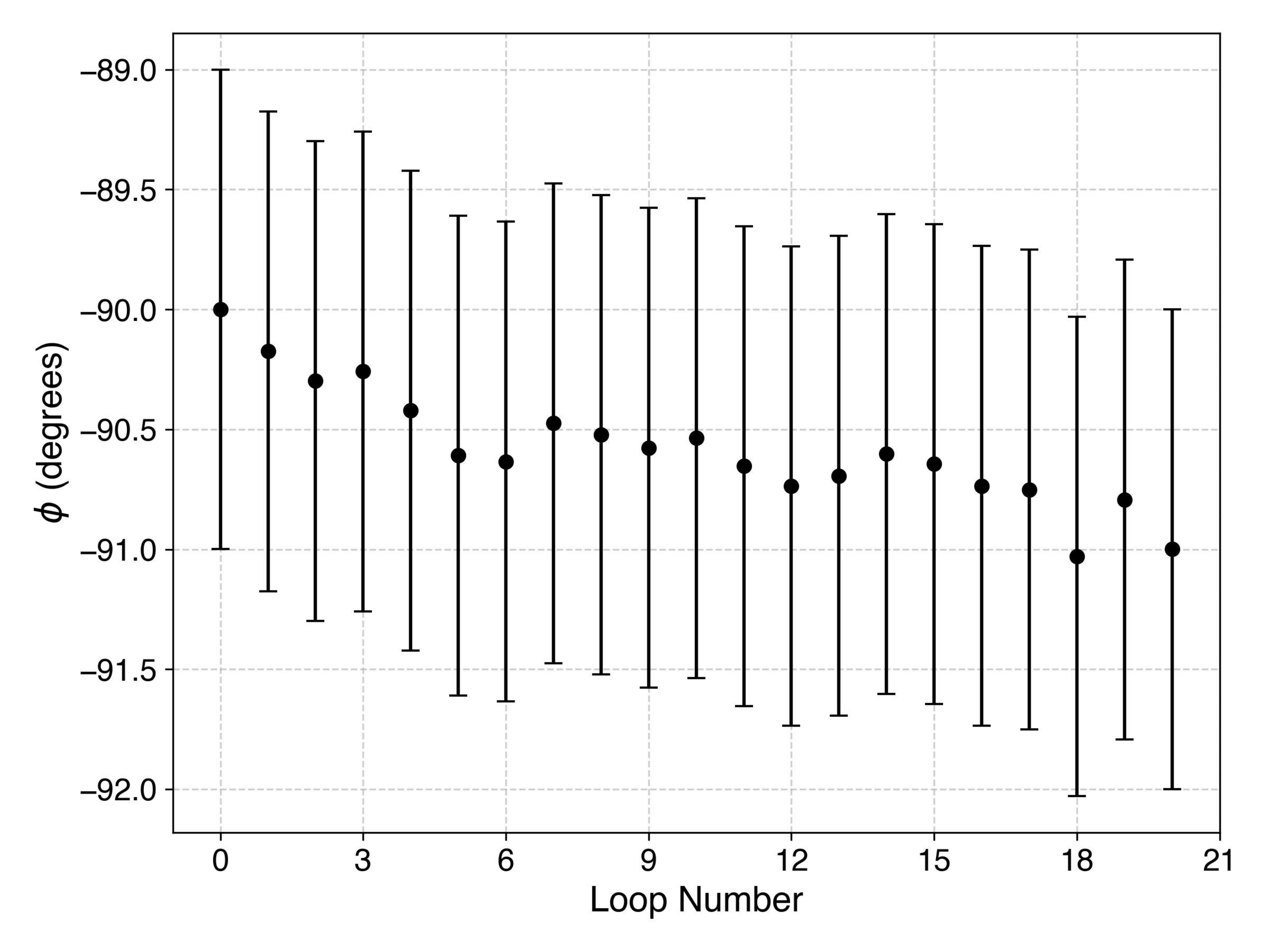}

\includegraphics[width=0.55\textwidth]{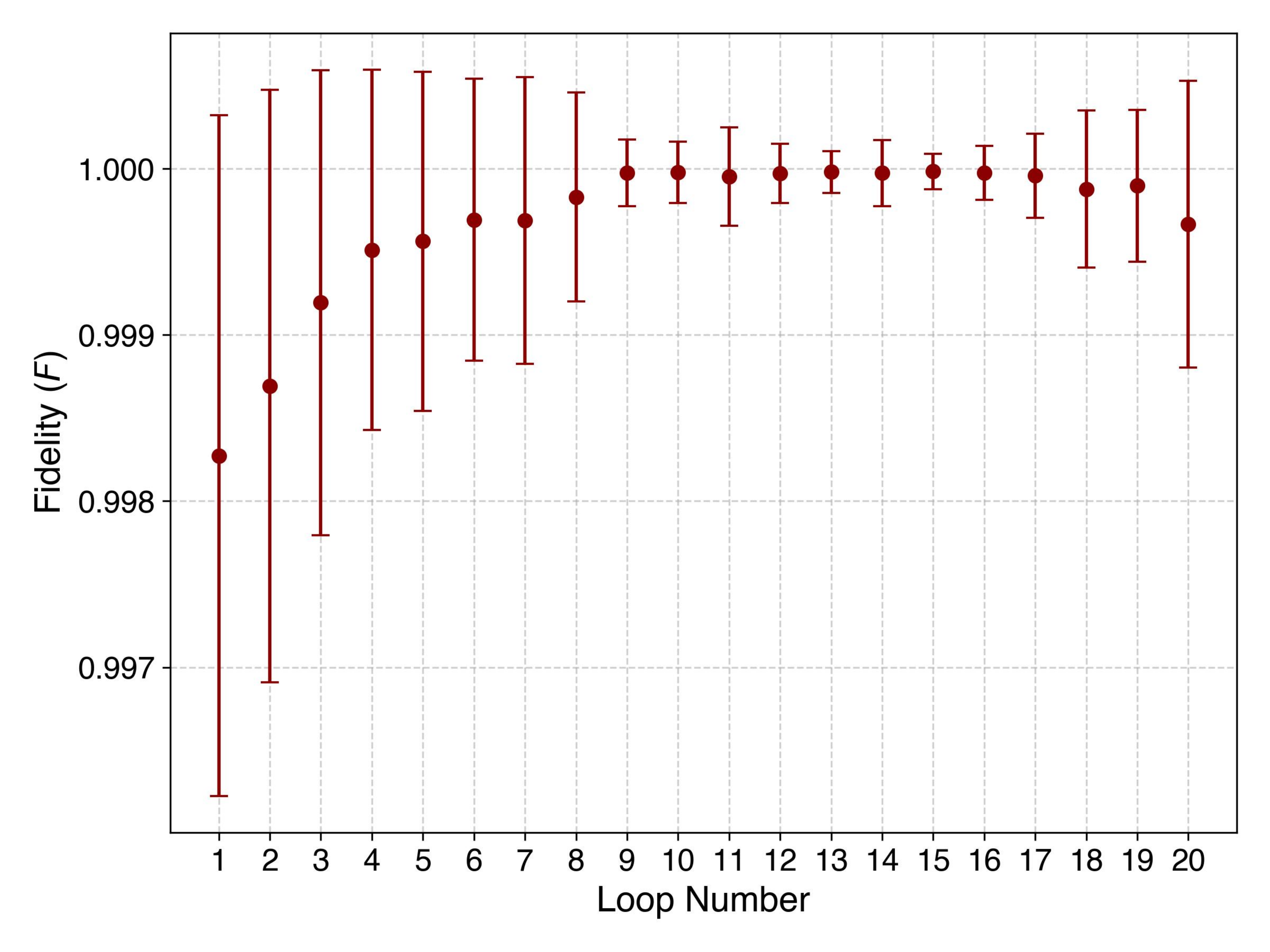}

\caption{\textbf{Polarization-bin state with nominal $\theta=70^\circ$ and $\phi=-90^\circ$.}
The panels show the reconstructed values of $\theta$, $\phi$, and $F$ with respect to the bypass reference state as a function of the memory loop number.}
\label{fig:pol_70deg_params}

\end{figure}

In this dataset, the input state was prepared with nominal amplitude angle $\theta=70^\circ$ and azimuthal angle $\phi=-90^\circ$. It therefore can be written as:
\begin{equation}
    \ket{\psi_\mathrm{in}} =
    \cos(70^\circ)\ket{H}
    + e^{-i90^\circ}\sin(70^\circ)\ket{V}
\end{equation}
The retrieved state was reconstructed after different numbers of memory round trips and compared with the bypass reference state. This allowed the evolution of $\theta$, $\phi$, and $F$ to be tracked as a function of storage time (Figs.~\ref{fig:pol_70deg_params} and~\ref{fig:pol_70deg_density}).

\begin{figure}[h!]
\centering

\includegraphics[width=1\textwidth]{fig_24_subplots_density_70_deg.pdf}

\caption{\textbf{Density matrices for the polarization-bin state with nominal $\theta=70^\circ$ and $\phi=-90^\circ$.}
Density matrices are shown for the bypass reference state and for the states retrieved after different memory loop numbers. The reconstruction is performed in the polarization basis $\{\ket{H},\ket{V}\}$.}
\label{fig:pol_70deg_density}

\end{figure}

\subsubsection{Polarization-bin input state with nominal $\theta=60^\circ$}

A second polarization-bin dataset was acquired for a different input state to evaluate the robustness of polarization preservation under a different preparation condition. 

\begin{figure}[h!]
\centering

\includegraphics[width=0.55\textwidth]{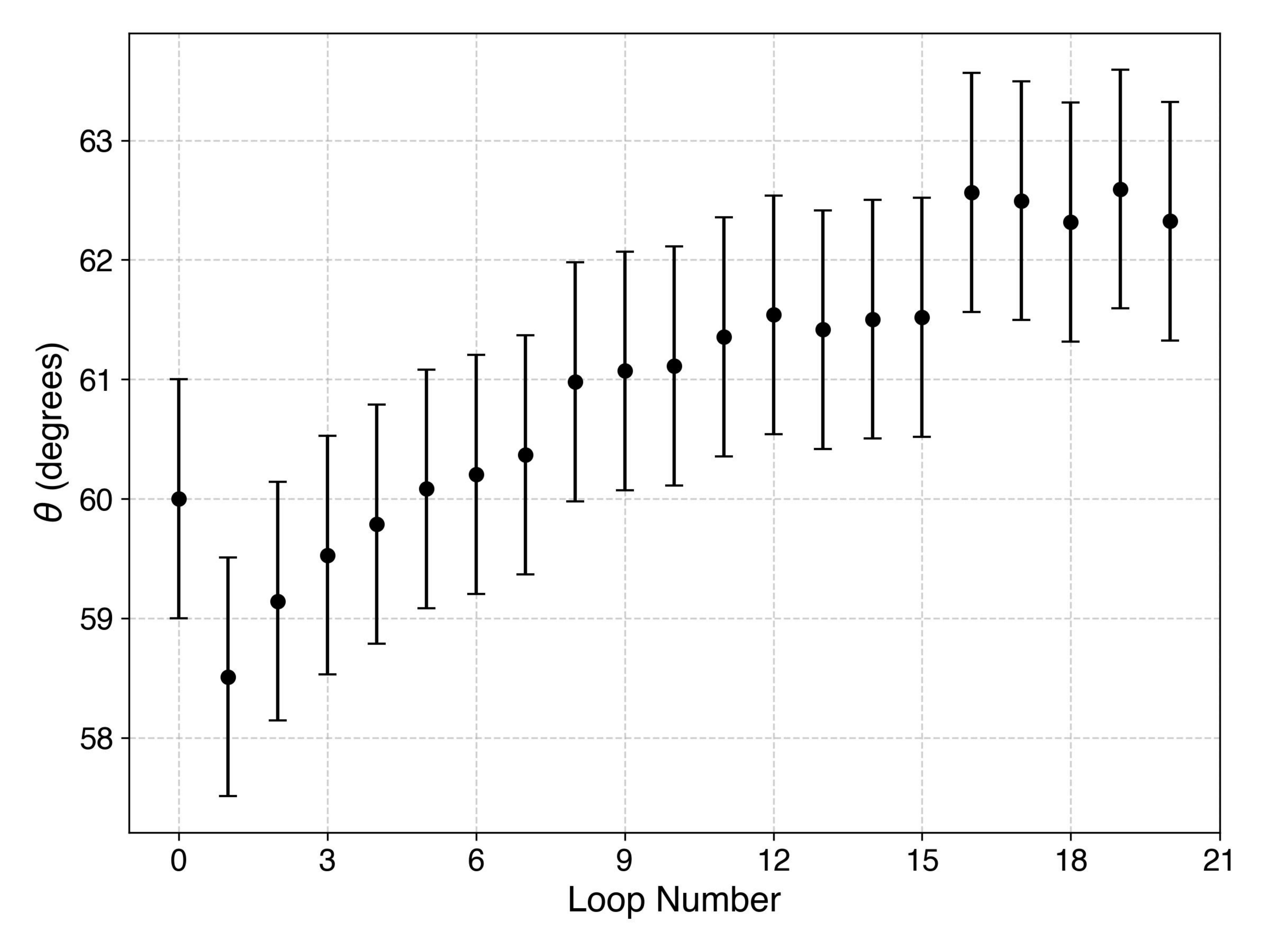}

\includegraphics[width=0.55\textwidth]{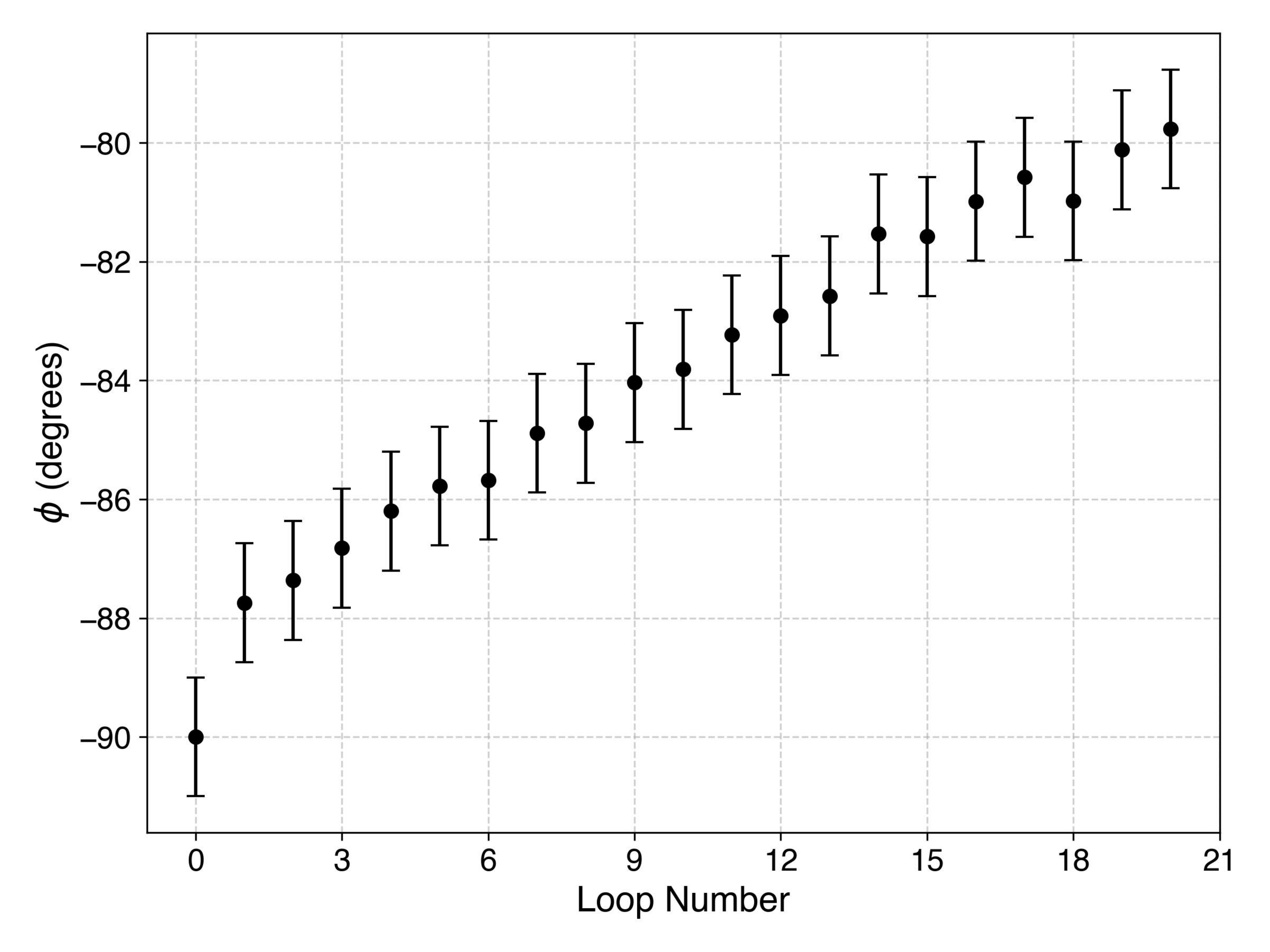}

\includegraphics[width=0.55\textwidth]{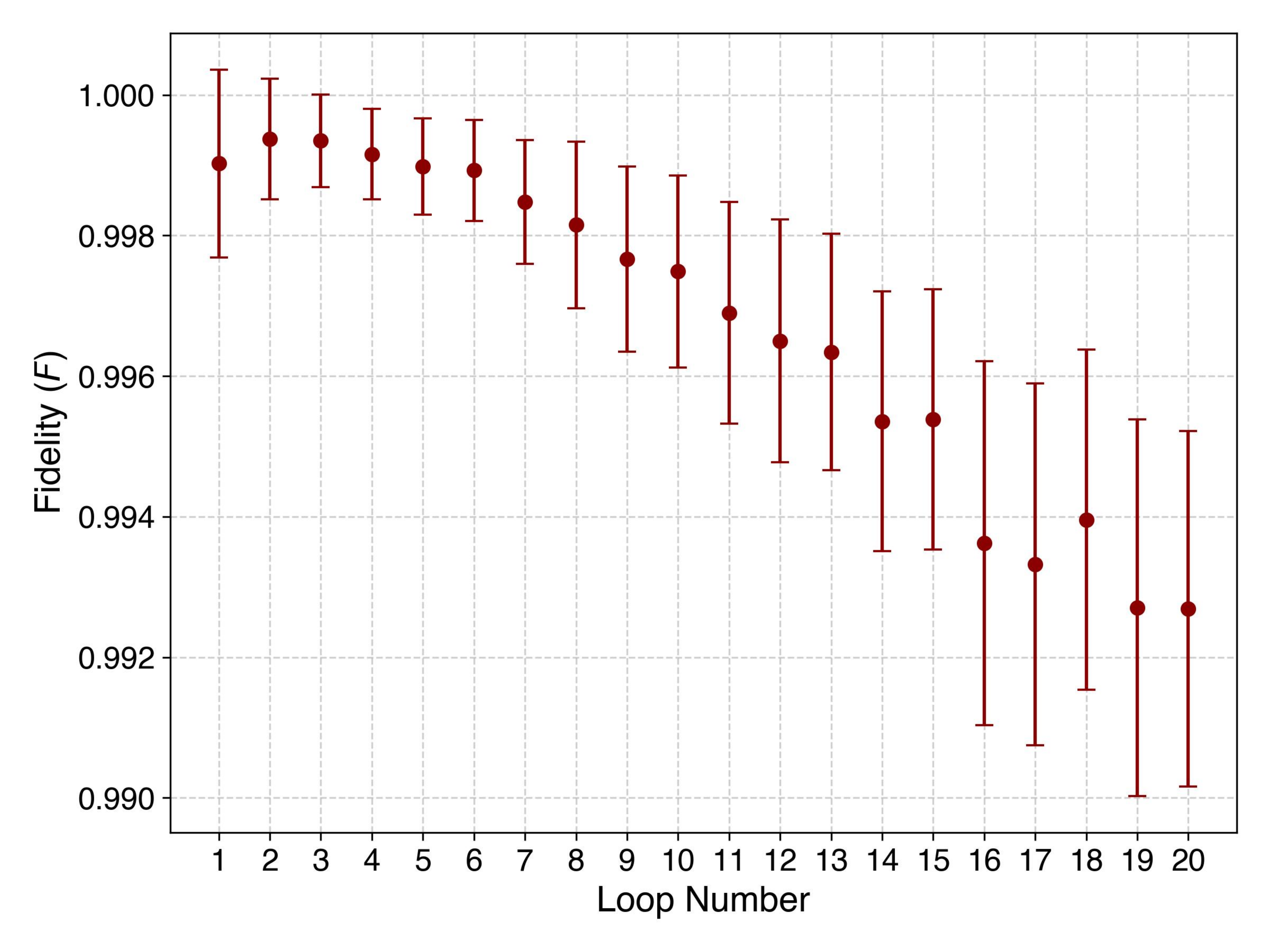}

\caption{\textbf{Polarization-bin state with nominal $\theta=60^\circ$ and $\phi=-90^\circ$.}
The panels show the reconstructed values of $\theta$, $\phi$, and $F$ with respect to the bypass reference state as a function of the memory loop number.}
\label{fig:pol_60deg_params}

\end{figure}

The state was prepared with a nominal amplitude angle $\theta=60^\circ$ and azimuthal angle $\phi=-90^\circ$. It therefore can be written as:
\begin{equation}
    \ket{\psi_\mathrm{in}} =
    \cos(60^\circ)\ket{H}
    +e^{-i90^\circ}\sin(60^\circ)\ket{V}
\end{equation}
As for the $\theta=70^\circ$ dataset, the retrieved state was reconstructed after different numbers of memory round trips and compared with the bypass reference state. This allowed the evolution of $\theta$, $\phi$, and $F$ to be tracked as a function of storage time (Figs.~\ref{fig:pol_60deg_params} and~\ref{fig:pol_60deg_density}).

The small systematic drift observed to this second input state is likely due to polarization rotation in the buffer, as single‐mode fiber is typically not depolarizing. Furthermore, none of the optical elements in the buffer are known to be depolarizing.

\begin{figure}[h!]
\centering

\includegraphics[width=1\textwidth]{fig_24_subplots_density_60deg.pdf}

\caption{\textbf{Density matrices for the polarization-bin state with nominal $\theta=60^\circ$ and $\phi=-90^\circ$.}
Density matrices are shown for the bypass reference state and for the states retrieved after different memory loop numbers. The reconstruction is performed in the polarization basis $\{\ket{H},\ket{V}\}$.}
\label{fig:pol_60deg_density}

\end{figure}

\subsubsection{Polarization-bin input state with nominal $\theta=45^\circ$}

A third polarization-bin dataset was acquired for a different input state to evaluate the robustness of polarization preservation under a different preparation condition. 

\begin{figure}[h!]
\centering

\includegraphics[width=0.55\textwidth]{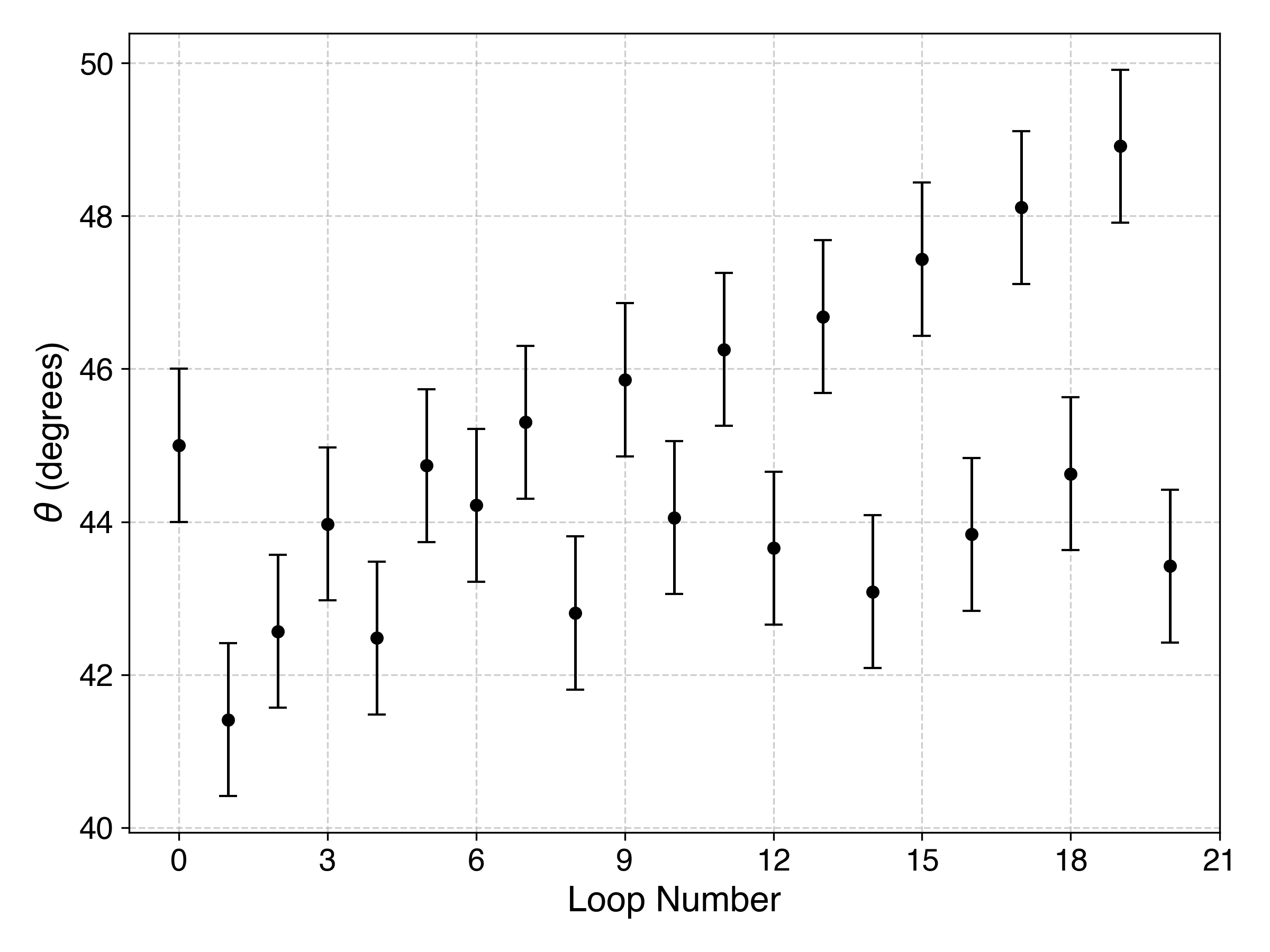}

\includegraphics[width=0.55\textwidth]{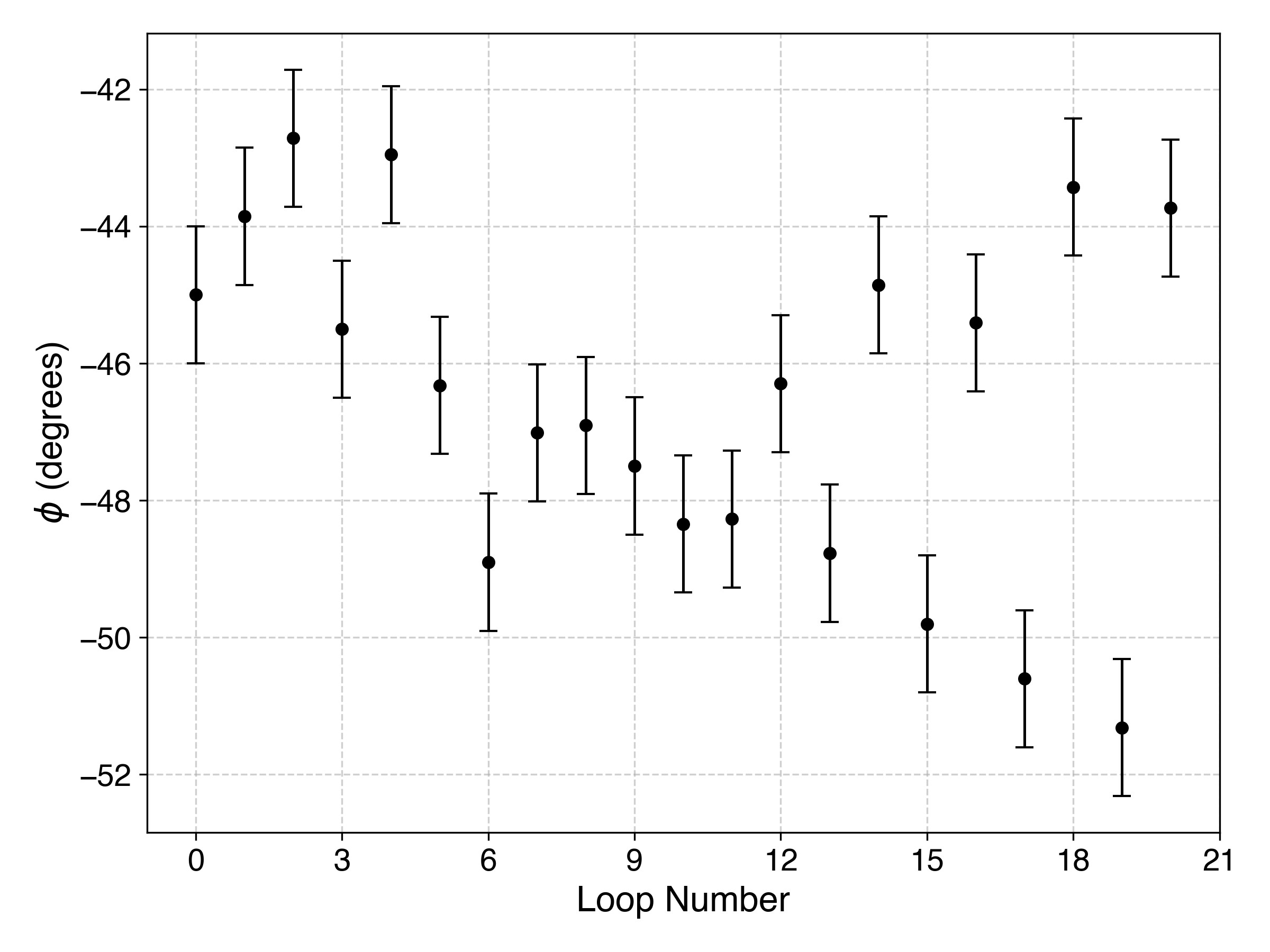}

\includegraphics[width=0.55\textwidth]{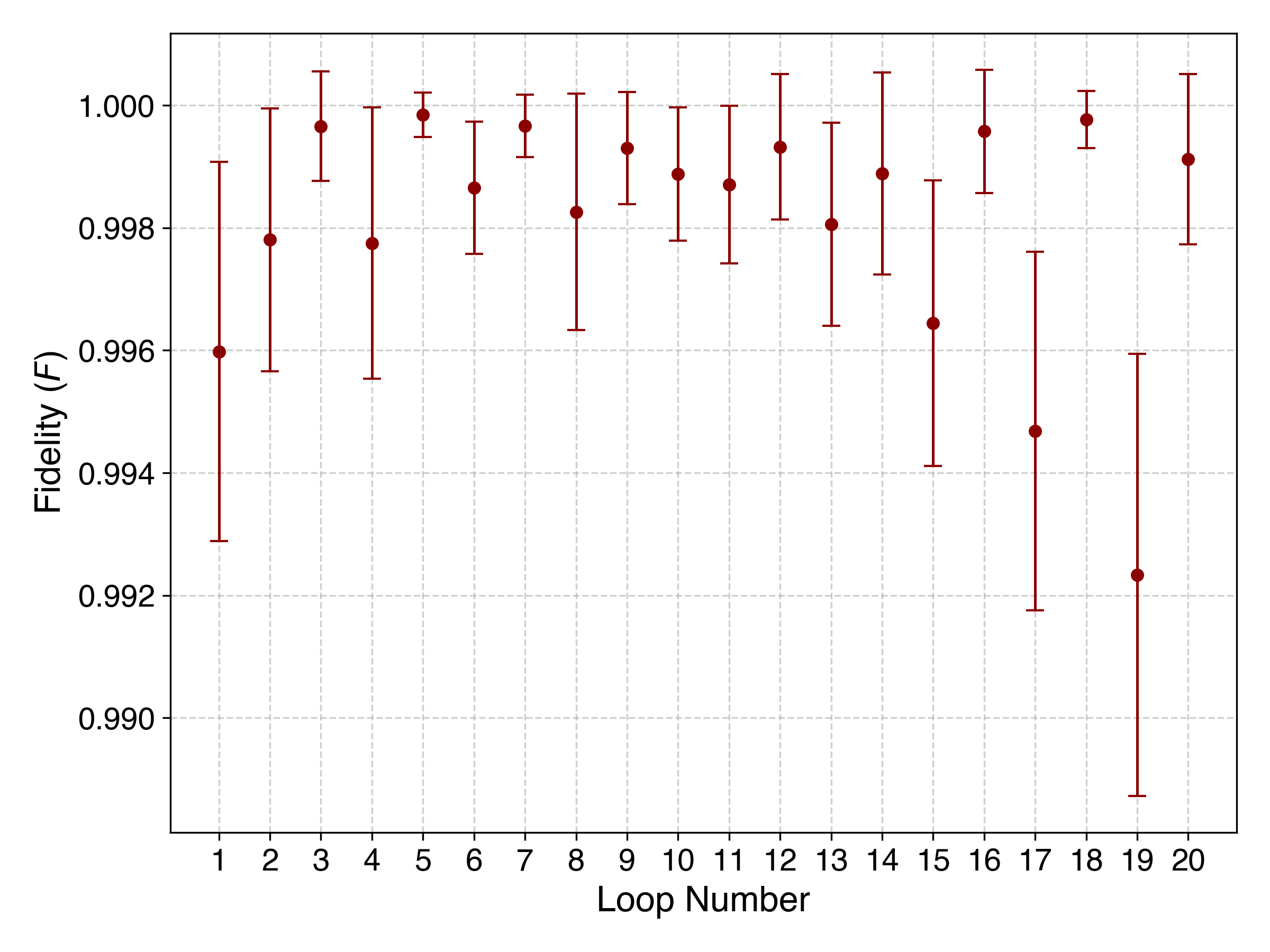}

\caption{\textbf{Polarization-bin state with nominal $\theta=45^\circ$ and $\phi=-45^\circ$.}
The panels show the reconstructed values of $\theta$, $\phi$, and $F$ with respect to the bypass reference state as a function of the memory loop number.}
\label{fig:pol_45deg_params}

\end{figure}

\begin{figure}[h!]
\centering

\includegraphics[width=1\textwidth]{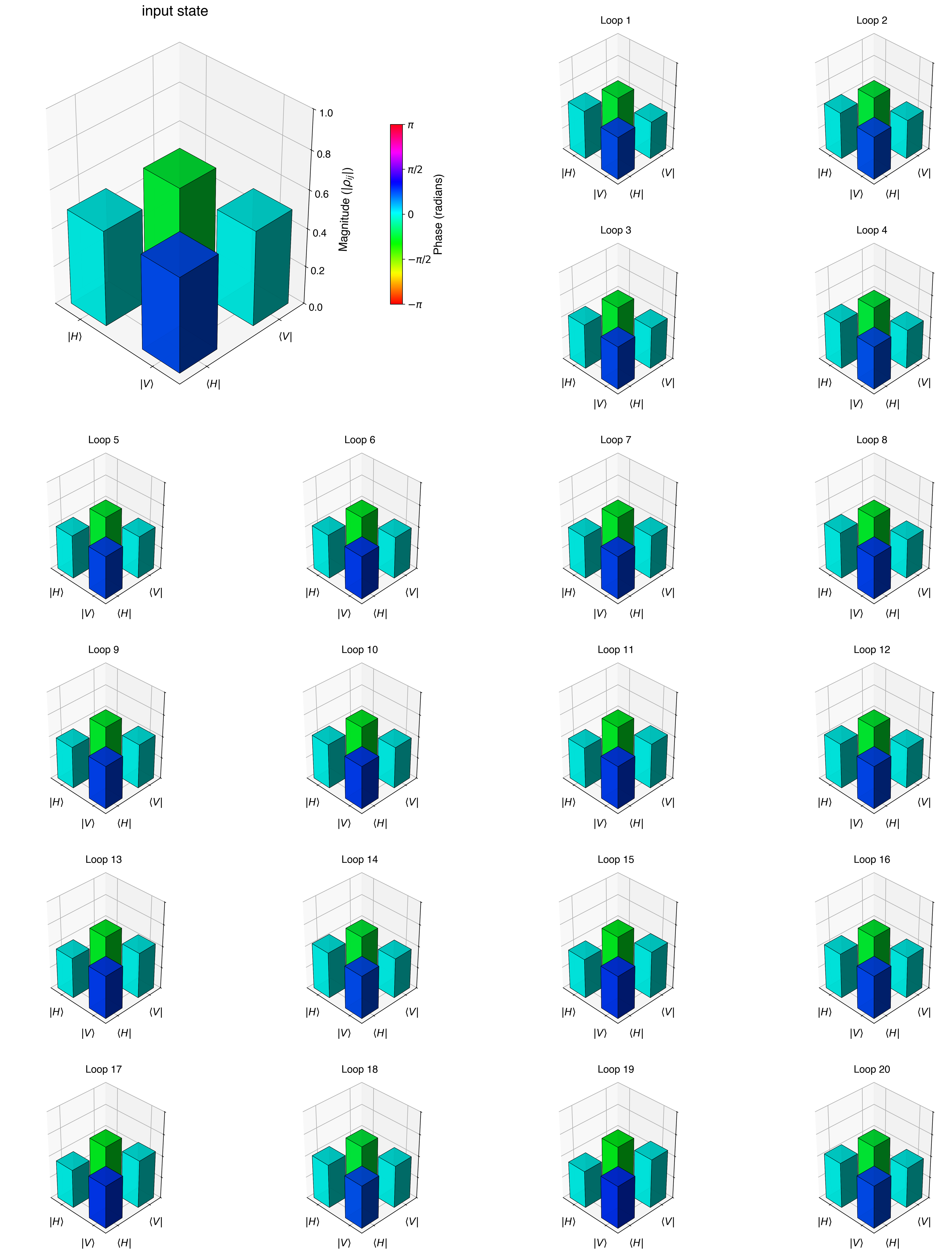}

\caption{\textbf{Density matrices for the polarization-bin state with nominal $\theta=45^\circ$ and $\phi=-45^\circ$.}
Density matrices are shown for the bypass reference state and for the states retrieved after different memory loop numbers. The reconstruction is performed in the polarization basis $\{\ket{H},\ket{V}\}$.}
\label{fig:pol_45deg_density}

\end{figure}

The state was prepared with a nominal amplitude angle $\theta=45^\circ$ and azimuthal angle $\phi=-45^\circ$. It therefore can be written as:
\begin{equation}
    \ket{\psi_\mathrm{in}} =
    \cos(45^\circ)\ket{H}
    +e^{-i45^\circ}\sin(45^\circ)\ket{V}
\end{equation}
As for the previous datasets, the retrieved state was reconstructed after different numbers of memory round trips and compared with the bypass reference state. This allowed the evolution of $\theta$, $\phi$, and $F$ to be tracked as a function of storage time (Figs.~\ref{fig:pol_45deg_params} and~\ref{fig:pol_45deg_density}).

\clearpage

\subsection{Frequency-bin states}

The frequency-bin dataset was acquired using the electro-optic-modulation configuration described in Sec.~4 of the main text. A continuous-wave laser tuned to ITU channel 21 was modulated using an electro-optic amplitude modulator driven at $f_{\mathrm{RF}} = 7~\mathrm{GHz}$. The phase modulation produces coherent frequency components separated by the modulation frequency. The frequency-bin basis states $\ket{R}$ and $\ket{B}$ are therefore defined by two optical components separated by $f_\mathrm{RF}$, whose mutual coherence gives rise to a temporal beating at $1~\mathrm{GHz}$.

The retrived state was passed through a second electro optical phase modulator to perform projective measurement and analyzed using an optical spectrum analyzer.

\begin{figure}[h!]
\centering

\includegraphics[width=0.55\textwidth]{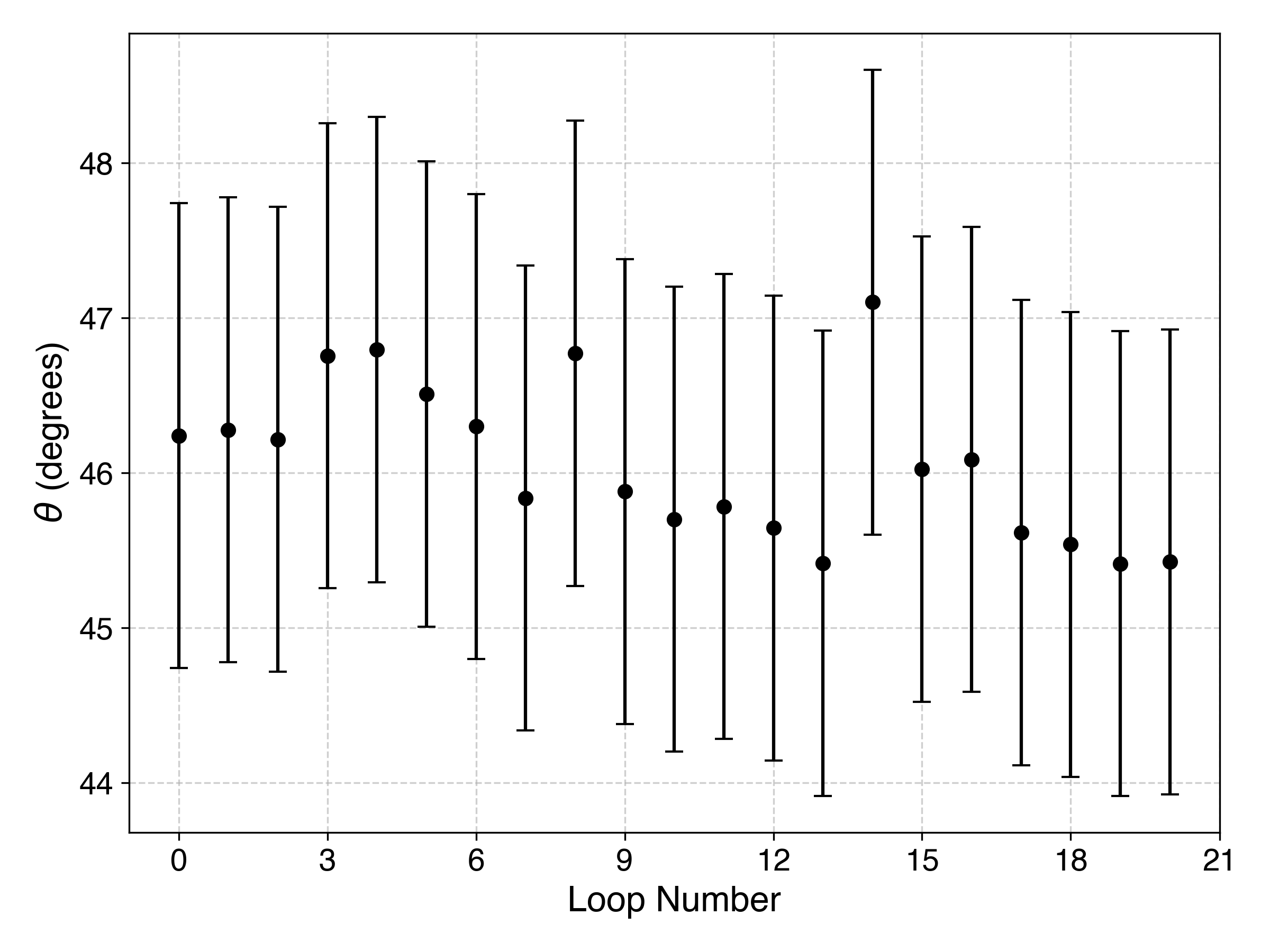}

\includegraphics[width=0.55\textwidth]{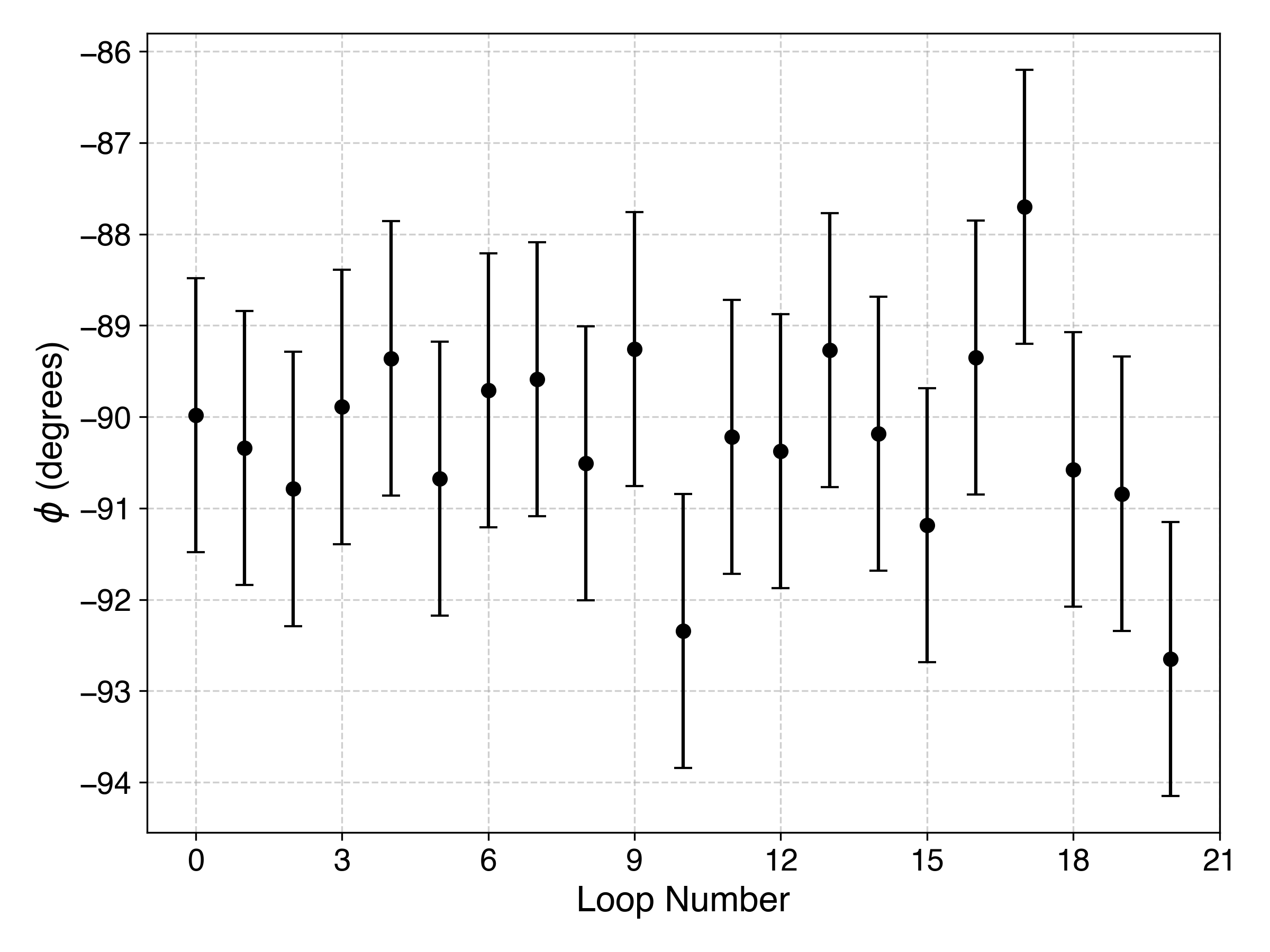}

\includegraphics[width=0.55\textwidth]{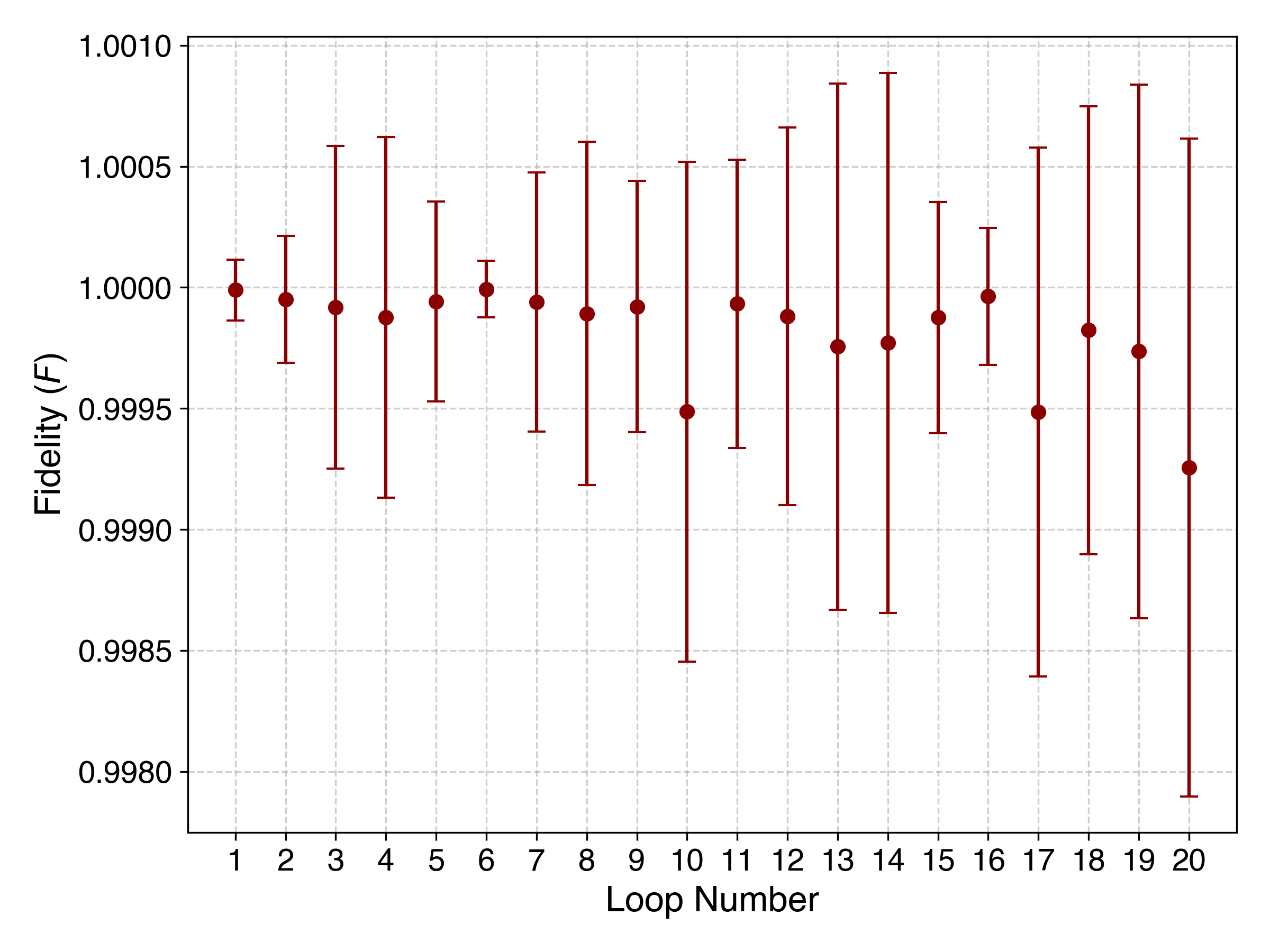}

\caption{\textbf{Frequency-bin state with nominal $\theta=46^\circ$ and $\phi=-90^\circ$.}
The panels show the reconstructed values of $\theta$, $\phi$, and $F$ with respect to the bypass reference state as a function of the memory loop number. The two frequency-bin basis states correspond to optical components separated by $7~\mathrm{GHz}$. }
\label{fig:freq_bin_params}

\end{figure}

\begin{figure}[h!]
\centering

\includegraphics[width=1\textwidth]{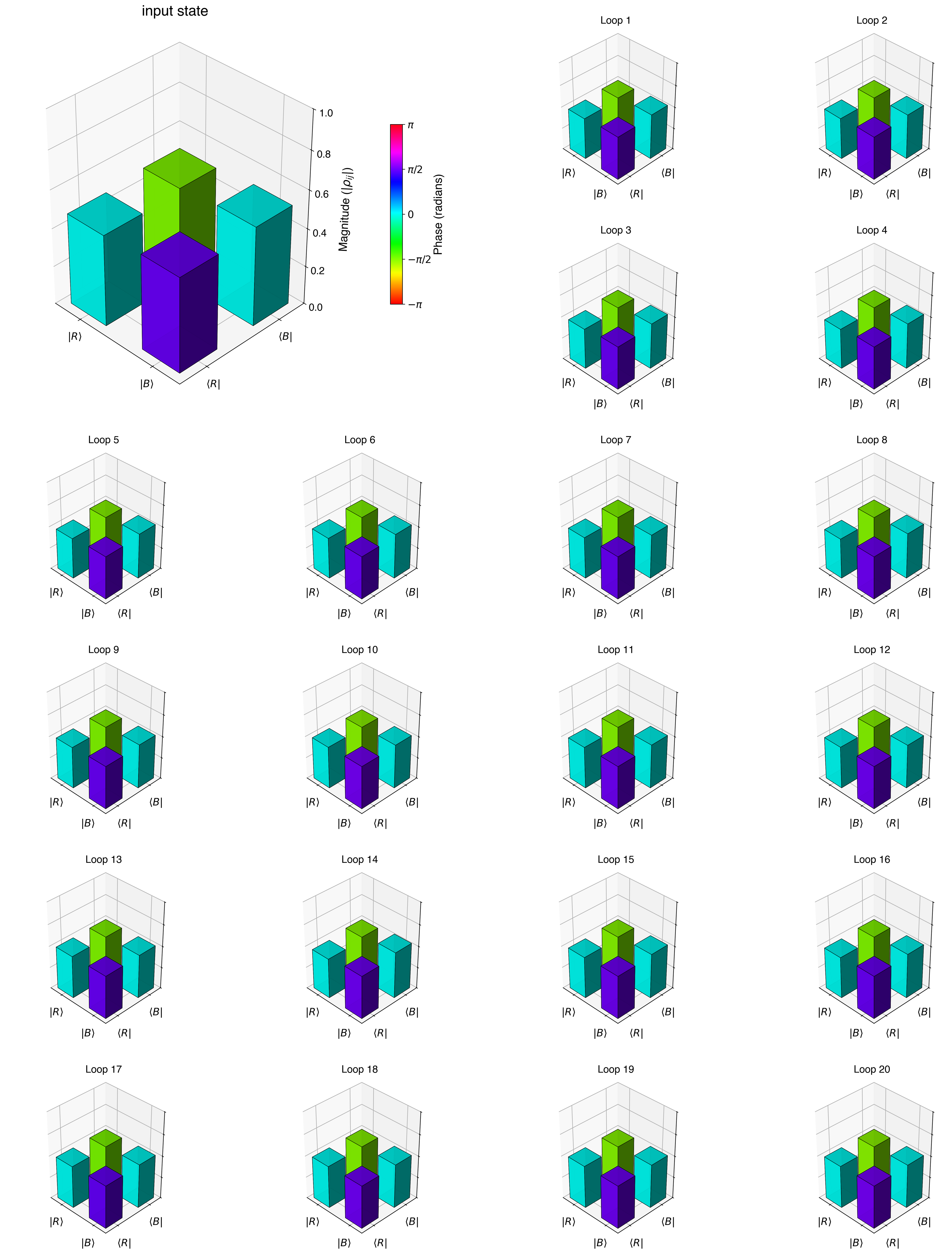}

\caption{\textbf{Density matrices for the frequency-bin state with with nominal $\theta=46^\circ$ and $\phi=-90^\circ$.}
Density matrices are shown for the bypass reference state and for the states retrieved after different memory loop numbers. The frequency-bin basis is defined by two optical components, $\ket{R}$ and $\ket{B}$, separated by $7~\mathrm{GHz}$}.
\label{fig:freq_bin_density}

\end{figure}

The nominal input state was prepared with amplitude angle $\theta=46^\circ$ and azimuthal angle $\phi=-90^\circ$ and can be written as:
\begin{equation}
    \ket{\psi_\mathrm{in}} =
    \cos(46^\circ)\ket{R}
    + e^{-i90^\circ}\sin(46^\circ)\ket{B}
\end{equation}
The state retrieved from the memory was reconstructed after different numbers of round trips and compared with the bypass reference state. The extracted values of $\theta$, $\phi$, and $F$ are reported as a function of the memory loop number in Fig.~\ref{fig:freq_bin_params}, while the corresponding reconstructed density matrices are shown in Fig.~\ref{fig:freq_bin_density}.

\clearpage

\subsection{Time-bin entangled photons}
\begin{figure}[h]
    \centering

    \begin{subfigure}[t]{0.47\textwidth}
        \centering
        \caption{Loop 1, maximum}
        \label{fig:tb_ent_loop1_max}
        \includegraphics[width=\linewidth]{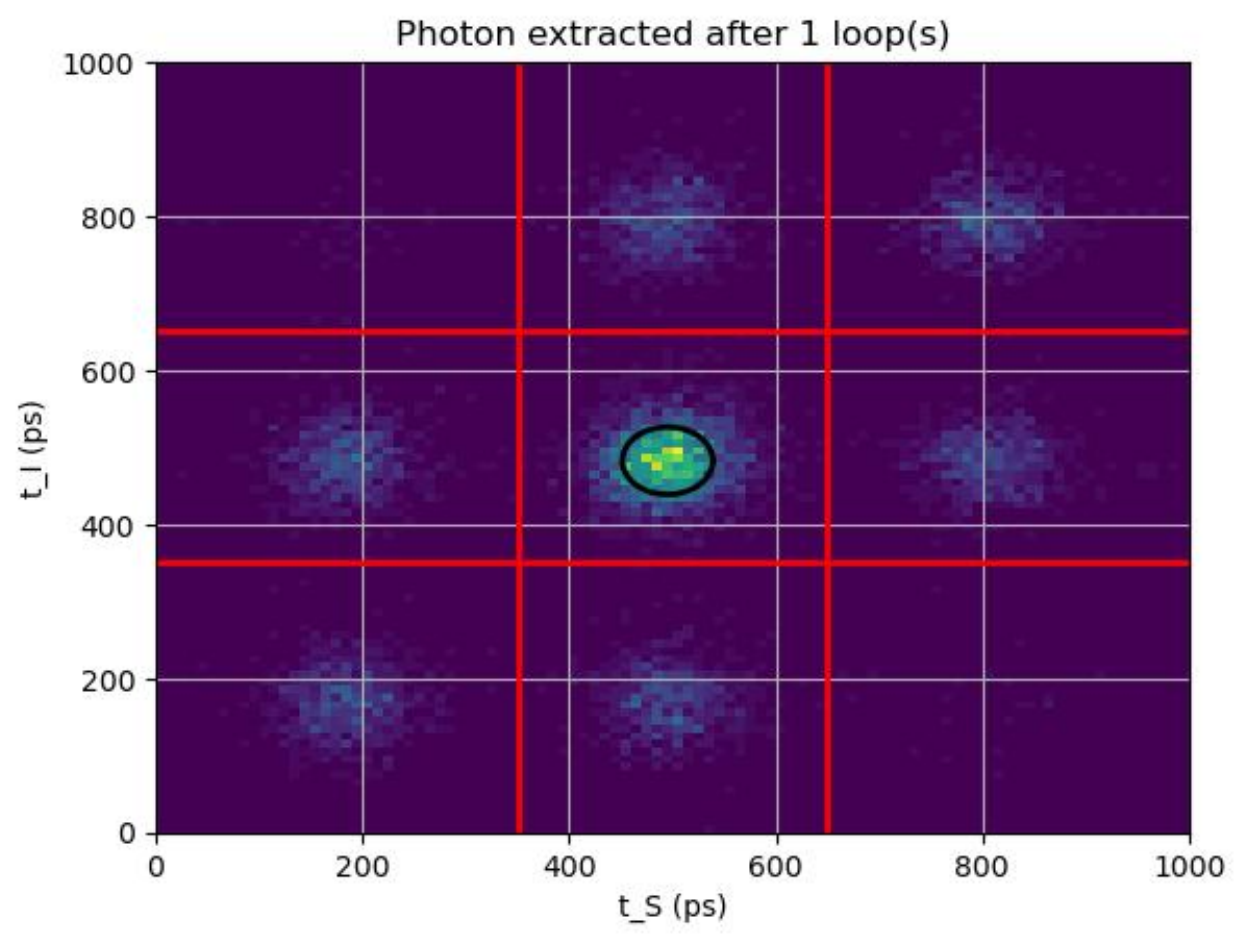}
    \end{subfigure}
    \hfill
    \begin{subfigure}[t]{0.47\textwidth}
        \centering
        \caption{Loop 1, minimum}
        \label{fig:tb_ent_loop1_min}
        \includegraphics[width=\linewidth]{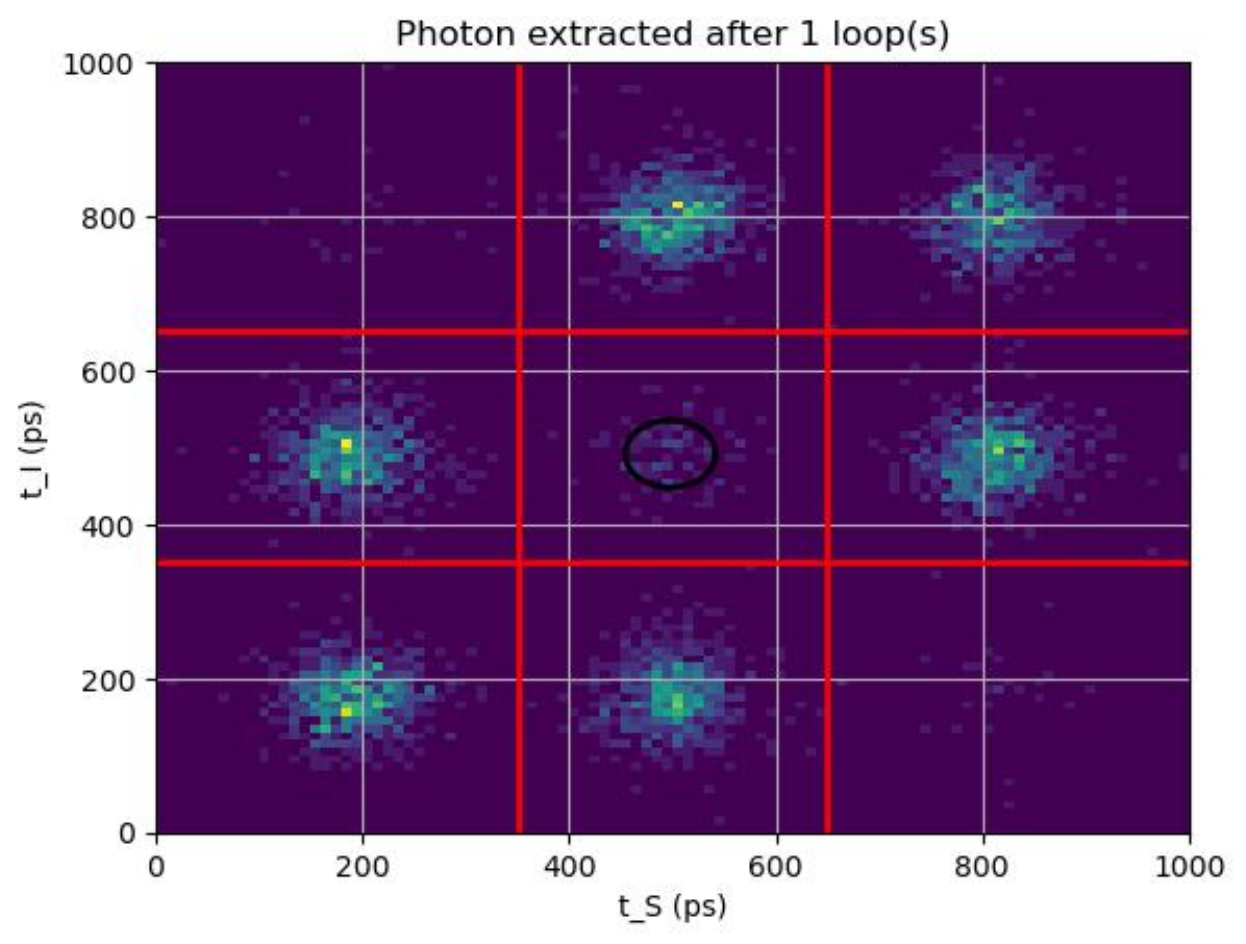}
    \end{subfigure}

    \vspace{0.3cm}

    \begin{subfigure}[t]{0.47\textwidth}
        \centering
        \caption{Loop 10, maximum}
        \label{fig:tb_ent_loop10_max}
        \includegraphics[width=\linewidth]{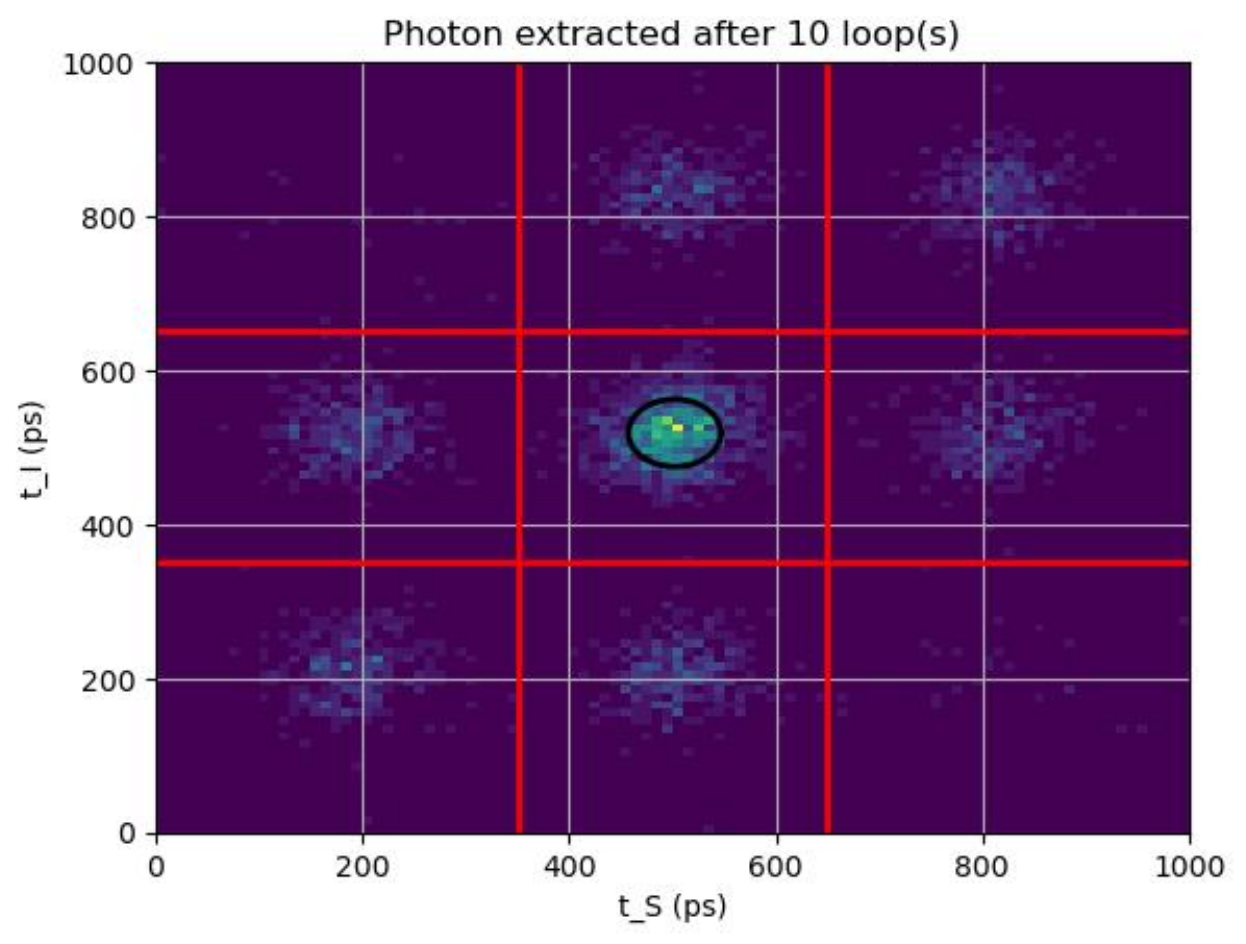}
    \end{subfigure}
    \hfill
    \begin{subfigure}[t]{0.47\textwidth}
        \centering
        \caption{Loop 10, minimum}
        \label{fig:tb_ent_loop10_min}
        \includegraphics[width=\linewidth]{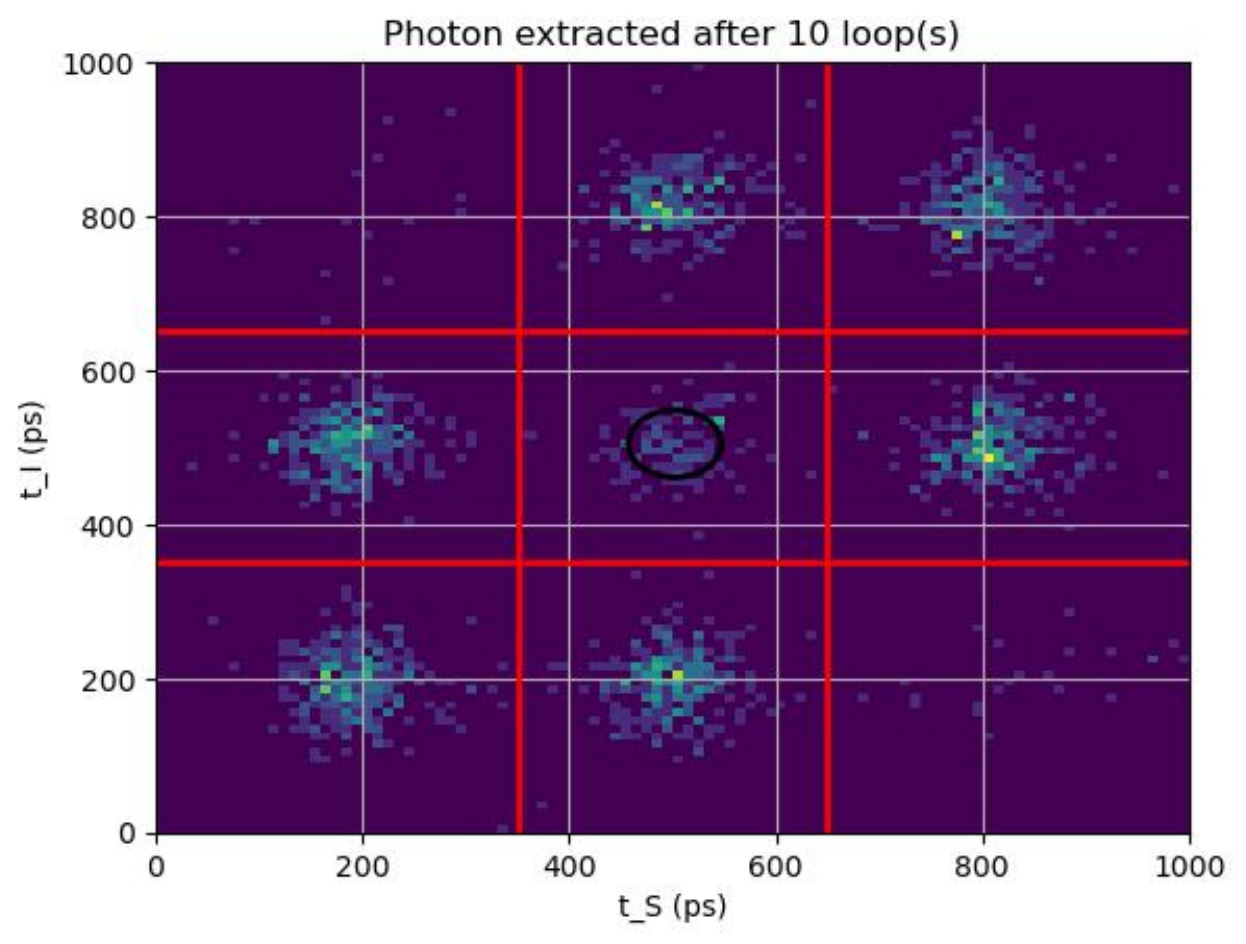}
    \end{subfigure}

    \vspace{0.3cm}

    \begin{subfigure}[t]{0.47\textwidth}
        \centering
        \caption{Loop 17, maximum}
        \label{fig:tb_ent_loop17_max}
        \includegraphics[width=\linewidth]{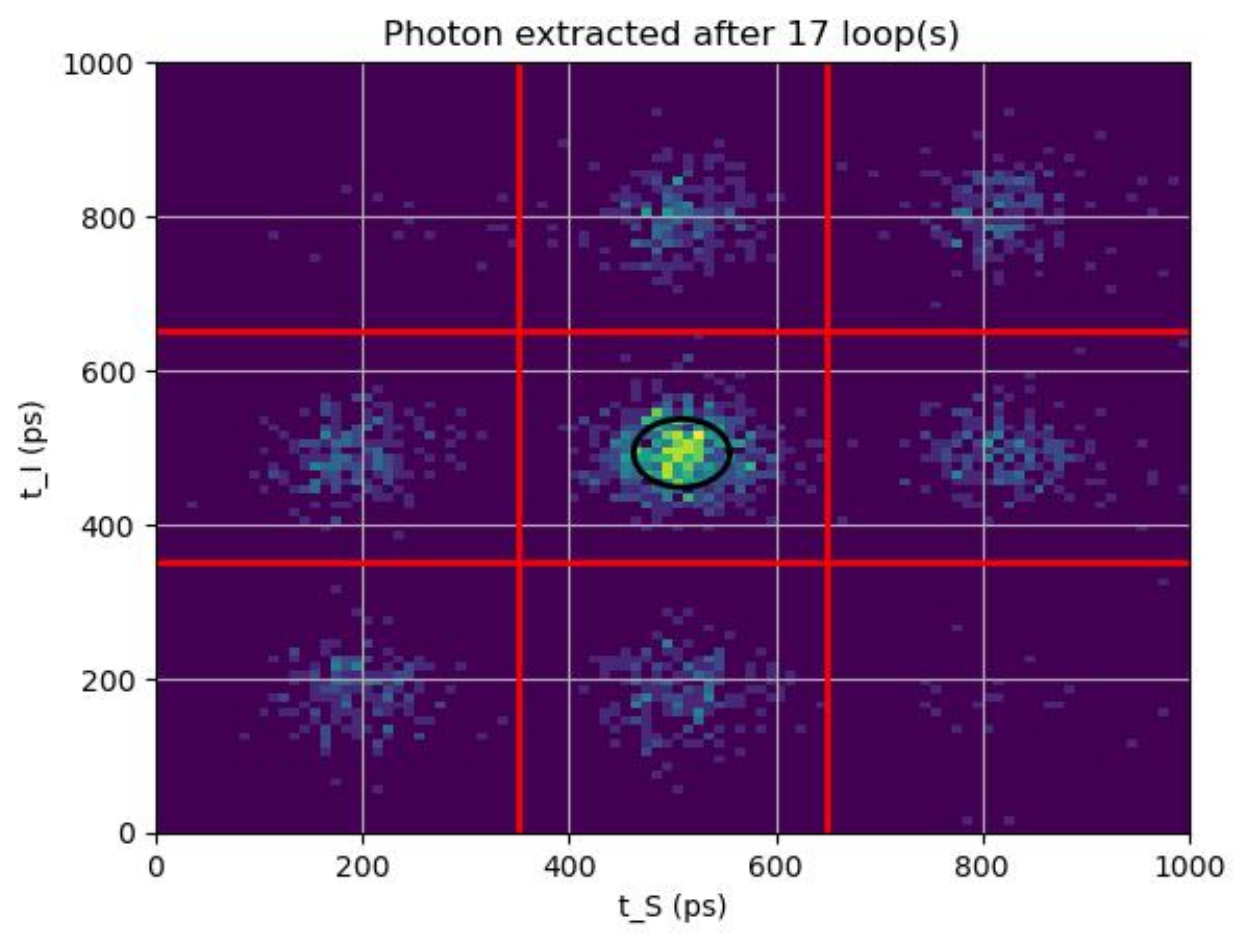}
    \end{subfigure}
    \hfill
    \begin{subfigure}[t]{0.47\textwidth}
        \centering
        \caption{Loop 17, minimum}
        \label{fig:tb_ent_loop17_min}
        \includegraphics[width=\linewidth]{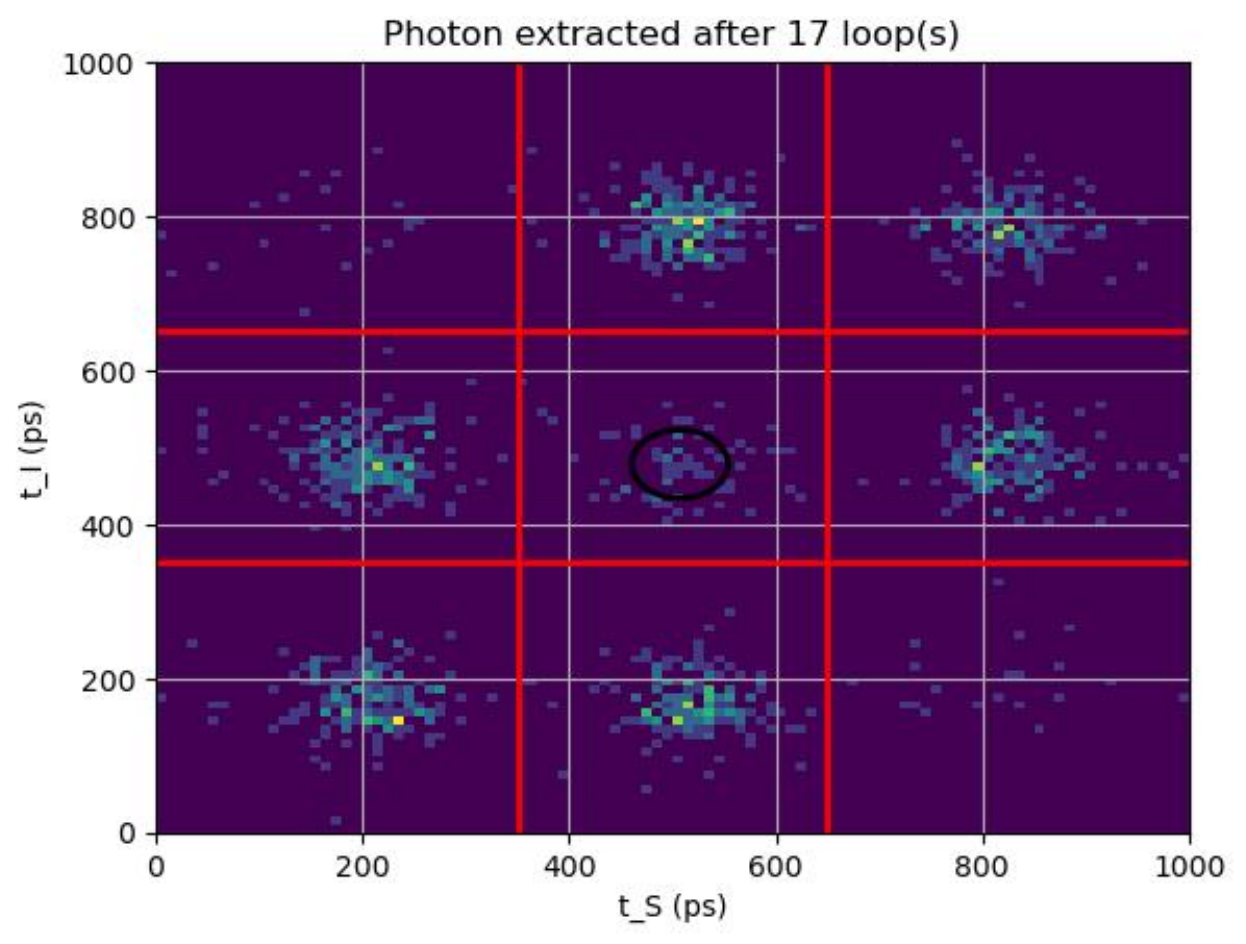}
    \end{subfigure}

    \caption{
    \textbf{Representative two-dimensional signal--idler coincidence histograms for
    time-bin entangled photons retrieved from the memory.}
    Panels (a,b), (c,d), and (e,f) correspond to signal-idler coincidences for photons retrieved
    after 1, 10, and 17 memory loops, respectively. The left column shows
    interferometer phases close to maxima of the two-photon interference fringe,
    while the right column shows phases close to minima. The horizontal and
    vertical axes correspond to the signal and idler detection times, $t_S$ and
    $t_I$, respectively. The highlighted central peak, marked by a black circle,
    corresponds to the indistinguishable two-photon quantum interference events,
    while the surrounding peaks originate from distinguishable early/late
    arrival-time combinations. The two-photon interference curves were reconstructed
    by integrating the coincidence counts within the selected central region
    marked by the red square for each value of the analysis phase.}
    \label{fig:time_bin_entanglement_maps}
\end{figure}

The time-bin entanglement dataset was acquired using the same photon-pair source and Mach--Zehnder interferometers described in the main text. In this configuration, the first interferometer defines the early and late time bins on the pump beam, while the second interferometer is used for the joint analysis of signal and idler photons. For each selected memory loop, signal--idler coincidences were recorded as a function of the relative phase between the preparation and analysis interferometers. The raw data associated with each phase setting consist of two-dimensional  coincidence histograms as a function of the signal $t_S$ and idler $t_I$ detection times.
Representative histograms are shown in Fig. \ref{fig:time_bin_entanglement_maps} for photons retrieved after 1, 10, and 17 memory loops. 
The central peak highlighted in the figures by the black circle corresponds to the indistinguishable two-photon alternatives that
give rise to Franson-type quantum interference.\footnote{\entangRefs} The other peaks arise from distinguishable combinations of early and late arrival times and are not used to extract the two-photon interference fringe.
The interference curves reported in Fig. 4 of the main text were obtained by integrating the coincidence counts within the selected region (red square) around the central peak for each value of the relative phase. To evaluate the visibility $V$ for each memory loop, the integrated central-peak coincidences were fitted as a function of the analysis phase using a sinusoidal function. The two-photon interference visibility was extracted as:
\begin{equation}
    V = \frac{C_{\mathrm{max}} - C_{\mathrm{min}}}{C_{\mathrm{max}} + C_{\mathrm{min}}}
\end{equation}
where $C_{\mathrm{max}}$ and $C_{\mathrm{min}}$ are the maximum and minimum fitted coincidence rates, respectively. Assuming the standard sinusoidal correlation expected for time-bin Franson interference, the corresponding CHSH parameter was estimated from the measured visibility as\footnote{\refSparam}:
\begin{equation}
    S = 2\sqrt{2}V
\end{equation}
As reported in the main text, the raw data were analyzed without subtracting background or accidental coincidences. Corrected visibilities were obtained only by accounting for the independently measured visibility of the analyzing interferometer.

\clearpage

\section{Temporal multimode storage}

The fiber-loop architecture naturally supports temporal multimode storage because several optical pulses separated in time can be simultaneously stored in the memory and retrieved after programmable numbers of round trips. In the main text, this capability was demonstrated by injecting a train of 25 temporal modes, generated by a pulsed laser operating at a repetition rate of 1~GHz, and selectively retrieving specific modes at different memory loops, while the remaining modes continued to circulate inside the memory. Here, we report a complementary time-domain measurement that directly visualizes selective input and delayed retrieval from the memory. In this measurement, attenuated laser pulses at ITU channel 21 were sent to the memory input with a 1~GHz repetition rate, while the input switching sequence was programmed to couple selected pulses into the fiber loop at different round trips.

\begin{figure}[h!]
    \centering

    \includegraphics[width=1\linewidth]{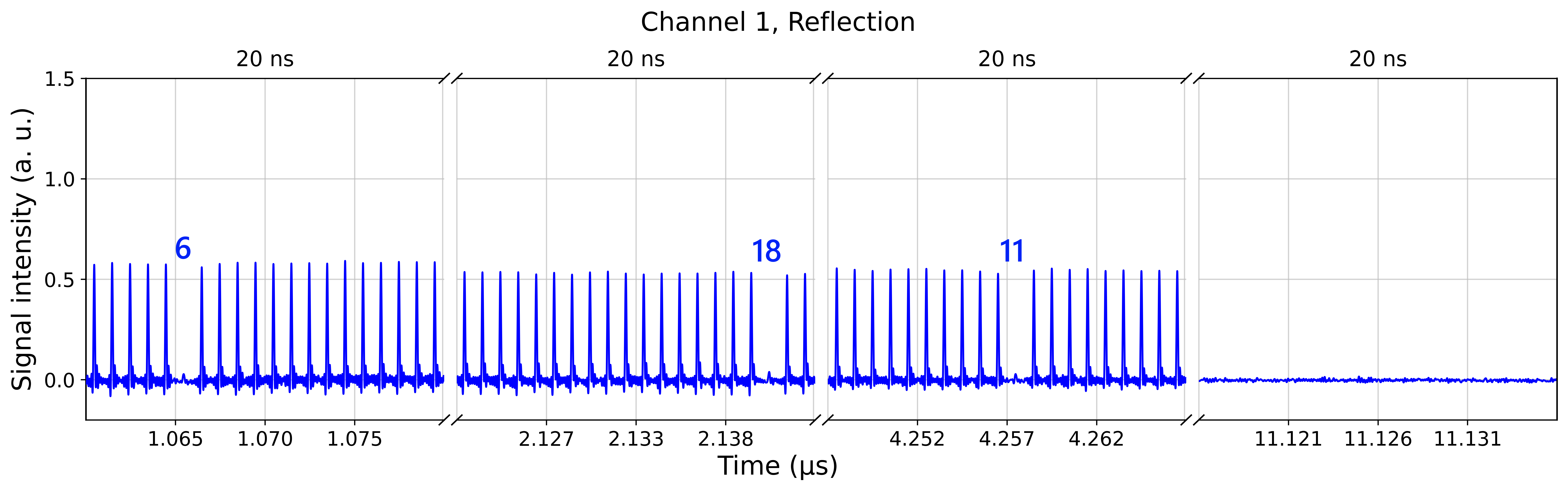}
    \hfill
    \includegraphics[width=1\linewidth]{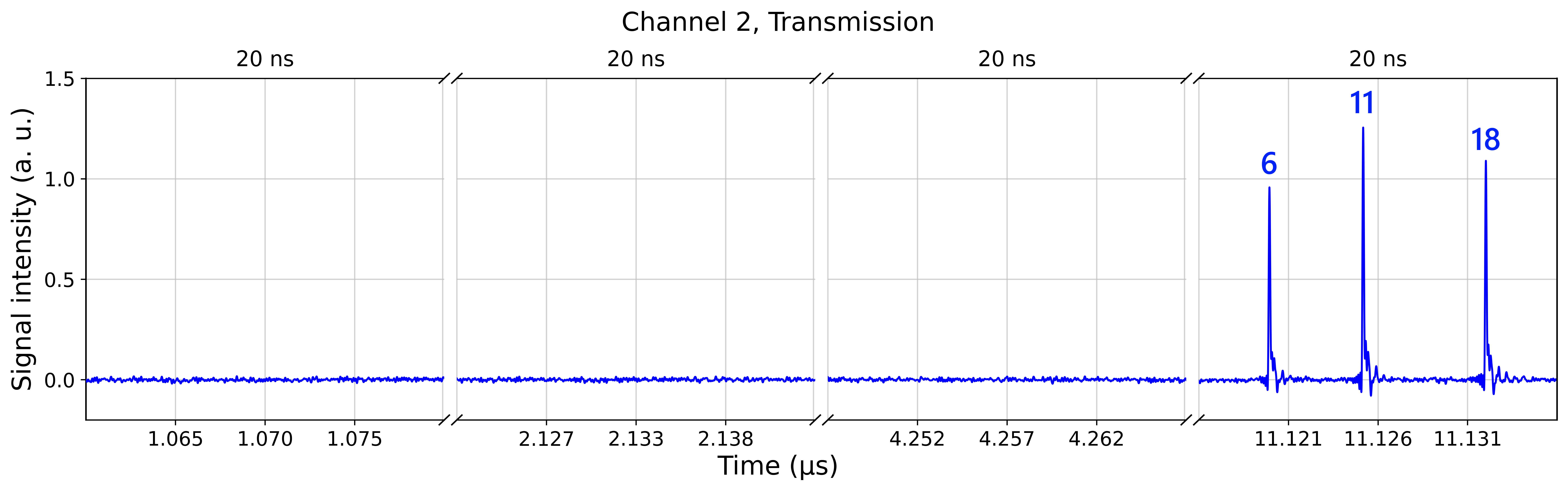}

    \caption{\textbf{Selective input and delayed retrieval of temporal modes.}
    Top: reflection trace measured at the input port of the memory during the selective loading sequence. Pulses that remain visible in reflection are not coupled into the memory, whereas missing pulses correspond to modes switched into the fiber loop. In this measurement, pulse 6 is loaded at loop 1, pulse 18 at loop 2, and pulse 11 at loop 4. Bottom: transmission trace measured at the output port. No signal is detected before the retrieval operation, while the three selected pulses are retrieved together at loop 11.}
    \label{fig:mux_loading_retrieval}
\end{figure}

The reflected and transmitted signals are shown in Fig.~\ref{fig:mux_loading_retrieval}. The pulses that were not coupled into the memory are shown in the reflection trace, whereas the selected pulses are missing because they were switched into the memory. In particular, pulse 6 was loaded at loop 1, pulse 18 was loaded at loop 2, and pulse 11 was loaded at loop 4. 
Therefore, these missing pulses identify the temporal modes selected by the input operation at different times. The corresponding transmission trace shows the delayed retrieval of the same selected modes. Before the retrieval operation, no signal is detected at the output port, confirming that the injected pulses remain stored inside the memory. When the output Sagnac interferometer S2 is opened at loop 11, the three previously loaded pulses are retrieved and appear as distinct peaks in the output trace. Their different peak amplitudes arise from the different storage times experienced inside the memory. This confirms that temporal modes loaded at different previous round trips can be stored simultaneously and retrieved during the same programmed output sequence.

This measurement complements the temporal temporal multimode storage experiment reported in the main text, where selected modes were individually retrieved at different memory loops while the remaining modes continued to circulate in the memory. Here, instead, the reflection trace directly shows the selective loading of individual temporal modes, while the transmission trace confirms their delayed retrieval after storage.

\subsection{Temporal Multimodality with  Entangled Photon Pairs}

Temporal multimodal storage was used also in the case of the experiments described in section 3 of the main text. We implement a dynamic control-pulse interleaving scheme to increase the measured coincidence rate up to 200 fold with respect to an experiment in which a single pulse was stored inside the memory. 

Table~\ref{tab_entanglemet_times} details the precise timing configurations of the injection and retrieval control pulses. 
The control pulses were configured with a duration of 20 ns to allow the storage or retrieval of 20 independent temporal modes in a single control pulse. 

\begin{figure}
    \centering
    \includegraphics[width=1\linewidth]{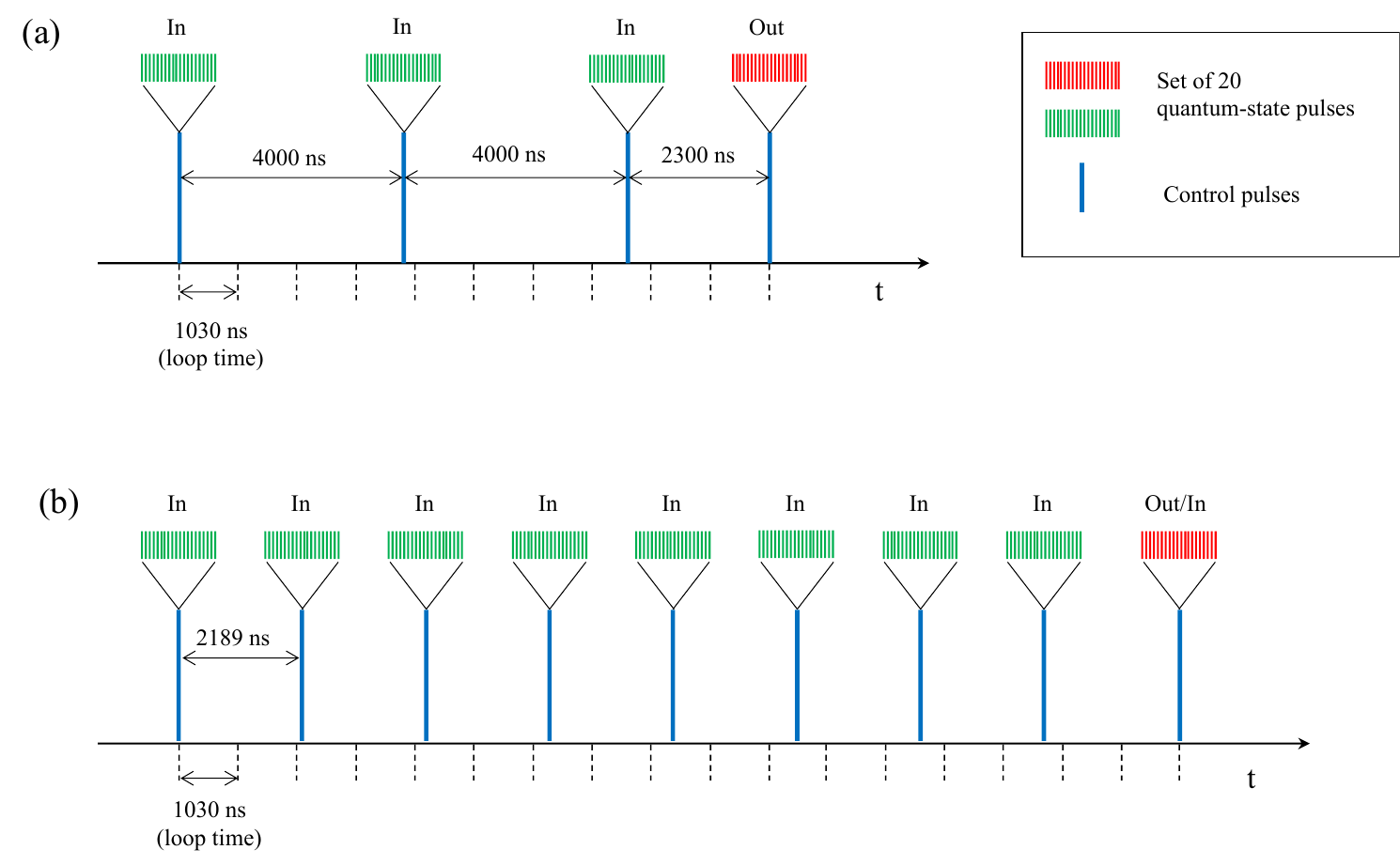}
    \caption{\textbf{Conceptual schemes for the temporal multimodality used in the entanglement experiments.} Schematic representation of the input and output of quantum state pulses as outlined in Table \ref{tab_entanglemet_times}. (a) Input/output scheme used for loop \# 10. A similar scheme was applied also in the case of loop \# 1. (b) Input/output scheme used for loop \# 17. A similar scheme was applied also in the case of loop \# 2, 5, 7, 13, 21.}
    \label{fig:Miltiplexing_scheme}
\end{figure}

We use two different storage/retrieval schemes depending on the number of loops, as schematically illustrated in Fig. \ref{fig:Miltiplexing_scheme}.
In the case of the 1st loop, a first control pulse was for storage of 20 temporal modes and a second control pulse delivered 1030 ns after the first was used for retrieval after 1 loop. This operation, which could in principle be repeated at each loop time, was repeated every 4 microseconds due to the maximum rate of pulses of our control laser.  In this case only 20 temporal modes circulate in the device.
A similar scheme was employed for the 10th loop. A first control pulse is used for storage every 4 microseconds and a second pulse is used for retrieval after 2300 ns. This 2300 ns delay is not in sync with the 20 temporal modes input by the immediately preceding control pulse. It is instead a delay commensurate with the temporal modes inserted in the memory 10300 ns before (4000 ns input pulse period times 2 + 2300 ns). This second pulse thus outputs the temporal modes that were stored for 10300 ns and have thus spent 10 loops inside the quantum memory, while the temporal modes that were stored in the two immediately preceeding input pulses continue to circulate. As a result there at least 40 temporal modes circulating in the memory at any time.

In all the other cases reported in Table \ref{tab_entanglemet_times} control pulses were sent to the memory with period $T$. If we want the quantum state pulses to be stored for $n$ loops, then the period is chosen so that $T$ is a submultiple of $n*1030$ ns. In this case a control pulse outputs from the memory quantum pulses stored $n$ loops before, and then inputs a new batch pf 20 quantum pulses. As an example in the case of the 21st loop the period is $T=$ 2163 ns, which is a submultiple of $21\times 1030$ ns and not a submultiple of $n\times 1030$ ns for $n$<21. In this case each control pulse outputs from the memory quantum pulses stored 21 loops before, and then inputs a new batch pf 20 quantum pulses. As a result 200 quantum pulses circulate within the memory at any time.

We emphasize that the capacities demonstrated in these measurements are strictly bounded by the modulation specifications of our current classical control system (timing precision, finite duration, and rise time of the control pulses), rather than by any structural limit of the memory itself. Temporal modes fully overlapping with the flat-top region of the control pulse are switched with maximum efficiency, whereas modes falling on the switching edges can be partially coupled. Therefore, by integrating multiple independent control lasers, the memory could be operated continuously with zero dead time, offering a genuinely high-capacity platform for high-speed quantum routing and network synchronization.

\begin{table}[hbt!]
\centering
\renewcommand{\arraystretch}{1.3}
\begin{tabular}{|l|l|l|l|}
\hline
\textbf{loop} & \textbf{pulse period (ns)} & \textbf{2nd pulse delay (ns)} & \textbf{stored temporal bins} \\
\hline
\textbf{1}  & 4000 & 1030 & 20 \\
\hline
\textbf{2}  & 2060 &  & 20    \\
\hline
\textbf{5}  & 2575 &  & 40    \\
\hline
\textbf{7}  & 3605 &  & 40    \\
\hline
\textbf{10} & 4000 & 2300 & 60 \\
\hline
\textbf{13} & 2232 &  & 100    \\
\hline
\textbf{17} & 2189 &  & 160    \\
\hline
\textbf{21} & 2163 &   & 200 \\
\hline
\end{tabular}
\caption{\textbf{Timing parameters of the entanglement experiments and multiplexing of the quantum memory.}
The table details the operation sequence for time-bin multiplexing at various storage loops. For each loop, the programmed control-pulse sequence defines packets of 20 usable quantum time bins, obtained after excluding the temporal modes partially affected by the finite switching edges. The "pulse period" represents the repetition interval at which groups of time bins are input to the memory. The "2nd pulse delay" dictates the extraction time for specific loop configurations where independent input/retrieve control pulses were utilized. The total number of circulating modes ("stored bins") is reported in the last column.}
\label{tab_entanglemet_times}
\end{table}

\clearpage

\end{document}